\documentclass[amsmath,amssymb,nobibnotes,prx,twocolumn,reprint]{revtex4-1}
\usepackage[utf8]{inputenc}
\usepackage[T1]{fontenc}
\usepackage{color}
\usepackage{hyperref}
\usepackage{geometry}
\usepackage{graphicx}
\usepackage{dcolumn}
\usepackage{float}
\usepackage{epsfig}
\usepackage{comment}
\usepackage{cleveref}
\usepackage{placeins}
\usepackage{balance}
\usepackage[percent]{overpic}

\newcommand{\added}[1]{\textcolor{black}{#1}}
\begin{document}

\title{Thermally Activated Non-Affine Rearrangements in Amorphous Glass: Emergence of Intrinsic Length Scales}
\author{Avinash Kumar Jha}
\affiliation{Department of Physics, Indian Institute of Technology Roorkee, Roorkee, Uttarakhand 247667, India}

\begin{abstract}

We present a systematic study of temperature-driven \emph{nonaffine} rearrangements in a model amorphous solid across the full thermodynamic range, from a high-temperature liquid, through supercooled and sub-glass regimes, into \added{deep} glassy states. The central result is a quantitative characterisation of the componentwise nonaffine residual displacements, obtained by subtracting local affine maps from particle displacements. For each state point the tails of the probability distributions of these nonaffine components display clear exponential decay; linear fits to the logarithm of the tail region yield characteristic nonaffine length scales \(\xi_{\mathrm{NA},x}\) and \(\xi_{\mathrm{NA},y}\), which quantify the spatial extent of purely nonaffine, local rearrangements. To compare with other length scales, we compute van Hove distributions \(G_x(u_x),\,G_y(u_y)\) which capture the full particle displacement field (coherent affine-like motion plus residuals). A robust, key finding is that the van Hove length scale consistently exceeds the filtered nonaffine length scale, i.e.\ $\xi_{\mathrm{VH}}>\xi_{\mathrm{NA}}$, across all temperatures, state points, and densities we studied. This ordering remains robust under temporal-sampling protocols used in this work. The nonaffine length $\xi_{\mathrm{NA}}$ quantifies the distance over which \emph{complex} deformation occurs, specifically nonlinear and anharmonic responses, irreversible (plastic) rearrangements, topological non-recoverable particle rearrangements, and other residual motions that cannot be represented by a local affine map. Because these mechanically relevant, higher-order contributions remain after subtracting the affine component, $\xi_{\mathrm{NA}}$ is a physically distinct (and systematically smaller) length scale than the van Hove measure, which contains both affine and nonaffine contributions. \added{Our companion analytical derivation explicitly verifies this inequality result \(\xi_{\mathrm{VH}}>\xi_{\mathrm{NA}}\) and therefore provides a direct, model-level explanation of the numerical observation}. Thus \(\xi_{\mathrm{NA}}\) provides a targeted, mechanically meaningful measure of the spatial extent of truly nonaffine, nonlinear events that are not captured by standard van Hove analyses. Moreover, near equality of $\xi_{\mathrm{NA},x}$ and $\xi_{\mathrm{NA},y}$ in all conditions provides further evidence that nonaffine rearrangements propagate isotropically under thermally driven deformation in contrast to externally driven shear. Overall, our work establishes a novel framework for quantifying temperature‐driven local rearrangements in amorphous materials and provides fresh insight into the interplay between thermal fluctuations, structural disorder, and mechanical stability.
\end{abstract}
\maketitle
\section{Introduction}

The microscopic dynamics of disordered materials (glasses, amorphous solids and dense liquids) are governed by heterogeneous, spatially correlated particle rearrangements that differ qualitatively from the simple oscillations about lattice sites found in crystals \cite{bennett2018liquid,kirchner2022beyond,ossi2024disordered}.  In addition to the well studied global measures such as mean square displacement (MSD), relaxation times and viscosity \cite{angell2000relaxation,debenedetti2001supercooled,angell1988structural,larini2008universal,wang2012elastic,mezard2002statistical}, componentwise displacement statistics provide a direct, local probe of how particles move and rearrange. A central object in this context is the (self) van Hove distribution of the total particle displacement $\mathbf{u}$, denoted by $G(\mathbf{u},t)$, and its Cartesian projections $G_x(u_x,t)$, $G_y(u_y,t)$, which have been extensively used to quantify heterogeneous dynamics and non-Gaussian tails in supercooled liquids and glasses \cite{van1954correlations,hansen2006theory,berthier2011theoretical,wang2012brownian,bhowmik2018non,chaudhuri2007universal}.

A complementary perspective focuses on \emph{nonaffine} displacements: the portion of particle motion that cannot be described by a local affine map (linear deformation) and therefore reflects genuinely local, plastic or anharmonic rearrangements \cite{falk1998dynamics,chikkadi2011long,nicolas2018deformation,jensen2014local}. Nonaffine measures are widely used in the mechanics literature to locate shear transformation zones, quantify local yield events, and characterize mechanical heterogeneity\cite{argon1979plastic,maloney2004subextensive,lemaitre2009rate,eshelby1957determination,lulli2018metastability,manning2011vibrational,jung2025roadmap}. Most prior work has considered nonaffinity under external deformation\cite{falk1998dynamics,maloney2004subextensive,lemaitre2009rate,shi2005strain,hatano2008scaling,bailey2007avalanche,langer2001microstructural,tanguy2002continuum,eshelby1957determination,spaepen1977microscopic,argon1979plastic,langer2003dynamics,picard2004elastic,maloney2004subextensive,maloney2006amorphous,lemaitre2009rate,baret2002extremal,nicolas2018deformation,chikkadi2011long,patinet2016connecting,barbot2018local,talamali2011avalanches,budrikis2017universal,lin2014scaling,demkowicz2005autocatalytic,liu2016driving}; less attention has been paid to thermally driven nonaffine rearrangements that arise solely from temperature and structural disorder in the absence of applied strain.

In this paper, we bring these two threads together. We analyze both (i) the componentwise nonaffine residuals distributions of \(u^{\mathrm{NA},x},\,u^{\mathrm{NA},y}\) obtained from local affine fits, measured across a wide range of temperatures and densities, and (ii) the signed componentwise van Hove distributions $G_x(u_x)$, $G_y(u_y)$. Methodologically, we make two deliberate choices that clarify the physical content of each measure: for the van Hove functions we keep the signed distributions so that coherent, affine-like motion and its directionality are retained; for the nonaffine residuals we analyze the \emph{folded} (absolute) distributions in order to quantify the magnitude of purely \emph{thermal} nonaffine rearrangements (direction is not the primary object of interest). Importantly, for symmetric distributions folding changes only the prefactor and not the exponential slope of the tails, so the extracted length scale is invariant under folding for symmetric data.

Our study uses two temporal-sampling protocols to demonstrate robustness. First, we analyze \emph{consecutive} frame pairs taken from trajectories stored in a log-linear frame scheme as described later. Second, in the supercooled regime we additionally select frame pairs separated by the structural relaxation time \(\tau_\alpha\). Both protocols produce consistent qualitative trends and allow us to test whether the extracted length scales depend sensitively on the sampling interval \(\Delta t\). We find that while details of the distributions change with \(\Delta t\) (as expected), the principal ordering of length scales is robust.

The central empirical finding is simple and reliably observed: the characteristic length scale extracted from the van Hove tails, \(\xi_{\mathrm{VH}}\), always exceeds the length scale extracted from the purely nonaffine, folded residuals, \(\xi_{\mathrm{NA}}\) (i.e. \(\xi_{\mathrm{VH}}>\xi_{\mathrm{NA}}\)) across all temperatures, densities and temporal sampling protocols we examined. Physically, this ordering reflects that the signed van Hove distribution contains both long-wavelength, coherent (affine-like) contributions and local, higher-order (nonaffine) contributions, whereas the nonaffine residual isolates the latter.

Temperature dependence of the nonaffine length scale \(\xi_{\mathrm{NA},x}\) and \(\xi_{\mathrm{NA},y}\) (the characteristic lengths extracted from the nonaffine \emph{x} and \emph{y} component distributions, respectively) reveals three distinct regimes: rapid growth at high temperature, a near-plateau in an intermediate range, and a marked decline in the deeply glassy state where particle caging dominates. Fitting the temperature dependence in semi-log space yields two interpretable parameters: baseline length \(\xi_{0,\mathrm{NA}}\) (extrapolated low-\(T\) limit) and a thermal sensitivity \(\alpha_{\mathrm{NA}}\), both of which systematically decrease with increasing density. We also verify isotropy of the nonaffine response, \(\xi_{\mathrm{NA},x} \!\approx\! \xi_{\mathrm{NA},y}\), across the state points examined.
Contributions of this work are therefore threefold. First, we introduce a careful, reproducible protocol to extract componentwise nonaffine length scales from unconstrained thermal trajectories and to compare them directly with van Hove length scales. Second, we demonstrate the robustness of the ordering \(\xi_{\mathrm{VH}}>\xi_{\mathrm{NA}}\) across temperatures, densities and sampling choices, and we provide analytical arguments to interpret this ordering in terms of affine versus higher-order non-linear (non-affine) contributions to particle motion. Third, by quantifying the temperature and density dependence of \(\xi_{\mathrm{NA}}\), we provide a new scalar measure of the spatial extent of mechanically relevant, nonlinear and irreversible rearrangements in disordered solids.\\
 The remainder of the paper is organized as follows. In Sec.~\ref{sec:methods} we describe the simulation details, the protocols used to identify characteristic temperatures at each density. Sec.~\ref{sec:theory_methods} presents the main analytical results: the physical significance of the nonaffine displacement, the mathematical formulation that relates length scales to displacement variances, and the definitions used in the computations. In Sec.~\ref{sec:nonaffine_results} we report the folded (absolute-value) Cartesian distributions of the nonaffine residuals, the extraction of the corresponding length scales and fit parameters, and their density dependence. Sec.~\ref{sec:vanhove_vs_nonaffine} contains the signed Cartesian van Hove distributions, the von Hove length-scale extraction, and a detailed comparison between von Hove and nonaffine length scales across time and temperature scales. We conclude in Sec.~\ref{sec:conclusion} with implications for experimental measurements and for the mechanical characterisation of amorphous solids.

\section{Simulation Details and Thermodynamic Characterization}
\label{sec:methods}
   \begin{align}
    U_{jk}(r)&=4\epsilon\left[\left(\frac{\sigma_{jk}}{r_{jk}}\right)^{12}-\left(\frac{\sigma_{jk}}{r_{jk}}\right)^{6} +\frac{1}{4}\right],\;\; \frac{r_{jk}}{\sigma_{jk}} < 2^{\frac{1}{6}} \nonumber\\
    &=0 \;\; \mbox{otherwise.}
\end{align}
This is the definition of Weeks-Chandler-Andersen potential \cite{tanaka2010critical,filion2011simulation}. Here $\epsilon$ gives the energy scale, lengths are reported in unit of the mean particle diameter $\langle\sigma\rangle$, temperature in unit of \(\epsilon/k_{\mathrm{B}}\) ( \(k_{\mathrm{B}}=1\)) and $\sigma_{jk}=(\sigma_j +\sigma_k)/2$ where $\sigma_j$ is the diameter of the $j^{th}$ particle.
We performed NVT molecular dynamics simulation using an isokinetic thermostat \cite{brown1984comparison} implemented by Brown and Clarke. We have chosen system size $N=2000$. Here we use the Gaussian distribution of particle size called polydisperse particles. Polydispersity is defines as $\Delta=\sqrt{(\langle \sigma ^2 \rangle-\langle \sigma \rangle^2)}/\langle \sigma \rangle$. Here we set average diameter size $\langle \sigma \rangle =1.0$ and $\Delta =0.11$. For each density, the initial condition is well-equilibrated high-temperature liquid, from there it is directly quenched to the desired temperature, and thereafter equilibration and production run was done. Equilibration run for supercooled to all lower temperatures is kept constant called waiting time $t_w=945000$. Here 10 -12 independent runs are taken for all analysis.
 Here we present a table \ref{tab:Temp} of characteristic temperatures found from different means at three different densities:
 \begin{table}[h!]
    \centering
    \begin{tabular}{|c|c|c|c|c|c|c|c|c|c|c|c|}
        \hline
        Density & $T_{g,\tau_\alpha}$ & $T_{g,D}$  & $T_{VFT}$& $T_{onset}$  \\
        \hline
        0.900     & 0.168 & 0.160     & 0.062    & 0.265  \\
        \hline
        0.950   & 0.343 & 0.331    & 0.178    & 0.679  \\
        \hline
        1.000   & 0.596 & 0.571      & 0.385    & 1.483   \\
        \hline
        
    \end{tabular}
    \caption{Density dependence of several characteristic temperatures for the 2DPC (two-dimensional polydisperse colloidal) system. Here $T_{g,\tau_{\alpha}}$ is thermodynamical glass transition temperature obtained from VFT(Vogel-Fulcher-Tammann) fit of alpha relaxation time obtained from self intermediate interaction function(FSKT), $T_{g,D}$ is glass transition temperature obtained from diffusion coefficient (D). $T_{VFT}$ is the temperature obtained from VFT fit. $T_{onset}$ is the onset temperature below which supercooled liquid exists.} 
    \label{tab:Temp}
\end{table}
 
 \section{Theoretical definitions and methodology}
\label{sec:theory_methods}

This section introduces notation, the per-particle nonaffine residual, the Taylor expansion that motivates the affine subtraction, the Cartesian van Hove distributions used for comparison, the folded (absolute-value) distribution of nonaffine field, and the variance–tail relations that explain why the van Hove length scale typically exceeds the nonaffine length scale. Full line-by-line algebra and detailed derivation are provided in  Appendix \ref{app:math}--\ref{app:folded_gen}.

\subsection{Notation}
Reference positions are denoted by capital letters 
$\vec{X}_i = (X_i, Y_i)$, 
current (deformed) positions by lowercase letters 
$\vec{x}_i = (x_i, y_i)$, 
and displacements by 
$\vec{u}_i = \vec{x}_i - \vec{X}_i$ for partcle $i$. 
In all equations used in this paper, the Greek subscripts $\alpha,\beta,\gamma,\delta$ denote Cartesian components, and the comma notation (e.g., $u^{\alpha}{}_{,\beta}$) represents differentiation with respect to reference position $X^{\beta}$. Also for notational brevity we use two equivalent forms for componentwise affine and nonaffine fields: $u_{i,x}^{(\mathrm{A})}$ and $u_i^{\mathrm{A},x}$ denote the same affine $x$-component for particle $i$, and likewise $u_{i,x}^{(\mathrm{NA})}$ and $u_i^{\mathrm{NA},x}$ denote the same nonaffine $x$-component (analogous substitutions apply for the $y$-component).
Thus, the Cartesian components of the displacement vector $\vec{u}_i$ of particle $i$ are defined as
\begin{equation}
\label{eq:component_def}
u_{i,x} \equiv x_i - X_i, 
\qquad 
u_{i,y} \equiv y_i - Y_i
\end{equation}

The reference and current separations between two paticles $i$ and $j$ are written in index notation as
\begin{equation}
\label{eq:Delta_def}
\Delta X_{ij}^{\alpha} \equiv X_{j}^{\alpha} - X_{i}^{\alpha}, 
\qquad 
\Delta x_{ij}^{\alpha} \equiv x_{j}^{\alpha} - x_{i}^{\alpha}
\end{equation}
where the Greek index $\alpha$ denotes the Cartesian components 
($\alpha = 1,2$ in two dimensions).

The corresponding relative displacement differences are
\begin{equation}
\label{eq:pair_differences}
\begin{aligned}
\delta u_{ij,x} &\equiv u_{j,x} - u_{i,x}\\[4pt]
\delta u_{ij,y} &\equiv u_{j,y} - u_{i,y}
\end{aligned}
\end{equation}

Hence, the current (deformed) separation is related to the reference separation and displacement difference as
\begin{equation}
\label{eq:relation_def}
\Delta x_{ij}^{\alpha} = \Delta X_{ij}^{\alpha} + \delta u_{ij}^{\alpha}
\end{equation}

\subsection{Local affine fit and the nonaffine residual}
We compute the best local affine map $\mathbf{H}$ (a $2\times2$ matrix) that maps reference separations to observed relative displacements in a least-squares sense ~\cite{jha2025thermal}. The componentwise nonaffine residual used in this work is 
\begin{align}
\label{eq:d2min_component_def}
u_i^{\mathrm{NA},x}
&=
\frac{1}{N_i}\sum_{j\in N_i}
\Big[\,\Delta x_{ij}
- \big(H_{xx}\,\Delta X_{ij} + H_{xy}\,\Delta Y_{ij}\big)\,\Big] \nonumber\\[6pt]
u_i^{\mathrm{NA},y}
&=
\frac{1}{N_i}\sum_{j\in N_i}
\Big[\,\Delta y_{ij}
- \big(H_{yx}\,\Delta X_{ij} + H_{yy}\,\Delta Y_{ij}\big)\,\Big]
\end{align}

where $x,y$ represent Cartesian coordinates and $H_{xx}$, $H_{xy}$, $H_{yx}$, and $H_{yy}$ denote the components of $\mathbf{H}$ and $N_i$ is the number of neighbors of particle $i$ upto second nearest neighbors, calculated from the pair correlation function $g(r)$.
By construction, this quantity vanishes if the local displacement field is exactly affine to the deformation gradient tensor $\mathbf{H}$; the computation of \( \mathbf{H} \) follows the procedure of Ref.~\cite{jha2025thermal}. 

\subsection{Taylor expansion of the displacement field (up to cubic order)}
Assume the displacement field $\vec{u}(\vec{X})$ is smooth on the neighbour scale. Expanding $u_\alpha(\vec{X}_j)$ (the $\alpha$-component of the displacement of particle $j$) about $\vec{X}_i$ gives (indices $\alpha,\beta,\gamma,\delta\in\{X,Y\}$ and repeated indices summed):
\begin{align}
\label{eq:taylor_expansion}
u_\alpha(\vec{X}_j) = u_\alpha(\vec{X}_i)
+ u_{\alpha,\beta}(\vec{X}_i)\,\Delta X_{ij}^{\beta}\nonumber\\[4pt]
+ \tfrac{1}{2}\,u_{\alpha,\beta\gamma}(\vec{X}_i)\,\Delta X_{ij}^{\beta}\Delta X_{ij}^{\gamma}\nonumber\\[4pt]
+ \tfrac{1}{6}\,u_{\alpha,\beta\gamma\delta}(\vec{X}_i)\,\Delta X_{ij}^{\beta}\Delta X_{ij}^{\gamma}\Delta X_{ij}^{\delta}
+ O(\|\Delta X_{ij}\|^4)
\end{align}

Taking the difference between $j$ and $i$ using equation~\ref{eq:pair_differences} and ~\ref{eq:relation_def} for the $x$-component:
\begin{equation}
\label{eq:difference_taylor}
\begin{aligned}
\Delta x_{ij} - \Delta X_{ij} 
= u_{x,\beta}\,\Delta X_{ij}^{\beta}
+ \tfrac{1}{2}\,u_{x,\beta\gamma}\,\Delta X_{ij}^{\beta}\Delta X_{ij}^{\gamma}\\[4pt]
\quad + \tfrac{1}{6}\,u_{x,\beta\gamma\delta}\,\Delta X_{ij}^{\beta}\Delta X_{ij}^{\gamma}\Delta X_{ij}^{\delta}
+ O(\|\Delta X_{ij}\|^4)
\end{aligned}
\end{equation}

so in component form we can write
\begin{equation}
\label{eq:difference_taylor2}
\begin{aligned}
\Delta x_{ij}^{\alpha} 
&= (\delta^{\alpha\beta} + u^{\alpha}{}_{,\beta})\,\Delta X_{ij}^{\beta}
+ \tfrac{1}{2}\,u^{\alpha}{}_{,\beta\gamma}\,\Delta X_{ij}^{\beta}\Delta X_{ij}^{\gamma}\\[4pt]
&\quad + \tfrac{1}{6}\,u^{\alpha}{}_{,\beta\gamma\delta}\,\Delta X_{ij}^{\beta}\Delta X_{ij}^{\gamma}\Delta X_{ij}^{\delta}
+ O(\|\Delta X_{ij}\|^4)
\end{aligned}
\end{equation}
because the deformation gradient tensor is defined \cite{jha2025thermal} as
\begin{equation}
H^{\alpha\beta} = \delta^{\alpha\beta} + u^{\alpha}{}_{,\beta}
\end{equation}

Putting this into the above equation ~\ref{eq:difference_taylor2}
\begin{equation}
\label{eq:difference_taylor3}
\begin{aligned}
\Delta x_{ij}^{\alpha} - H^{\alpha\beta}\,\Delta X_{ij}^{\beta}
= \tfrac{1}{2}\,u^{\alpha}{}_{,\beta\gamma}\,\Delta X_{ij}^{\beta}\Delta X_{ij}^{\gamma}\\[4pt]
\quad + \tfrac{1}{6}\,u^{\alpha}{}_{,\beta\gamma\delta}\,\Delta X_{ij}^{\beta}\Delta X_{ij}^{\gamma}\Delta X_{ij}^{\delta}+ O(\|\Delta X_{ij}\|^4)
\end{aligned}
\end{equation}
where all derivatives are evaluated at $\vec{X}_i$.
Substituting equation~\eqref{eq:difference_taylor3} into the nonaffine definition given in equation~\eqref{eq:d2min_component_def} yields the following expression (for detailed derivation, see Appendix~\ref{app:aff_nf_sec}).

\begin{equation}
\label{eq:nonaffine_series}
\begin{aligned}
\sum_{j\in N_i}\left(\Delta x_{ij}^{\alpha} - H^{\alpha\beta}\,\Delta X_{ij}^{\beta}\right)
=\sum_{j\in N_i}\Bigl[\tfrac{1}{2}\,u^{\alpha}{}_{,\beta\gamma}\,\Delta X_{ij}^{\beta}\Delta X_{ij}^{\gamma}\\
\quad + \tfrac{1}{6}\,u^{\alpha}{}_{,\beta\gamma\delta}\,\Delta X_{ij}^{\beta}\Delta X_{ij}^{\gamma}\Delta X_{ij}^{\delta}
+ O(\|\Delta X_{ij}\|^4)\Bigr]
\end{aligned}
\end{equation}

From Eq.~\ref{eq:nonaffine_series}, it is explicitly evident that, once the best local linear (affine) mapping is subtracted, the remaining term represents the \emph{nonlinear, higher-order nonaffine} component of the local displacement field. This residual contains second- and higher-order spatial derivatives of the displacement field, corresponding to anharmonic, asymmetric, and topological rearrangements that cannot be captured by deterministic affine maps. Consequently, the probability distribution of these residuals directly reflects the magnitude and spatial extent of purely nonaffine, unsystematic motions
(for which no single local deformation matrix \(\mathbf{H}\) exists (unlike affine deformations that can be described by a well-defined single \(\mathbf{H}\)), and which arise from structural disorder and thermal fluctuations. The characteristic length scale extracted from the exponential tail of this distribution quantifies the spatial range over which complex, irreversible rearrangements occur. Since this exponential tail represents the dominant region of the deformation, our analysis specifically focuses on this tail portion. \emph{The equation \ref{eq:nonaffine_series} thus forms the central theoretical foundation of the present work: it defines the nonaffine residual that filters out the affine (linear) response, isolates higher-order, nonlinear dynamics, and serves as the basis for constructing the nonaffine length scale analyzed across all temperatures, densities, and thermodynamic regimes in this study}.

\subsection{Cartesian van Hove self-distribution}
The signed Cartesian van Hove self-distribution for the $x$-component, which captures the complete spectrum of particle displacements (including both affine and nonaffine contributions), is defined as 
\begin{equation}
\label{eq:van_hove_x}
\begin{split}
  G_x(u_x,t) &= \Big\langle \frac{1}{N}\sum_{i=1}^N 
\delta\!\big(u_x - [x_i(t)-X_i]\big)\Big\rangle \\
&= \Big\langle \frac{1}{N}\sum_{i=1}^N 
\delta\!\big(u_x - u_{i,x}(t)\big)\Big\rangle
\qquad u_x\in\mathbb{R}
\end{split}
\end{equation}
and analogously for the $y$-component,
\begin{equation}
\label{eq:van_hove_y}
\begin{split}
G_y(u_y,t) &= \Big\langle \frac{1}{N}\sum_{i=1}^N 
\delta\!\big(u_y - [y_i(t)-Y_i]\big)\Big\rangle \\
&= \Big\langle \frac{1}{N}\sum_{i=1}^N 
\delta\!\big(u_y - u_{i,y}(t)\big)\Big\rangle
\qquad u_y\in\mathbb{R}
\end{split}
\end{equation}
 Here, $N$ denotes the total number of particles, and the angular brackets indicate ensemble averaging over all temporal origins and independent realizations. The van Hove self-distribution $G_x(u_x,t)$ and $G_y(u_y,t)$ thus provide a complete measure of particle displacement statistics along the corresponding Cartesian directions over a time lag $t$, encompassing both coherent (affine-like) motion and localized nonaffine rearrangements. It serves as the reference framework in this literature for characterizing dynamical heterogeneity in amorphous materials, \emph{against which we compare our newly defined nonaffine residual distributions to isolate the purely nonlinear, irreversible component of motion}. 
\subsection{Folded (magnitude) distribution}
We denote by \(u\) the signed displacement (\(u\in\mathbb{R}\)). For the folded (absolute-value) distribution, we introduce the nonnegative magnitude variable
\[
r \equiv |u|,\qquad r \ge 0
\]
and write the folded PDF as \(p_{\mathrm{abs}}(r)\). With this convention, the folded distribution is given by the standard change of variables
\[
p_{\mathrm{abs}}(r) = p(r) + p(-r),\qquad r \ge 0
\]
and, if \(p(u)\) is symmetric, \(p_{\mathrm{abs}}(r)=2p(r)\) \emph{\added{in the absence of any external loading or any directional bias}}; folding therefore changes only the multiplicative prefactor and not the exponential tail slope or length scale. \added{The excellent numerical agreement between the length scales obtained from signed-tail fits and the folded-tail fit shown explicitly in Figs.~\ref{fig:signed} and~\ref{fig:folded} and therefore confirms this analytical expectation}. Appendix~\ref{app:fold_sec} gives a step-by-step normalisation check. After this explicit definition of \(r\equiv|u|\), all existing expressions that use \(p_{\mathrm{abs}}(\cdot)\) or \(p_{\mathrm{abs}}(r)\) remain valid without further modification.

\subsection{Laplace-tail hypothesis and variance-to-length mapping}
Our numerical results indicate that the tails of both the folded nonaffine and the van Hove distributions are well described by a symmetric Laplace (double-exponential) form for the single component displacement $u$:
\begin{equation}
\label{eq:laplace_tail}
p(u)\approx \frac{1}{2\xi}\exp\!\big(-|u|/\xi\big),\qquad |u|\gg\text{(core radius)}
\end{equation}

The variance (second central moment) of the Laplace distribution \eqref{eq:laplace_tail} is
\begin{equation}
\label{eq:laplace_variance}
\langle u^2\rangle \;=\; 2\xi^{2}
\end{equation}
where we note that, because the Laplace distribution is symmetric about zero, its mean is \(\langle u\rangle=0\). Hence the variance \(\mathrm{Var}(u)=\big\langle\big(u-\langle u\rangle\big)^2\big\rangle\) coincides with the second moment \(\langle u^2\rangle\).
So, the tail length is related to the variance by

\begin{equation}
\label{eq:xi_def}
\xi \;=\; \sqrt{\frac{\langle u^2\rangle}{2}}
\end{equation}
The integral that yields \eqref{eq:laplace_variance} is given in Appendix~\ref{app:laplace}.

\subsection{Variance decomposition and why $\xi_{\mathrm{VH}}>\xi_{\mathrm{NA}}$ typically}

Decompose a single-component displacement $u$ as 
\begin{equation} \label{eq:u_decomp}
u = u^{(A)} + u^{(NA)} 
\end{equation} 
Here \(u^{(A)}\) is the affine component and \(u^{(NA)}\) is the nonaffine component of the displacement, since any displacement can be decomposed into affine and nonaffine parts.

Under the Laplace mapping \eqref{eq:laplace_variance}, the squared length scales are proportional to the corresponding variances of the displacement. Thus
\begin{equation}
\label{eq:xi_definitions}
\begin{aligned}
\xi_{\mathrm{VH}}^{2} &= \frac{\langle u^{2}\rangle}{2}\\[4pt]
\xi_{\mathrm{A}}^{2}  &= \frac{\langle (u^{(A)})^{2}\rangle}{2}\\[4pt]
\xi_{\mathrm{NA}}^{2} &= \frac{\langle (u^{(\mathrm{NA})})^{2}\rangle}{2}
\end{aligned}
\end{equation}
Here, \(\xi_{\mathrm{VH}}\) denotes the van Hove length scale corresponding to the total displacement field,  
\(\xi_{\mathrm{A}}\) is the affine length scale associated with the affine component of the displacement field, and  
\(\xi_{\mathrm{NA}}\) is the nonaffine length scale capturing fluctuations arising from nonaffine rearrangements.
From this relation, the ratio between the van Hove and nonaffine length scales can be expressed as shown below. The detailed derivation of this expression is provided in Appendix~\ref{app:affine_nonaffine}.

\begin{equation}
\label{eq:xi_ratio}
\frac{\xi_{\mathrm{VH}}}{\xi_{\mathrm{NA}}}
= \sqrt{\frac{\langle (u^{(A)})^{2}\rangle + \langle (u^{(\mathrm{NA})})^{2}\rangle
}{\langle (u^{(\mathrm{NA})})^{2}\rangle}}
\end{equation}
\emph{
Equation~\ref{eq:xi_ratio} represents a key analytical result that substantiates our computational finding that the van Hove length scale, $\xi_{\mathrm{VH}}$, is greater than the nonaffine length scale, $\xi_{\mathrm{NA}}$. This follows directly from Equation~\ref{eq:xi_ratio}, which shows that the van Hove measure encompasses both affine and nonaffine components of displacement, whereas the nonaffine measure isolates only the nonlinear residual part. Consequently, the numerator in Equation~\ref{eq:xi_ratio} is larger, leading to the inequality $\xi_{\mathrm{VH}} > \xi_{\mathrm{NA}}$. This analytical relation provides a firm theoretical basis for the observed hierarchy of length scales in our results.
}

\subsection{Temporal sampling protocols}
\label{subsec:temporal_sampling}

\subsubsection{Block-based log--linear saving scheme of configurations.}
Trajectories are recorded using a block-based, log--linear saving scheme. Each trajectory is divided into contiguous time blocks; within each block we save a fixed number \(M\) of configurations. The \(m\)-th saved configuration in block \(b\) is written at an MD-step offset (measured from the start of that block) as
\[
S_m = 2^{\,m-1}\qquad (m=1,\ldots,M)
\]
i.e. at \(1,2,4,8,\dots\) MD-step offsets from the block start. Denoting the MD-step index where block \(b\) begins by \(T_b\) (blocks are contiguous so \(T_{b+1}=T_b+L_b\), with \(L_b\) the duration of block \(b\) in MD steps). The absolute MD-step index of the \(m\)-th saved configuration in block \(b\) is therefore
\[
\mathcal{S}_{b,m} \;=\; T_b + (S_m - 1)
\]
The gap (in MD steps) between consecutive saved frames inside a block is
\[
\Delta S_m = S_{m+1}-S_m = 2^{\,m-1}
\]
so the physical time gap is \(\Delta t_m=\Delta S_m\,dt = 2^{\,m-1}dt\), where \(dt\) is the MD timestep. Because saved offsets follow a power-of-two spacing within blocks and blocks are contiguous, the global saved-sequence provides dense sampling at short times and exponentially increasing gaps at longer times; consecutive saved frames may lie inside the same block or straddle a block boundary.

All displacement statistics reported here are formed from \emph{frame pairs}. We employ two complementary temporal-sampling protocols to ensure robustness to the choice of lag:
\paragraph{Default (consecutive-frame) protocol.}
We form pairs from \emph{consecutive} entries of the global, log--linear saved-sequence (so pairs that straddle block boundaries are included). This parameter-free protocol automatically samples a continuous, non-uniform family of \(\Delta t\) values (dense at short times, extending to very long lags) and therefore (i) applies uniformly across temperatures and densities (including deeply glassy states where a single \(\tau_\alpha\) may be ill-defined) and (ii) captures both thermal motion and rare rearrangements without the need to choose an external lag.

\paragraph{Fixed-lag protocol.} 
The second protocol targets a fixed physical time lag $\Delta t$ in the supercooled regime: we select all saved frame pairs whose MD-step separation matches the structural relaxation time $\tau_{\alpha}$. Here $\tau_{\alpha}$ is determined using the standard dynamical criterion based on the decay of the self-intermediate scattering function, defined as the time at which it decays to $1/e$ of its initial value, following established practice in glass-forming systems \cite{kob1995testing,berthier2011theoretical}. In some studies, $\tau_{\alpha}$ is alternatively extracted from stretched-exponential (Kohlrausch--Williams--Watts) fits to the same correlation function; both procedures yield comparable relaxation times in the supercooled regime \cite{alvarez1991relationship,williams1970non,kawasaki2014structural,van1994glass,glotzer2000spatially}. When the $\tau_{\alpha}$-matched protocol is used, it is stated explicitly in the text and figure captions; otherwise, consecutive (log-linear) frame spacing is assumed.

\par\medskip  

\noindent\textit{\small
Note: Unless otherwise stated, figures and numerical results use the default
consecutive (log–linear) frame spacing.}

\begin{figure}[htb]
  \centering
  \begin{overpic}[width=\columnwidth,keepaspectratio]{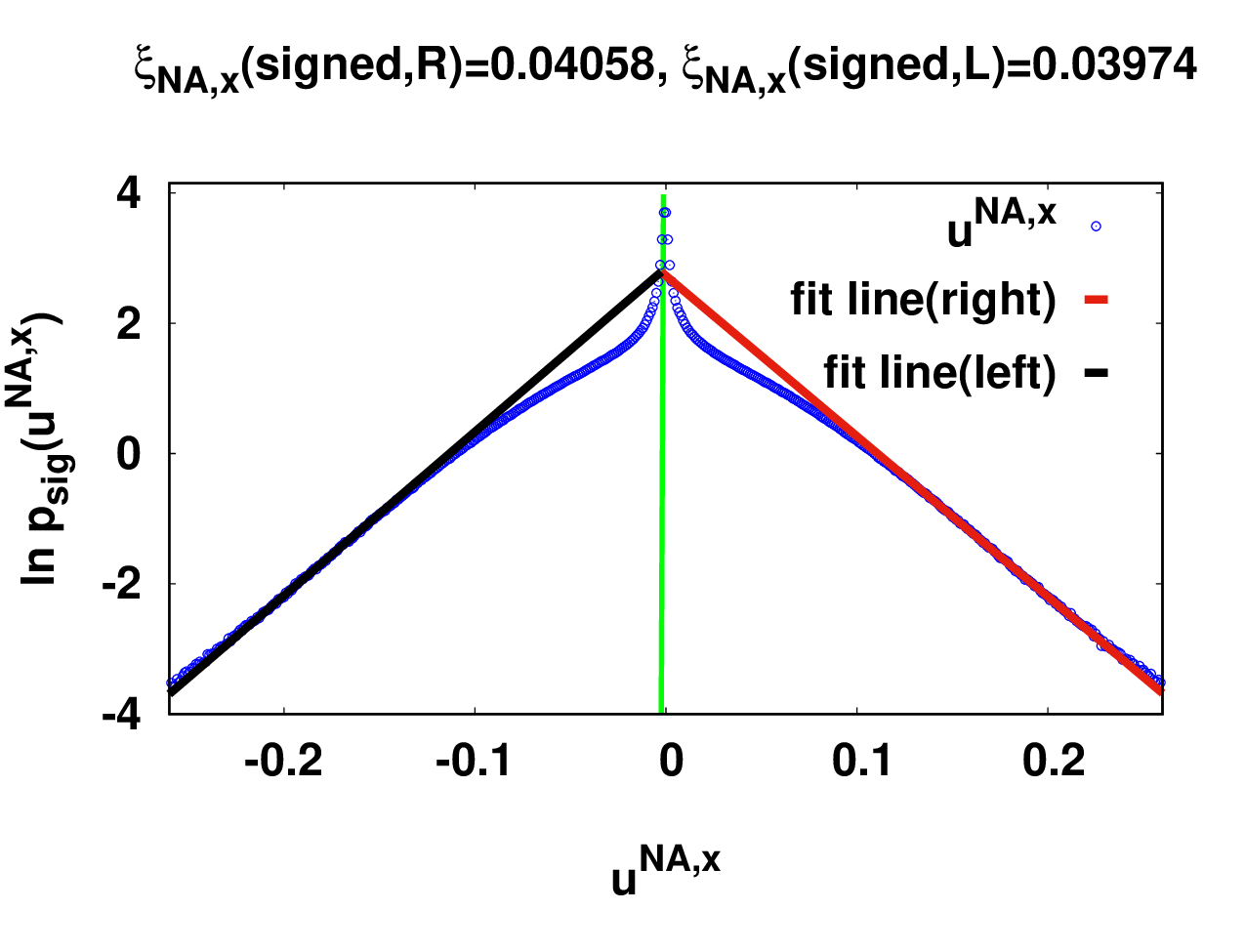}
    \put(15,50){\(\mathbf{\rho=0.900}\)}
    \put(15,42){\(\mathbf{T=0.120}\)}
  \end{overpic}
  \caption{Fitting protocol for the signed (full, positive and negative) distribution of the \(x\) component of non-affine at density $\rho=0.900$, and temperature $T=0.120$. Length scale obtained from left and right tail fitting is $\xi_{NA,x}\text{\textbf{(signed,\,L)}}=0.03974$ and $\xi_{NA,x}\text{\textbf{(signed,\,R})}=0.04058$ respectively. For clarity and readability the plotted quantity is \(\ln p_{\mathrm{sig}}(u^{\mathrm{NA},x})\), where \(p_{\mathrm{sig}}\) denotes the signed PDF of the \(x\)-component.}
  \label{fig:signed}
\end{figure}

\begin{figure}[htb]
  \centering
  \begin{overpic}[width=\columnwidth,keepaspectratio]{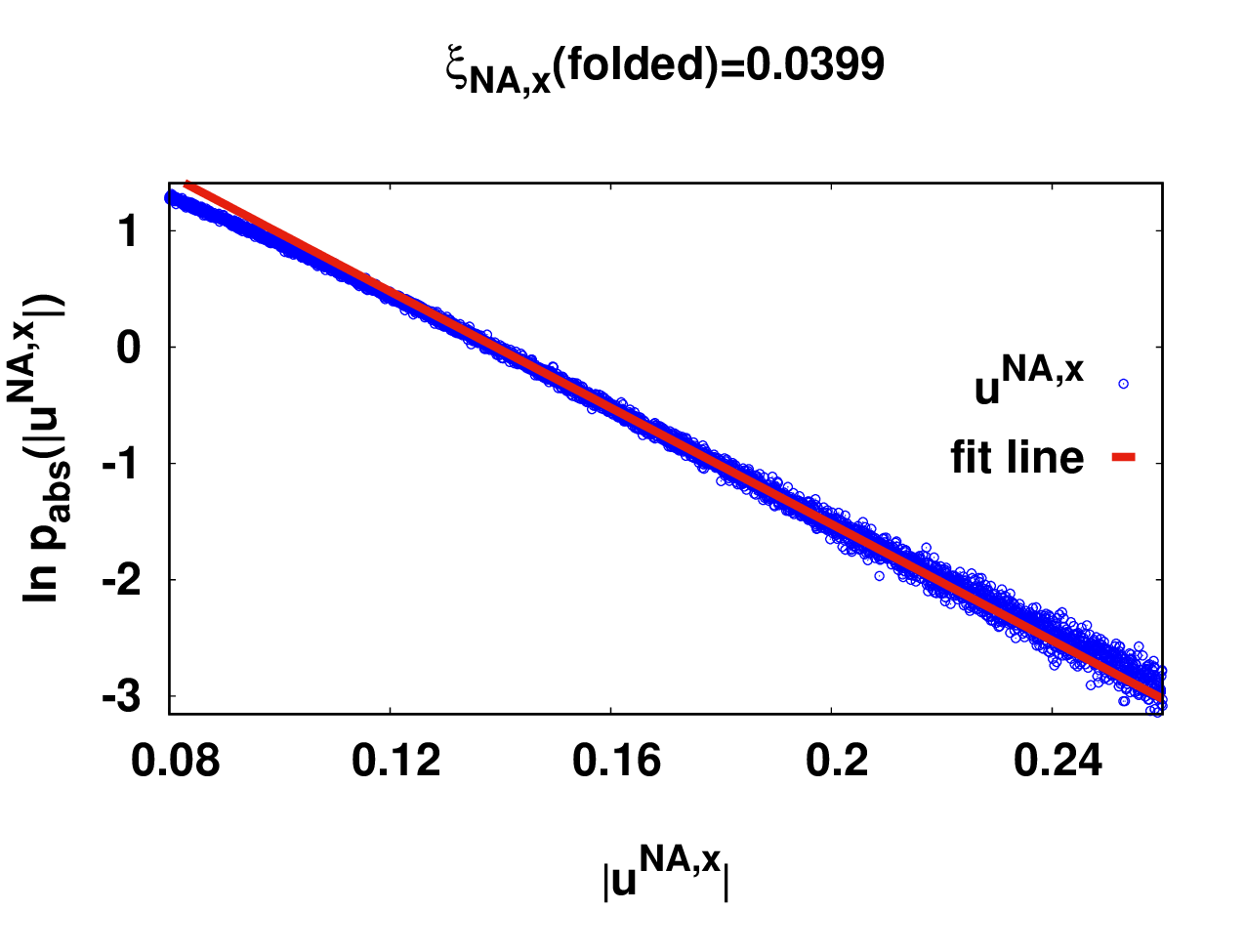}
\put(30,36){\(\mathbf{\rho=0.900}\)}
\put(30,30){\(\mathbf{T=0.120}\)}
  \end{overpic}
  \caption{Fitting protocol for the folded distribution of same \(x\) component of non-affine at density $\rho=0.900$, and temperature $T=0.120$ as in Fig.~\ref{fig:signed}. Length scale obtained from tail fitting is $\xi_{NA,x}=0.0399$. This value coincides with the length scale obtained from the signed distribution of \(x\) component of non-affine tails for a symmetric (zero-mean) distribution in Fig.~\ref{fig:signed}. For clarity the plotted quantity is \(\ln p_{\mathrm{abs}}(|u^{\mathrm{NA},x}|)\), where \(\ln p_{\mathrm{abs}}(|u^{\mathrm{NA},x}|)\) denotes the folded (absolute-value) PDF. }
  \label{fig:folded}
\end{figure}

\section{Characteristic Length Scale of Thermal Nonaffine Rearrangements}
\label{sec:nonaffine_results}

\begin{figure*}[htbp!]
  \centering
  \begin{tabular}{cc}
    \includegraphics[width=0.48\textwidth,height=0.48\textwidth,keepaspectratio]{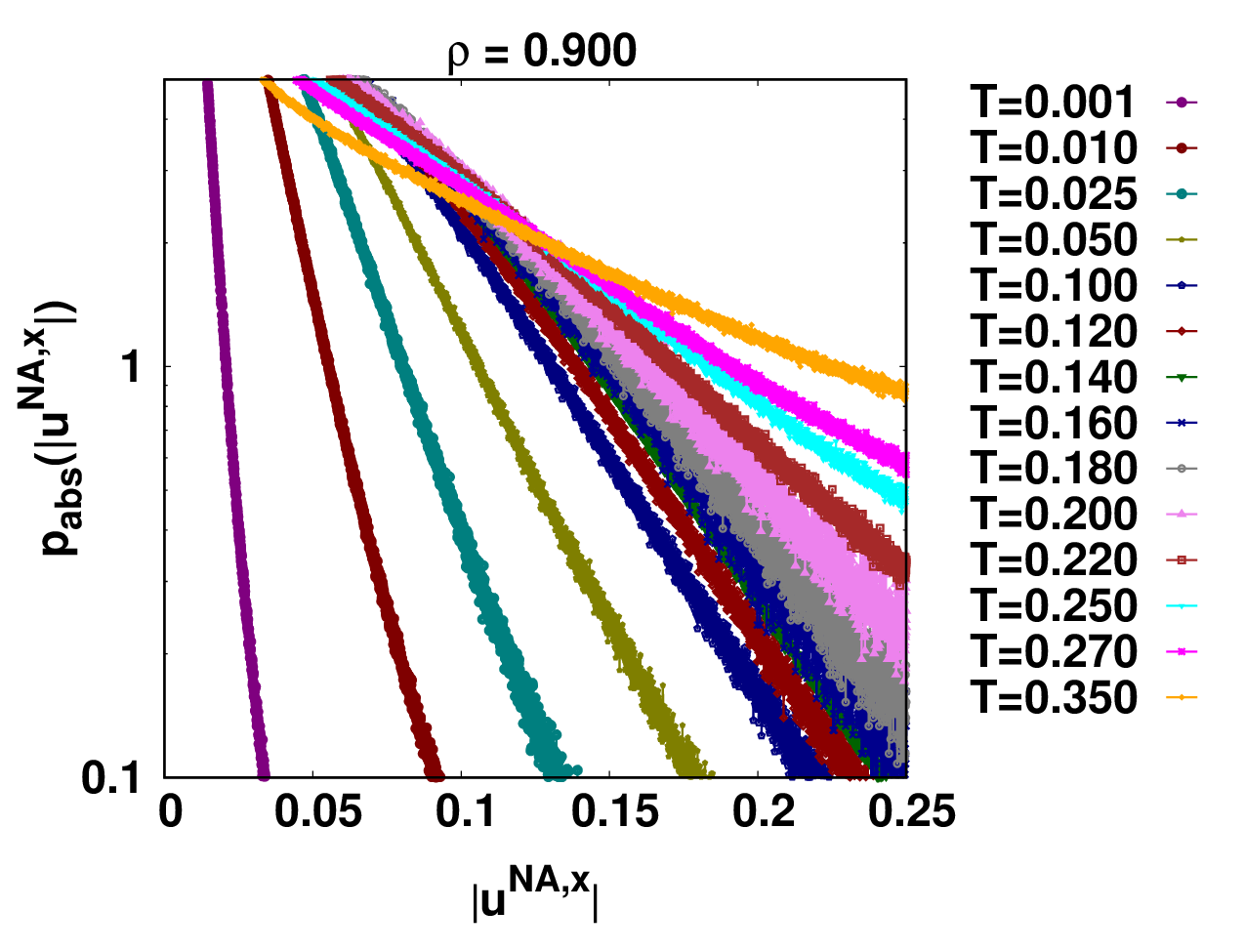}
    \includegraphics[width=0.48\textwidth,height=0.48\textwidth,keepaspectratio]{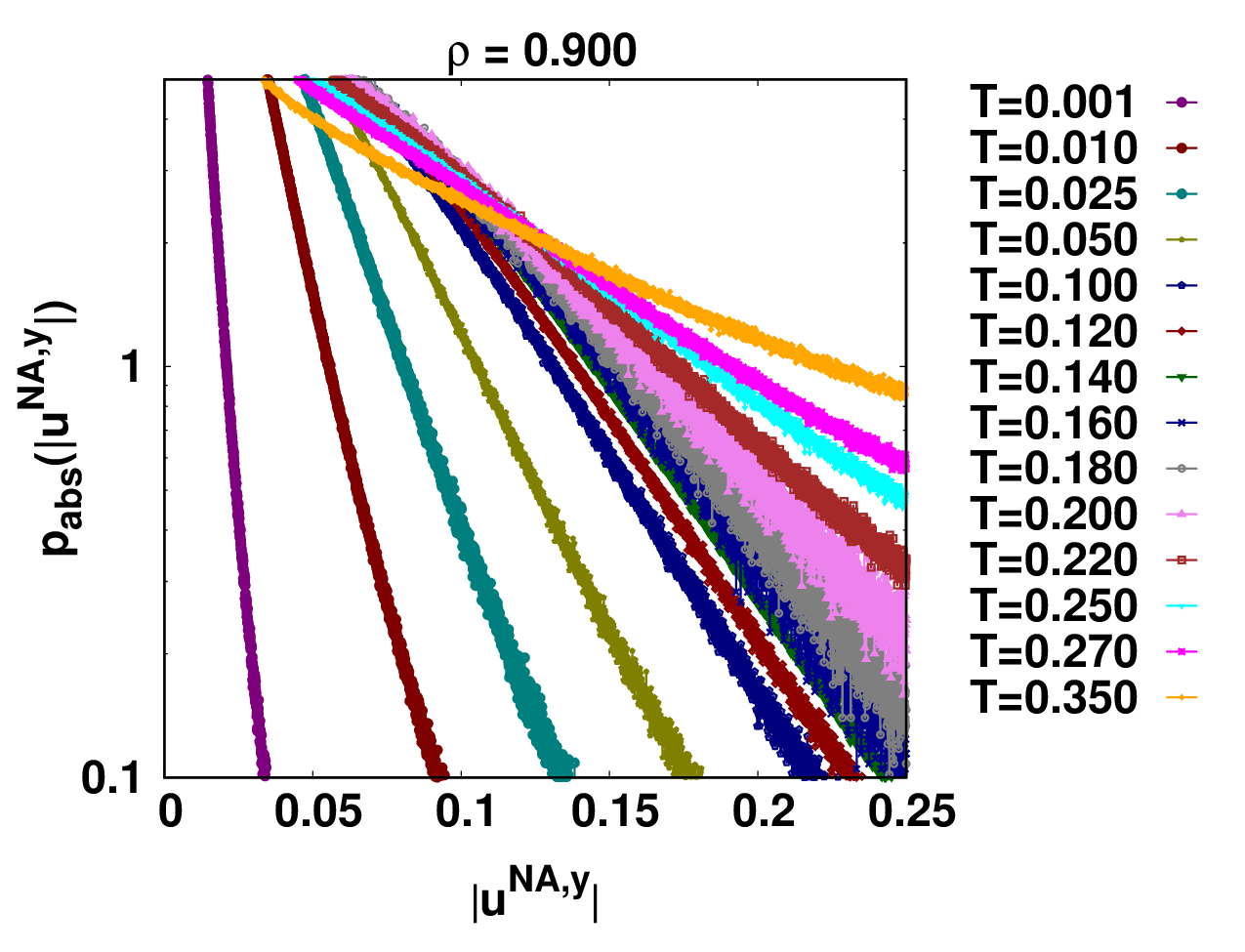} \\
    \includegraphics[width=0.48\textwidth,height=0.48\textwidth,keepaspectratio]{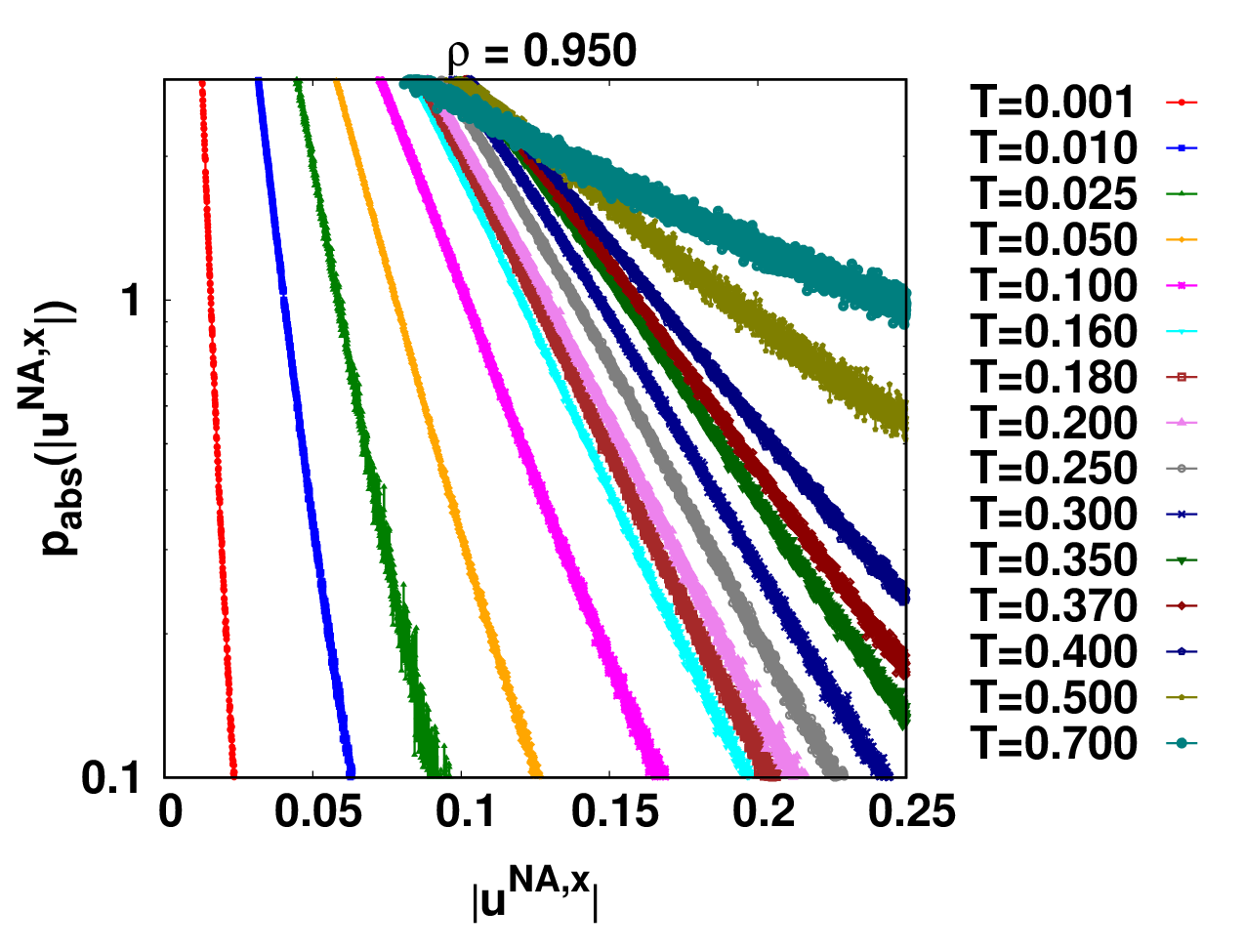} 
    \includegraphics[width=0.48\textwidth,height=0.48\textwidth,keepaspectratio]{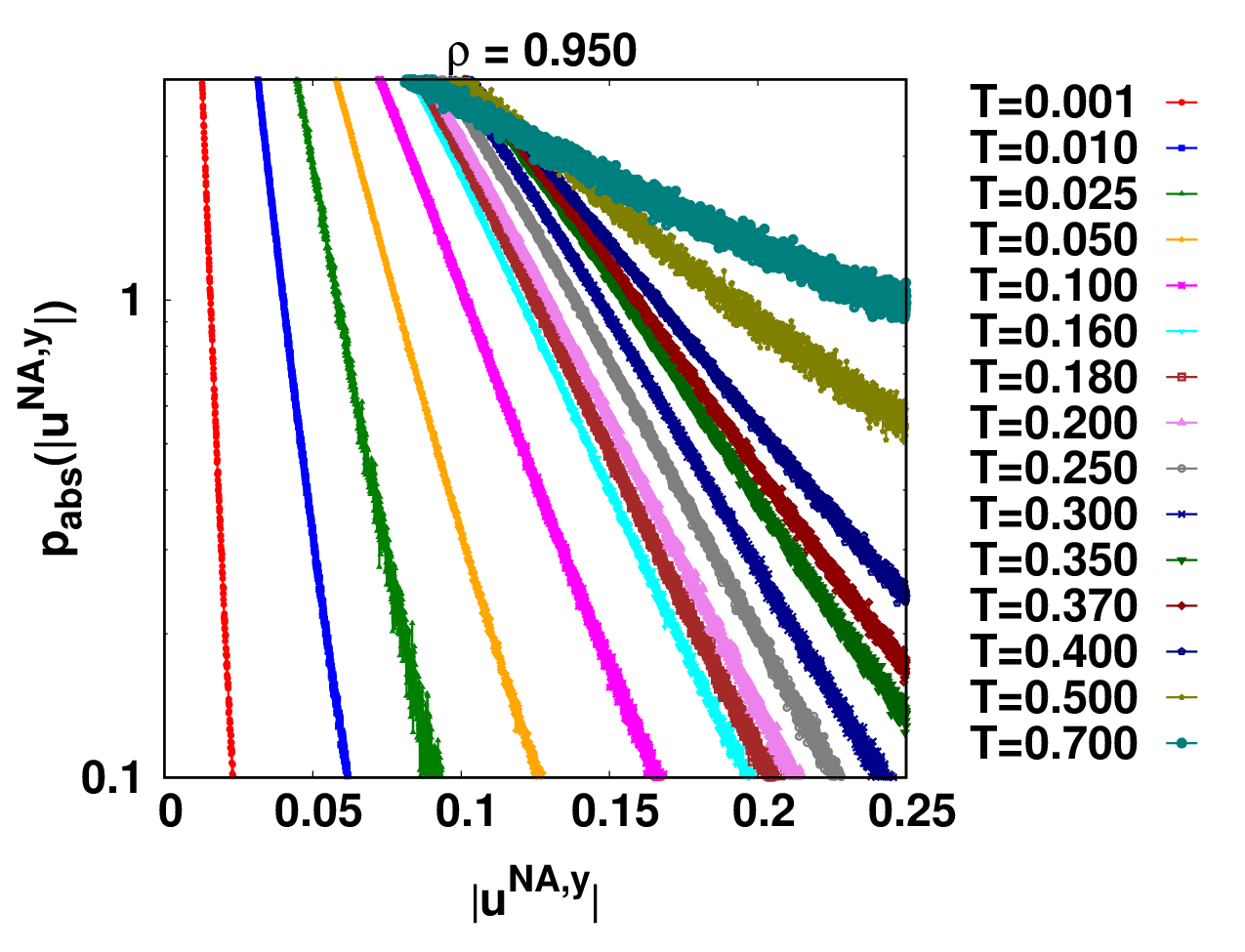} \\
    \includegraphics[width=0.48\textwidth,height=0.48\textwidth,keepaspectratio]{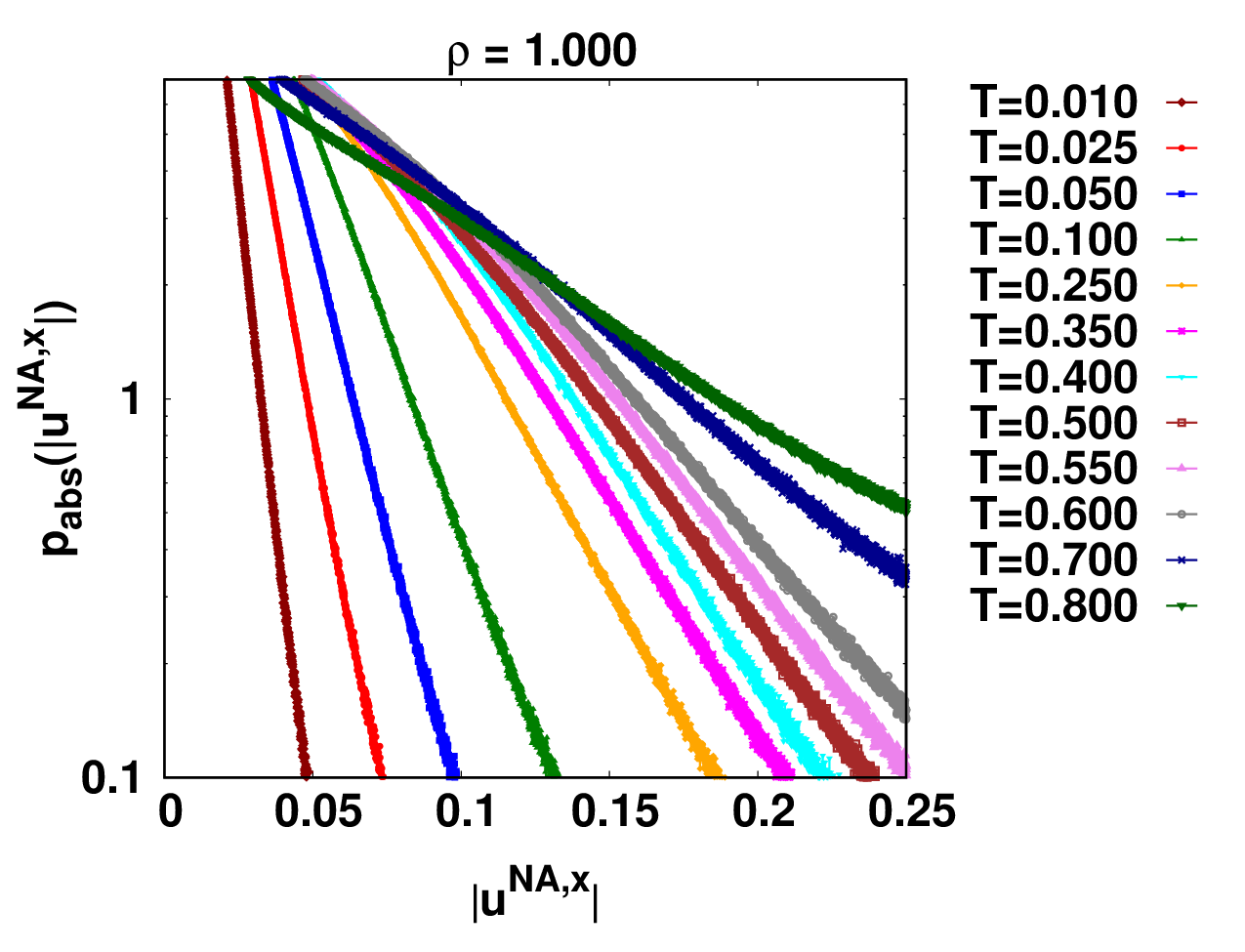} 
    \includegraphics[width=0.48\textwidth,height=0.48\textwidth,keepaspectratio]{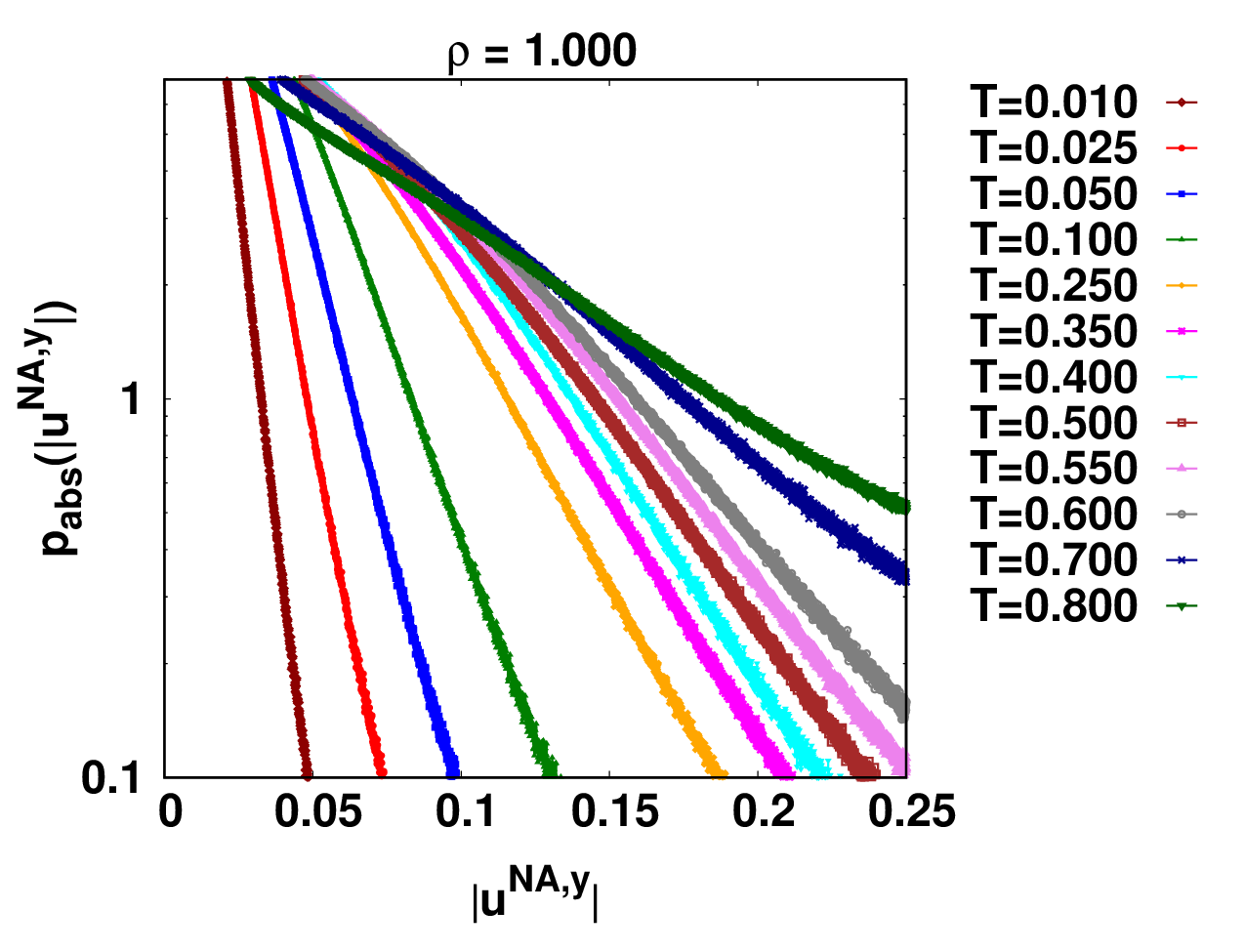}
  \end{tabular}
  \caption{Absolute-value distributions (semilog plots) of the nonaffine displacement components for three densities. Each row corresponds to a single density: $\rho=0.900$ (top), $\rho=0.950$ (middle), and $\rho=1.000$ (bottom). Left column: $|u_{\mathrm{NA},x}|$ distributions; right column: $|u_{\mathrm{NA},y}|$ distributions. Each curve in a panel is for a distinct temperature (labels shown inside the panels). The vertical axis is logarithmic so that exponential tails appear as straight lines; the fitted slope is $-1/\xi_{\mathrm{NA},\alpha}$ (component $\alpha=x,y$), where larger $\xi_{\mathrm{NA},\alpha}$ indicates more spatially extended nonaffine rearrangements. The three-row layout places identical plot types for the three densities together for direct comparison.}
  \label{fig:len_all}
\end{figure*}
\begin{figure*}
  \centering
  \begin{tabular}{@{}c@{\hspace{5mm}}c@{\hspace{5mm}}c@{}}
    \begin{overpic}[width=0.32\textwidth,keepaspectratio]{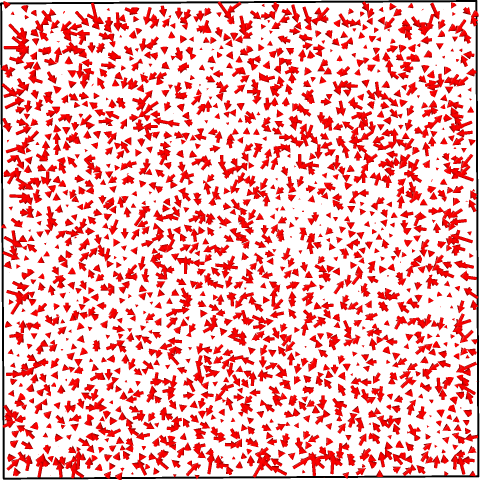}
      \put(2,105){\large\textbf{(a)}}
      \put(35,105){\large\textbf{T=0.350}}
    \end{overpic} &
    \begin{overpic}[width=0.32\textwidth,keepaspectratio]{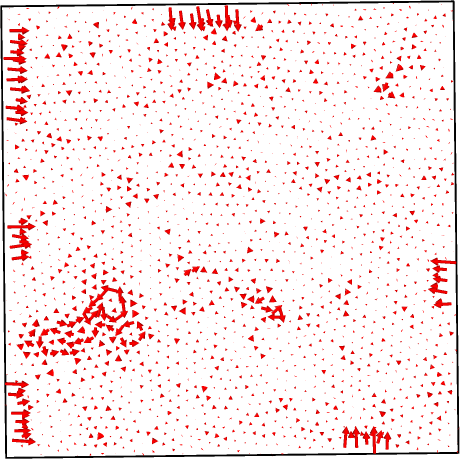}
      \put(2,105){\large\textbf{(b)}}
      \put(35,105){\large\textbf{T=0.140}}
    \end{overpic} &
    \begin{overpic}[width=0.32\textwidth,keepaspectratio]{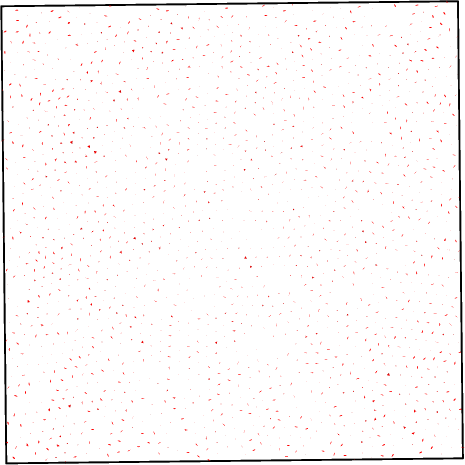}
      \put(2,105){\large\textbf{(c)}}
      \put(35,105){\large\textbf{T=0.010}}
    \end{overpic} \\[20pt]
    \begin{overpic}[width=0.32\textwidth,keepaspectratio]{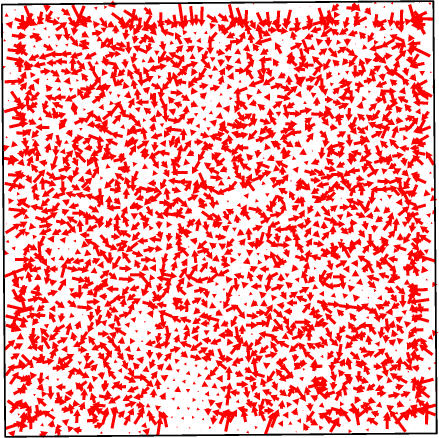}
      \put(2,105){\large\textbf{(d)}}
      \put(35,105){\large\textbf{T=0.700}}
    \end{overpic} &
    \begin{overpic}[width=0.32\textwidth,keepaspectratio]{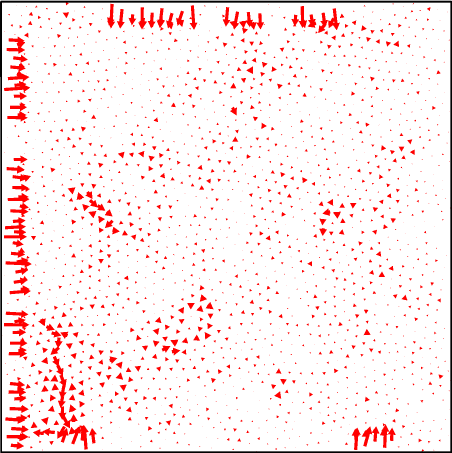}
      \put(2,105){\large\textbf{(e)}}
      \put(35,105){\large\textbf{T=0.180}}
    \end{overpic} &
    \begin{overpic}[width=0.32\textwidth,keepaspectratio]{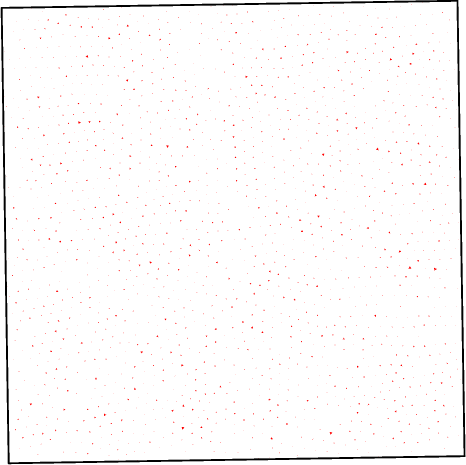}
      \put(2,105){\large\textbf{(f)}}
      \put(35,105){\large\textbf{T=0.050}}
    \end{overpic} \\[20pt]
    \begin{overpic}[width=0.32\textwidth,keepaspectratio]{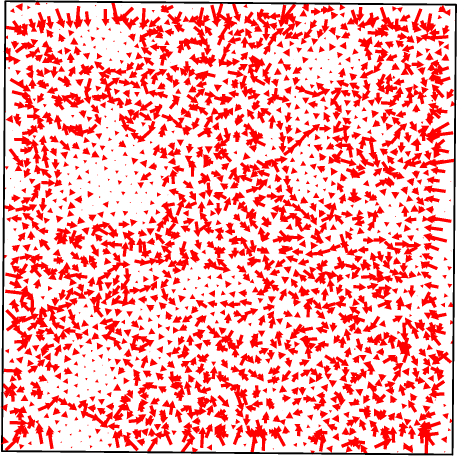}
      \put(2,105){\large\textbf{(g)}}
      \put(35,105){\large\textbf{T=0.800}}
    \end{overpic} &
    \begin{overpic}[width=0.32\textwidth,keepaspectratio]{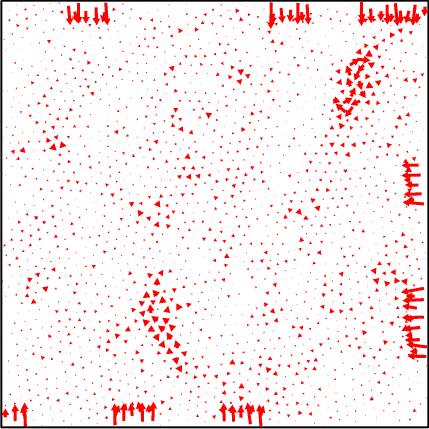}
      \put(2,105){\large\textbf{(h)}}
      \put(35,105){\large\textbf{T=0.350}}
    \end{overpic} &
    \begin{overpic}[width=0.32\textwidth,keepaspectratio]{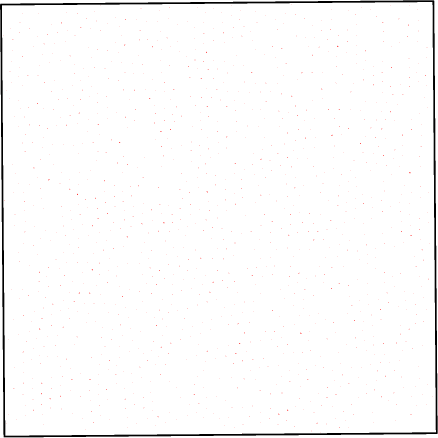}
      \put(2,105){\large\textbf{(i)}}
     \put(35,105){\large\textbf{T=0.100}}
    \end{overpic}
  \end{tabular}
  \caption{Visualization of nonaffine displacement fields arranged in a 3×3 grid (rows = densities, left-to-right = temperatures). Top row: $\rho=0.900$ at $T=0.350,0.140,0.010$ (panels a–c). Middle row: $\rho=0.950$ at $T=0.700,0.180,0.050$ (panels d–f). Bottom row: $\rho=1.000$ at $T=0.800,0.350,0.100$ (panels g–i). Each subpanel shows the per-particle nonaffine displacement field as vector arrows; higher-temperature panels display more extensive nonaffine activity, while low-temperature panels show fainter, localized arrows consistent with stronger caging and suppressed higher-order nonaffine dynamics.}
  \label{fig:nonaffine_fields_all}
\end{figure*}
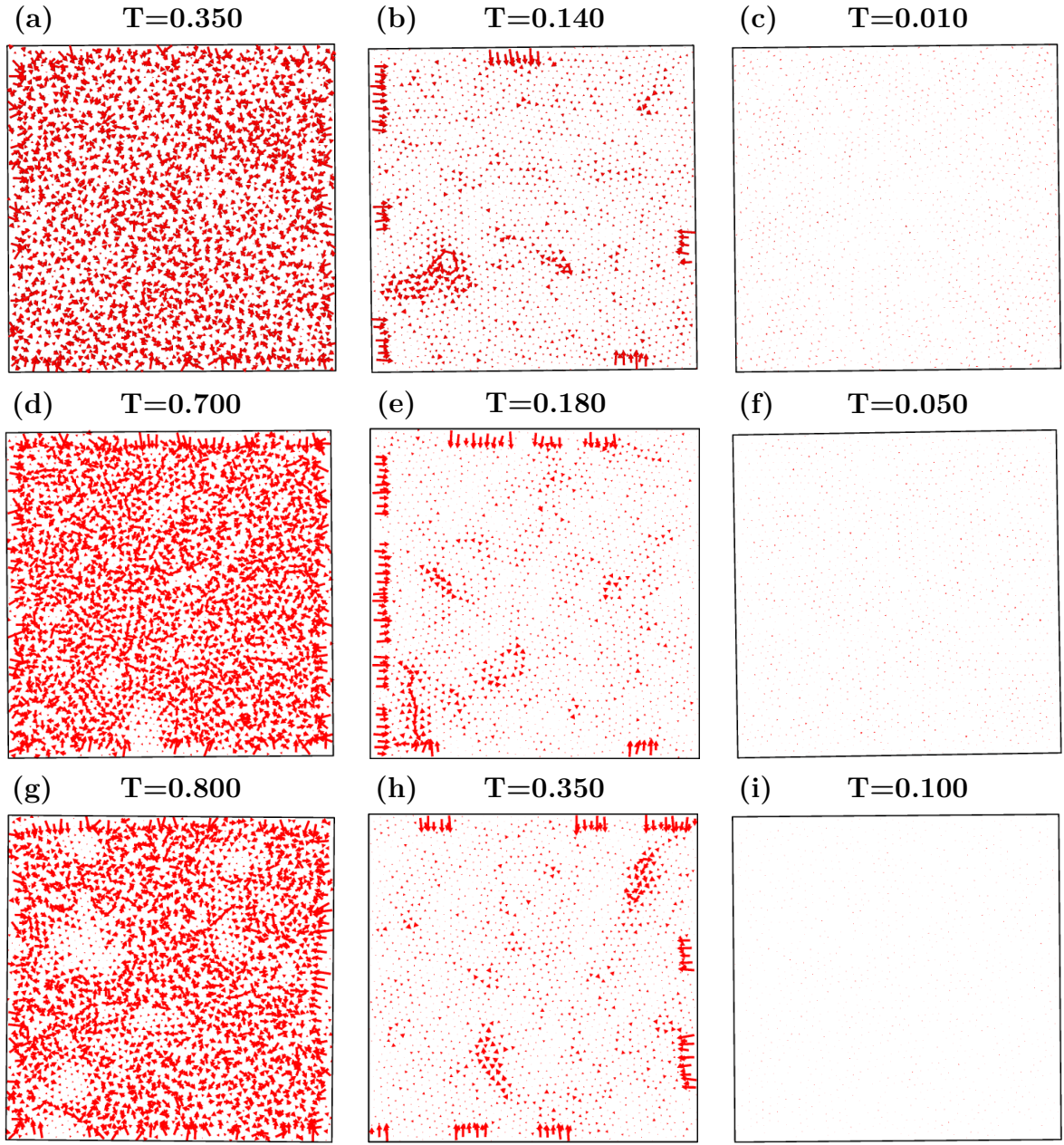

The componentwise nonaffine residual defined in equation \eqref{eq:d2min_component_def} provides the per-particle nonaffine displacements that we analyse statistically. In the following we denote the $x$- and $y$-components of the nonaffine residual by $u^{\mathrm{NA},x}$ and $u^{\mathrm{NA},y}$, respectively. \added{Although the distributions of the nonaffine displacement components are symmetric about zero and folding only produces a uniform upward shift of the entire log-PDF(probability distribution function) curve, the length scale $\xi$ is identical for signed and folded tails(as verified in Figs.~\ref{fig:signed} and~\ref{fig:folded})}, we deliberately analyse their folded (absolute-value) distributions of both components to focus only on its magnitude, instead of direction or signed distribution. we want to see how intensively this residual nonlinear deformation fluctuates under purely thermal driving. Empirically, the folded distributions display clear exponential tails. Restricting our analysis to these tails (the semilog plots exhibit an extended linear regime) allows a robust exponential fit and the extraction of characteristic length scales \(\xi_{\mathrm{NA},x}\) and \(\xi_{\mathrm{NA},y}\). The tail subrange is chosen by visual inspection of all panels of Figs.~\ref{fig:len_all} so that the fitted region reflects the thermal-activation dominated regime and excludes (i) the small-displacement core (often near-Gaussian) and (ii) any far-tail anomalies or finite-sample noise. 
Accordingly, we model the folded (absolute-value) distributions by 
\begin{subequations}\label{eq:abs_fit}
\begin{align}
p_{\mathrm{abs}}(|u^{\mathrm{NA},x}|) &= D_{0x}\,\exp\!\left(-\frac{|u^{\mathrm{NA},x}|}{\xi_{\mathrm{NA},x}}\right)\\[4pt]
p_{\mathrm{abs}}(|u^{\mathrm{NA},y}|) &= D_{0y}\,\exp\!\left(-\frac{|u^{\mathrm{NA},y}|}{\xi_{\mathrm{NA},y}}\right)
\end{align}
\end{subequations}
where $|u^{\mathrm{NA},x}|$ and $|u^{\mathrm{NA},y}|$ denote the absolute nonaffine displacements in the $x$ and $y$ directions, respectively. 
$D_{0x}$ and $D_{0y}$ are prefactors, and $\xi_{\mathrm{NA},x}$ and $\xi_{\mathrm{NA},y}$ are the fitted characteristic length scales for the $x$- and $y$-components. 
(In the main text, we used $\xi_{\mathrm{NA}}$ to denote the overall nonaffine length scale when component indices are not explicitly required.)
Taking the natural logarithms of \eqref{eq:abs_fit} gives the linear relations used for the exponential-tail fitting:
\begin{subequations}\label{eq:abs_fit_log}
\begin{align}
\ln p_{\mathrm{abs}}(|u^{\mathrm{NA},x}|) &= \ln D_{0x} - \frac{|u^{\mathrm{NA},x}|}{\xi_{\mathrm{NA},x}}\\[4pt]
\ln p_{\mathrm{abs}}(|u^{\mathrm{NA},y}|) &= \ln D_{0y} - \frac{|u^{\mathrm{NA},y}|}{\xi_{\mathrm{NA},y}}
\end{align}
\end{subequations}
Thus, in a semilog plot, the slope of the linear regime is $-1/\xi_{\mathrm{NA},x}$ (or $-1/\xi_{\mathrm{NA},y}$). \added{Here $\xi_{\mathrm{NA},\alpha}$ is the lengthscale which measures how far the influence of a strong nonaffine event spreads in space before it dies out. If we take a strong nonaffine jump as a small local disturbance, neighbors close to the center move noticeably, but that motion fades away with distance. The length scale quantifies how far that influence reaches. So, it’s the radius of the rearranging zone.} Larger values of $\xi_{\mathrm{NA},\alpha}$ correspond to slower decay and therefore spatially more extended nonaffine displacements in component $\alpha$, while smaller $\xi_{\mathrm{NA},\alpha}$ indicate more strongly localized rearrangements. \added{This interpretation is corroborated by particle-level nonaffine visualizations in Fig.~\ref{fig:nonaffine_fields_all}}.
The pronounced linear regimes of the semilog tails in Fig.~\ref{fig:len_all} demonstrate that our procedure for extracting the nonaffine length scale \(\xi_{\mathrm{NA}}\) is robust. The temperature dependence of the tail slope maps directly onto the spatial extent of nonaffine activity: at the highest temperatures the tails are shallow (small absolute slope), corresponding to large \(\xi_{\mathrm{NA}}\) and thus extended, system-spanning nonaffine dynamics. In an intermediate temperature window the decay rate increases (the tails become steeper), indicating a reduction in the spatial reach of nonaffine rearrangements while nonaffine activity nevertheless remains appreciable. In the deeply glassy regime the tails are very steep (large absolute slope), signifying a dramatic suppression of nonaffine excursions and a strong localization of residual rearrangements, which show only faint, spatially sparse displacements as shown in Fig.~\ref{fig:nonaffine_fields_all}, and the system therefore behaves increasingly like a mechanically stable solid. Finally, the close agreement between the \(x\)- and \(y\)-component panels shows that these trends are isotropic: thermal driving deformation produces directionally unbiased unlikely to external loading deformation, magnitude-controlled nonaffine responses across the explored temperatures. This demonstrates that thermal activation manifests uniformly in all directions and is systematically modulated by both temperature and density.
\added{So take away message of Fig.~\ref{fig:len_all} is that tail slopes steepen monotonically with decreasing temperature, reflecting a crossover from extended, thermally activated nonaffine motion at high \(T\) to strongly localized, solid-like behavior in the deep glass.}
The particle-level displacement fields shown in Fig.~\ref{fig:nonaffine_fields_all} give a direct, local interpretation of these length-scale trends.  At the highest temperatures, the nonaffine vectors are large, widely distributed and spatially random, indicating pervasive thermal activation and percolating nonaffine motion.
In the intermediate regime lying below the glass transition but not yet in the deeply solid state, field breaks up into localized patches and loop-like (cage-like) structures, consistent with finite but spatially confined nonaffine rearrangements that produce the measurable, intermediate-length exponential tails. Finally, in the deep glassy regime, the nonaffine arrows are faint and scarce, reflecting strong particle caging and mechanical stability; correspondingly, the fitted nonaffine length scales decrease sharply.  Taken together, the visualization and the robust linear fits show that our particle-level nonaffine measure captures the transition from thermally dominated, spatially extended disorder to localized, rare rearrangements and, ultimately, to mechanically stable glassy configurations. \emph{This is new in this context of role of nonaffinity in determination of stability of different regimes controlled by thermal fluctuation only, and no external loading}.

\subsection{Temperature Dependence of Nonaffine Displacement Scales at Different Densities}
\begin{figure*}[htbp!]
    \centering
    \includegraphics[width=0.4\textwidth]{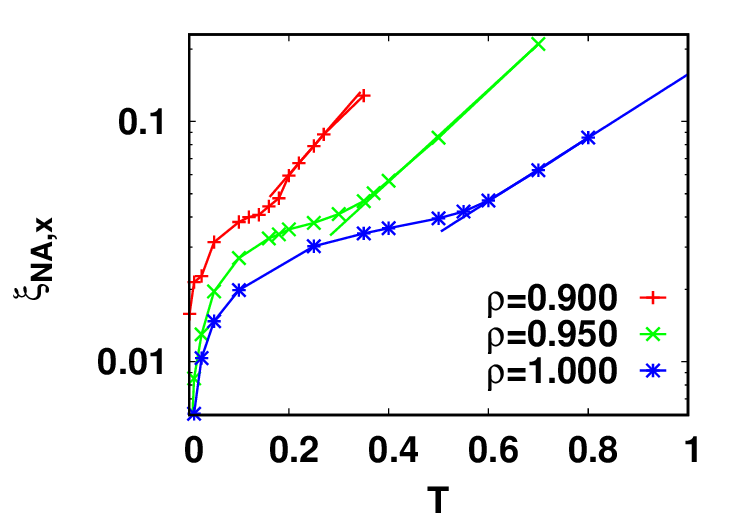}
    \includegraphics[width=0.4\textwidth]{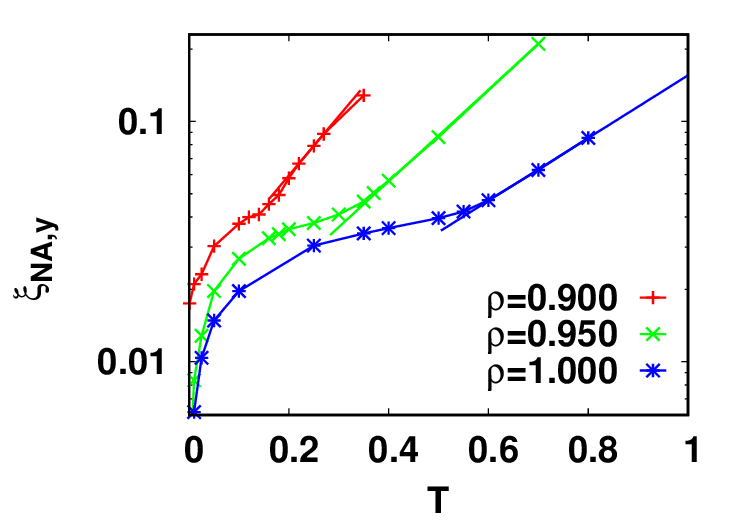}
    \includegraphics[width=0.4\textwidth]{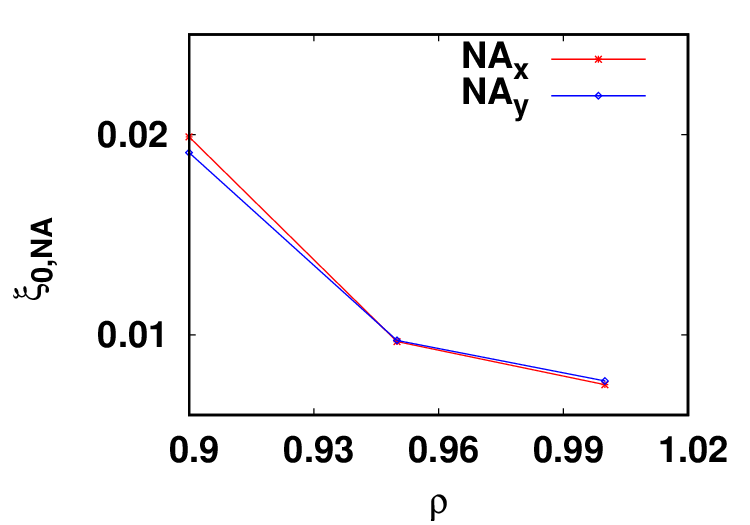}
    \includegraphics[width=0.4\textwidth]{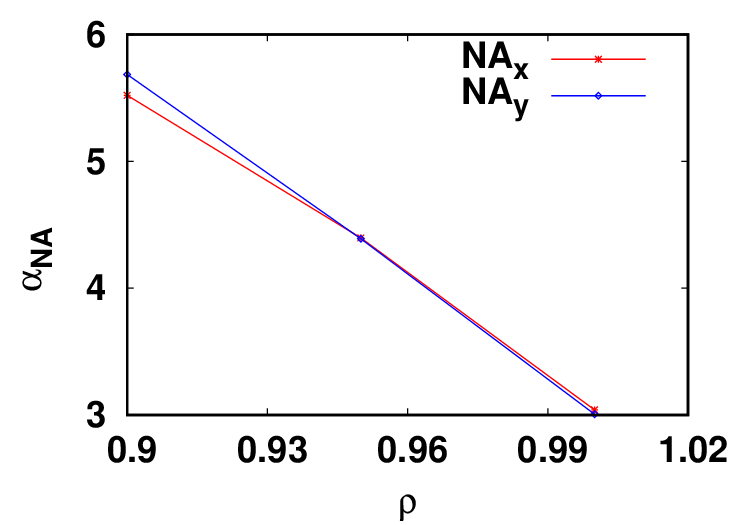}
  \caption{
\textbf{Top panels:} Semilog plots of the characteristic non-affine displacement scales $\xi_{\mathrm{NA},x}$ and $\xi_{\mathrm{NA},y}$ as functions of temperature $T$ for three densities, $\rho=0.900$, $0.950$, and $1.000$. The left panel shows $\xi_{\mathrm{NA},x}$ (derived from the $x$-component of non-affine displacements) and the right panel shows $\xi_{\mathrm{NA},y}$ (from the $y$-component), with each curve corresponding to a distinct temperature. Similar to MSD curves, which typically exhibit ballistic, plateau, and diffusive regimes, these plots reveal an initial rapid increase (ballistic-like), followed by a plateau, and then a diffusive-like region where the data are linear on a semilog scale with temperature $T$ instead of time. Linear fits in this diffusive-like regime yield two key parameters: the baseline displacement scale (extrapolated to $T=0$) and the thermal sensitivity exponent. Notably, with increasing density the plateau broadens and both the baseline value and sensitivity decrease, indicating that denser systems exhibit more constrained non-affine rearrangements. \textbf{Bottom panels:} Plots of the fitted parameters obtained from the exponential model $\xi_{\mathrm{NA}}(T)=\xi_{0,\mathrm{NA}}\,\exp(\alpha_{\mathrm{NA}} T)$. The left panel displays the extrapolated baseline displacement scale $\xi_{0,\mathrm{NA}}$, and the right panel shows the thermal sensitivity exponent $\alpha_{\mathrm{NA}}$, which quantifies the relative rate of increase in $\xi_{\mathrm{NA}}$ with temperature. Both $\xi_{0,\mathrm{NA}}$ and $\alpha_{\mathrm{NA}}$ decrease systematically with increasing density, demonstrating that higher-density systems have smaller inherent non-affine displacement scales and lower thermal sensitivity. Moreover, the close agreement between $\xi_{\mathrm{NA},x}$ and $\xi_{\mathrm{NA},y}$ across all densities confirms that the thermally induced non-affine rearrangements are isotropic in nature.
}
\label{fig:len4}
\end{figure*}

Building on the definitions and analysis presented in the previous section, we now examine how the characteristic nonaffine displacement length scales vary with temperature and density. In Fig.~\ref{fig:len4}, top panels display the extracted length scales $\xi_{\mathrm{NA},x}$ (Left) and $\xi_{\mathrm{NA},y}$ (Right) as functions of temperature for three different densities. These plots exhibit a behavior reminiscent of the ballistic–plateau–diffusive sequence observed in mean square displacement (MSD) curves; however, here the independent variable is temperature rather than time. At higher temperatures (above the glass transition $T_g$), both $\xi_{\mathrm{NA},x}$ and $\xi_{\mathrm{NA},y}$ increase rapidly, indicating more extended thermal activation and frequent nonaffine rearrangements. In an intermediate temperature range below $T_g$, the characteristic length scale $\xi_{\mathrm{NA}}$ of nonaffine displacements declines gradually, indicating that while nonaffine rearrangements persist, they become increasingly constrained by partial particle caging. In the deeply glassy regime representing stable glassy solids, $\xi_{\mathrm{NA}}$ decreases sharply, reflecting strong particle caging and minimal nonaffine motion. \emph{These trends imply that the thermal nonaffinity diagnostic successfully quantifies the mechanical stability of different thermodynamic regimes. Large values of the diagnostic identify activation-dominated, mechanically softer (less stable) states, while very small values identify strongly caged, mechanically stable glassy solids with only sparse nonaffine activity}. Although the overall shape of these spatial length scale plots appears similar to that of MSD curves (which are plotted as a function of time), the underlying physics are distinct. The MSD quantifies the cumulative displacement over time, whereas the spatial length scale obtained from the nonaffine displacement distributions reflects local rearrangements at a given temperature. This distinction underscores that \added{the thermally driven nonaffine rearrangements play a significant role in the particle's overall mobility}. Furthermore, our analysis indicates a significant effect of density. At lower densities, the nonaffine displacement scales are larger, suggesting that particles experience more extended rearrangements due to the looser packing. In contrast, at higher densities the plateau in the spatial length scale extends over a broader temperature range, reflecting that tighter particle packing effectively constrains thermal rearrangements so that the length scale remains nearly constant over that region. In the linear (diffusive-like) regime of the semilog plots in Fig.~\ref{fig:len4}, we fit the data with an exponential function of the form
\begin{align}
    \xi_{\mathrm{NA}} &= \xi_{0,\mathrm{NA}}\,\exp(\alpha_{\mathrm{NA}} T)
    \label{math3}
\end{align}
which transforms to a line in semi-log space as
\begin{align}
    \log(\xi_{\mathrm{NA}}) &= \log(\xi_{0,\mathrm{NA}}) + \alpha_{\mathrm{NA}}\, T
    \label{math4}
\end{align}
Here $\xi_{0,\mathrm{NA}}$ is the extrapolated nonaffine displacement length scale at $T=0$ (within the context of the diffusive-like regime), and $\alpha_{\mathrm{NA}}$ quantifies how rapidly the length scale increases with temperature (i.e. $\frac{d\xi_{\mathrm{NA}}}{dT}$). Because this fit is performed only over the temperature window where the data are linear in semilog space, the parameters $\xi_{0,\mathrm{NA}}$ and $\alpha_{\mathrm{NA}}$ should be interpreted as representative of that specific regime rather than the entire temperature range. This means they describe how the characteristic nonaffine length scale behaves when the system's response is dominated by diffusive-like thermal fluctuations, providing insight into the microscopic dynamics in that limited yet significant temperature interval. Importantly, we observe that both $\xi_{0,\mathrm{NA}}$ and $\alpha_{\mathrm{NA}}$ decrease with increasing density, indicating that denser systems not only have a smaller baseline nonaffine displacement scale but also exhibit a reduced sensitivity to temperature. Moreover, the nearly identical fitting parameters obtained for the $x$ and $y$ components in our analysis of the spatial length scale versus temperature (as shown in the bottom panels of Fig.~\ref{fig:len4}), together with the quantitative agreement \( \xi_{\mathrm{NA},x} \approx \xi_{\mathrm{NA},y} \) further support the isotropic propagation of nonaffine disturbances under thermally driven deformation. \emph{So, thermally driven nonaffine disturbances do not exhibit a preferred direction of propagation, unlike the shear-driven case where applied shear produces anisotropic, directionally biased nonaffine responses that are typically aligned with the shear direction}.
 
\section{Cartesian van Hove distributions and comparison with nonaffine length scales}
\label{sec:vanhove_vs_nonaffine}

\begin{figure*}[htbp!]
  \centering
  \begin{tabular}{cc}
    \includegraphics[width=0.48\textwidth,height=0.48\textwidth,keepaspectratio]{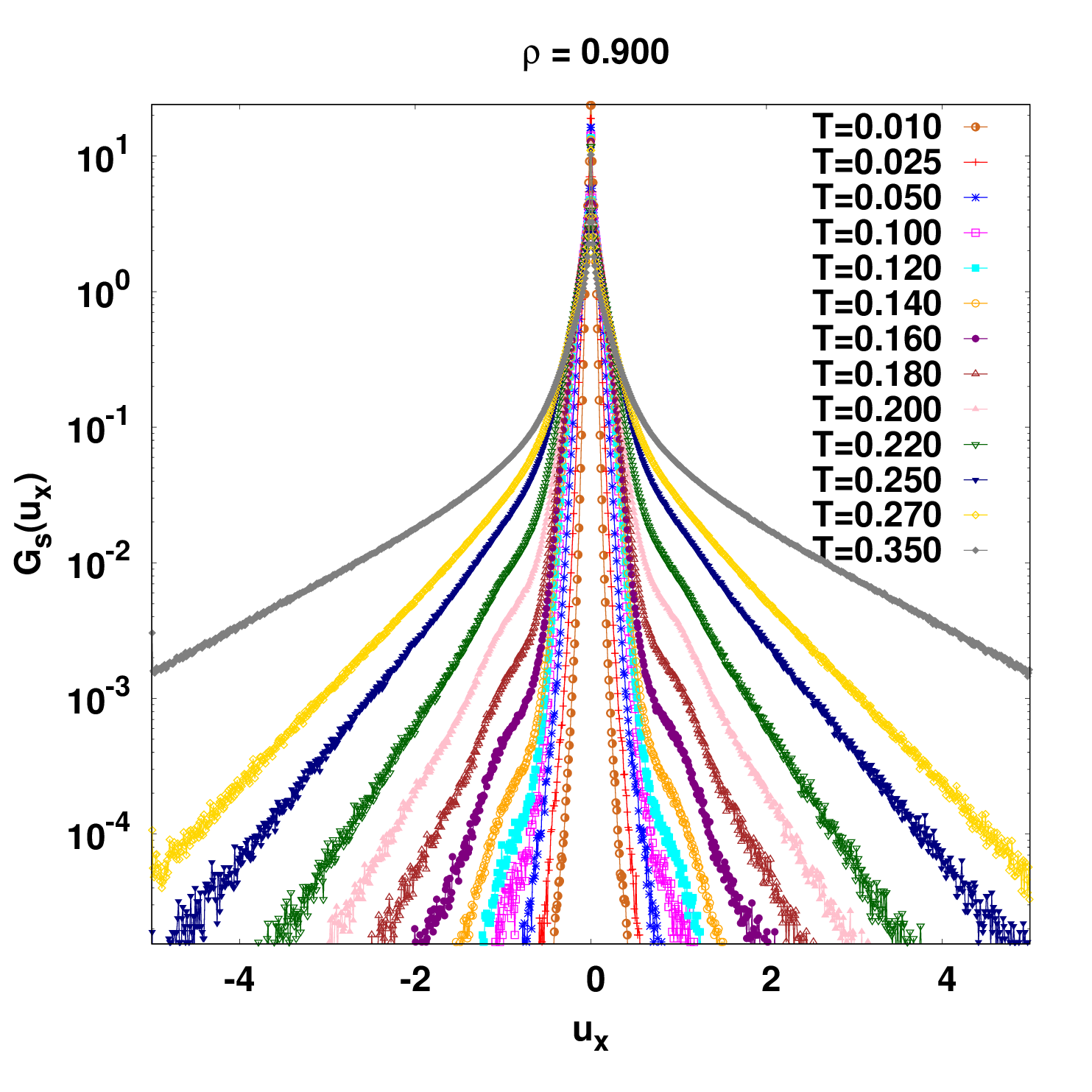}\hspace{8mm} 
    \includegraphics[width=0.48\textwidth,height=0.48\textwidth,keepaspectratio]{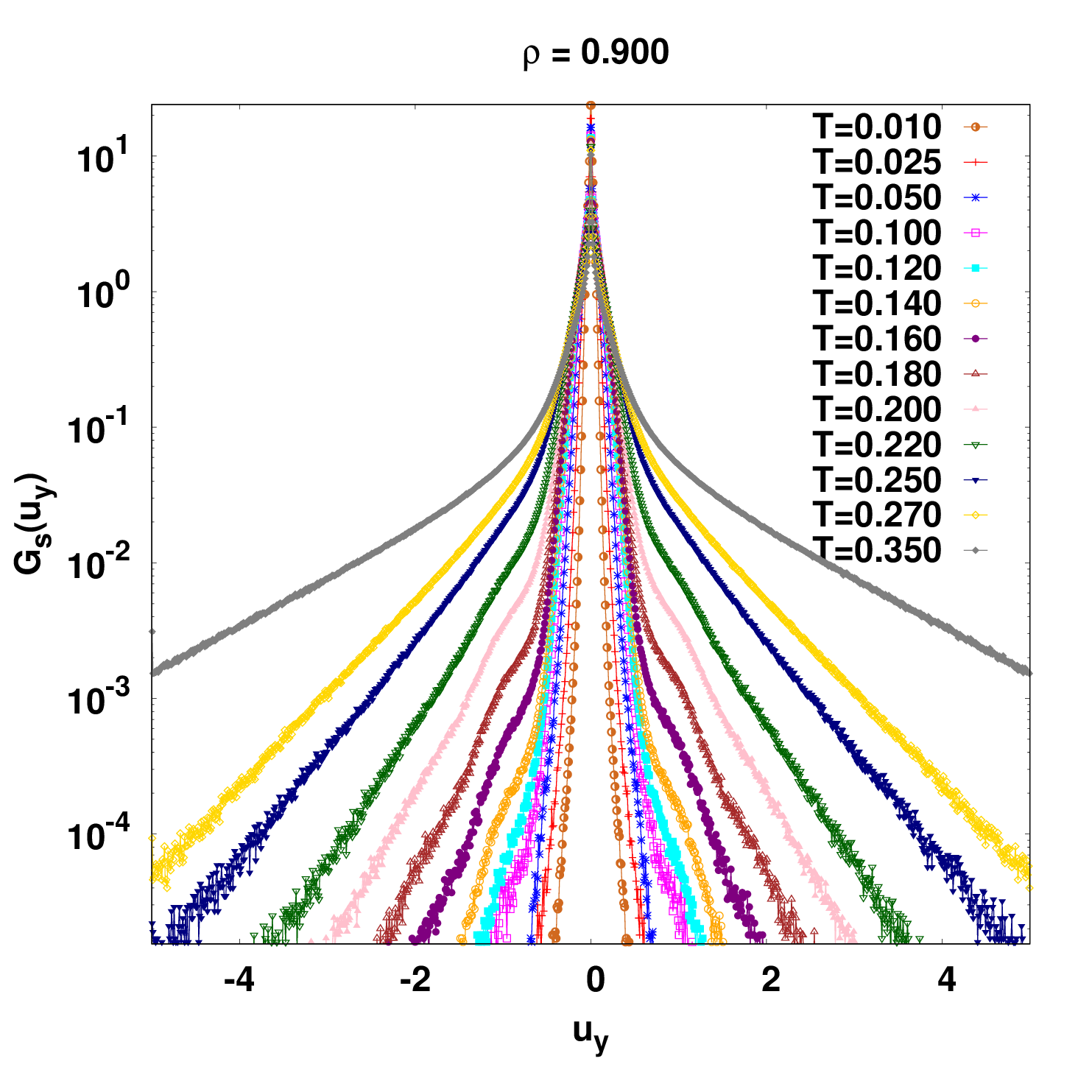} \\
    \includegraphics[width=0.48\textwidth,height=0.48\textwidth,keepaspectratio]{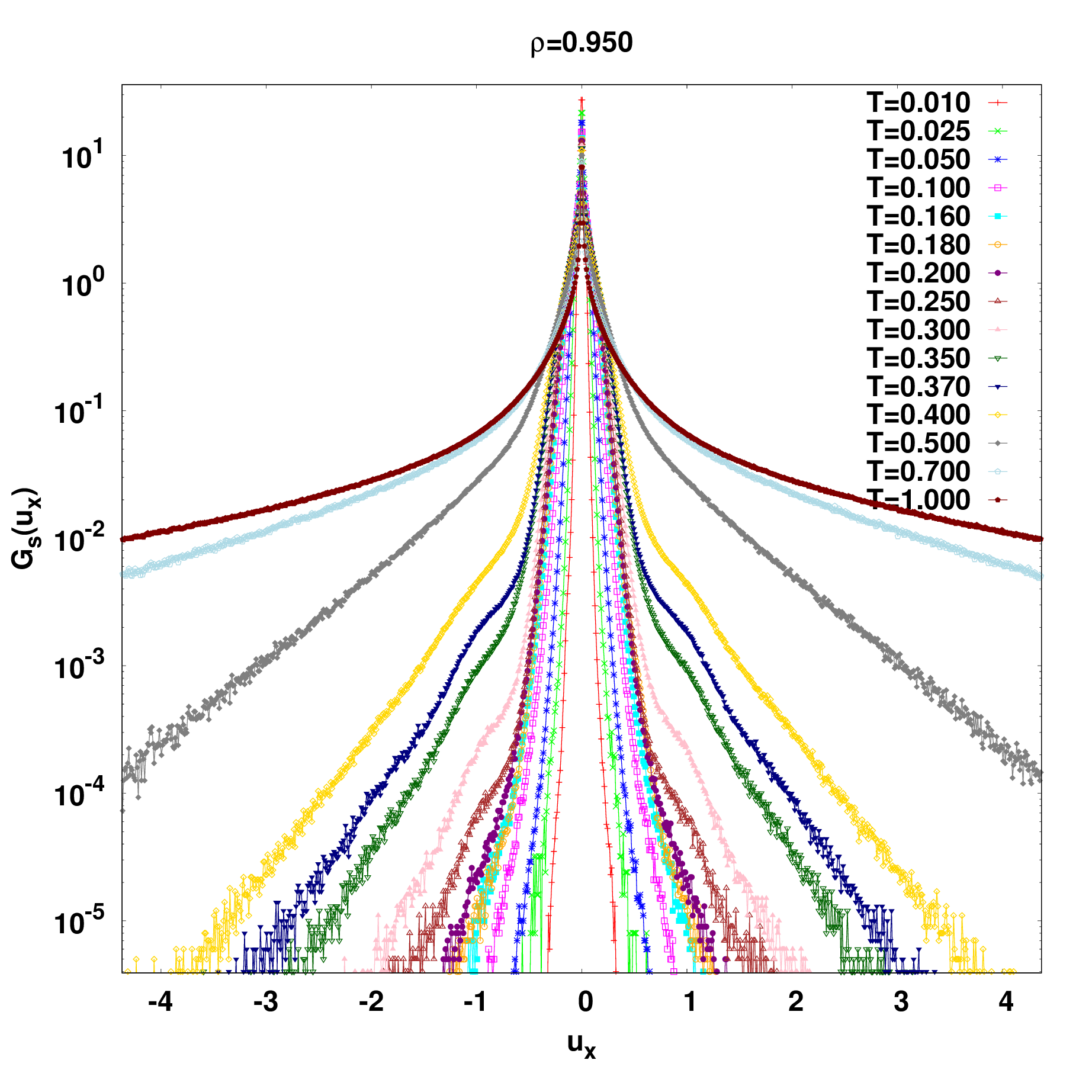}\hspace{8mm} 
    \includegraphics[width=0.48\textwidth,height=0.48\textwidth,keepaspectratio]{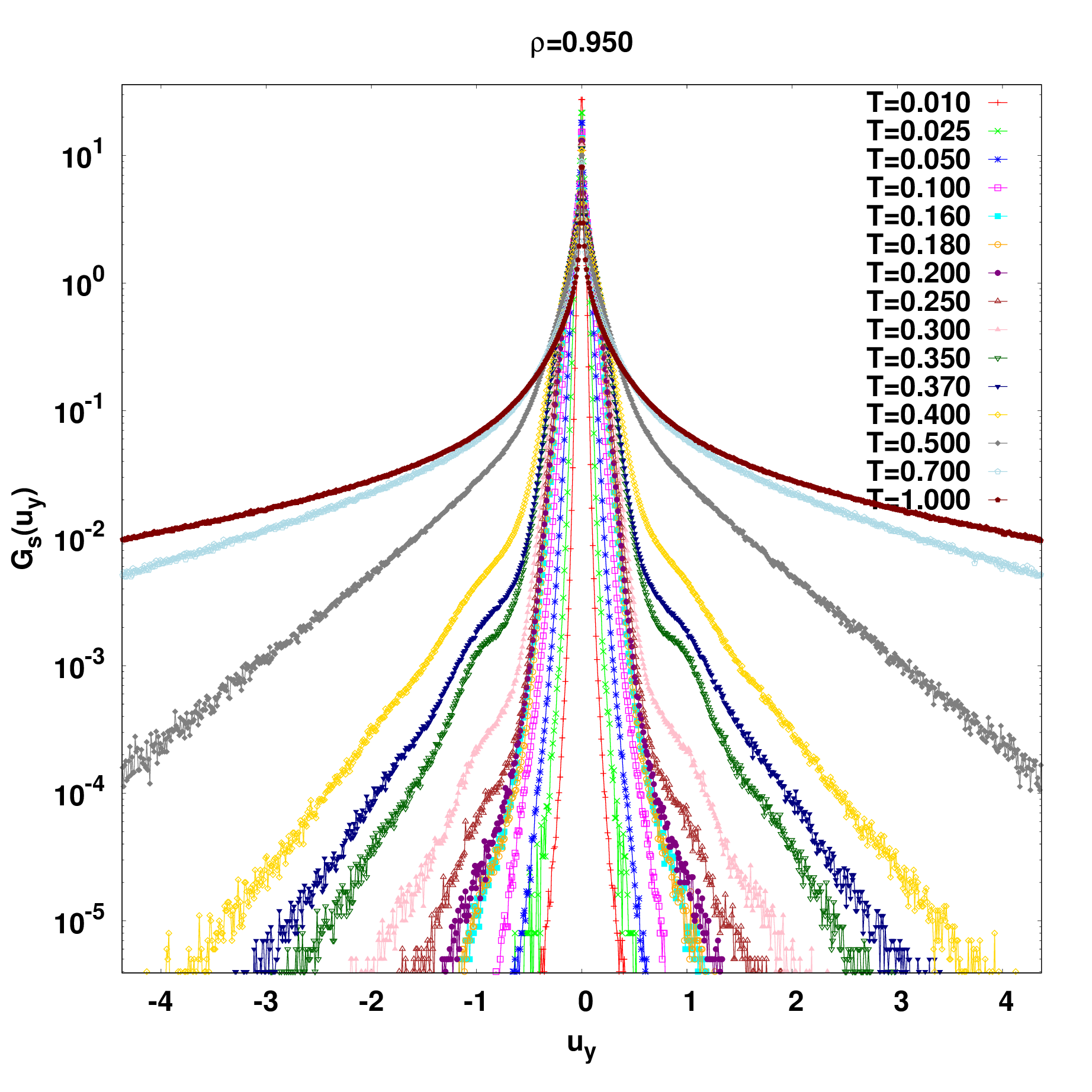} \\
    \includegraphics[width=0.48\textwidth,height=0.48\textwidth,keepaspectratio]{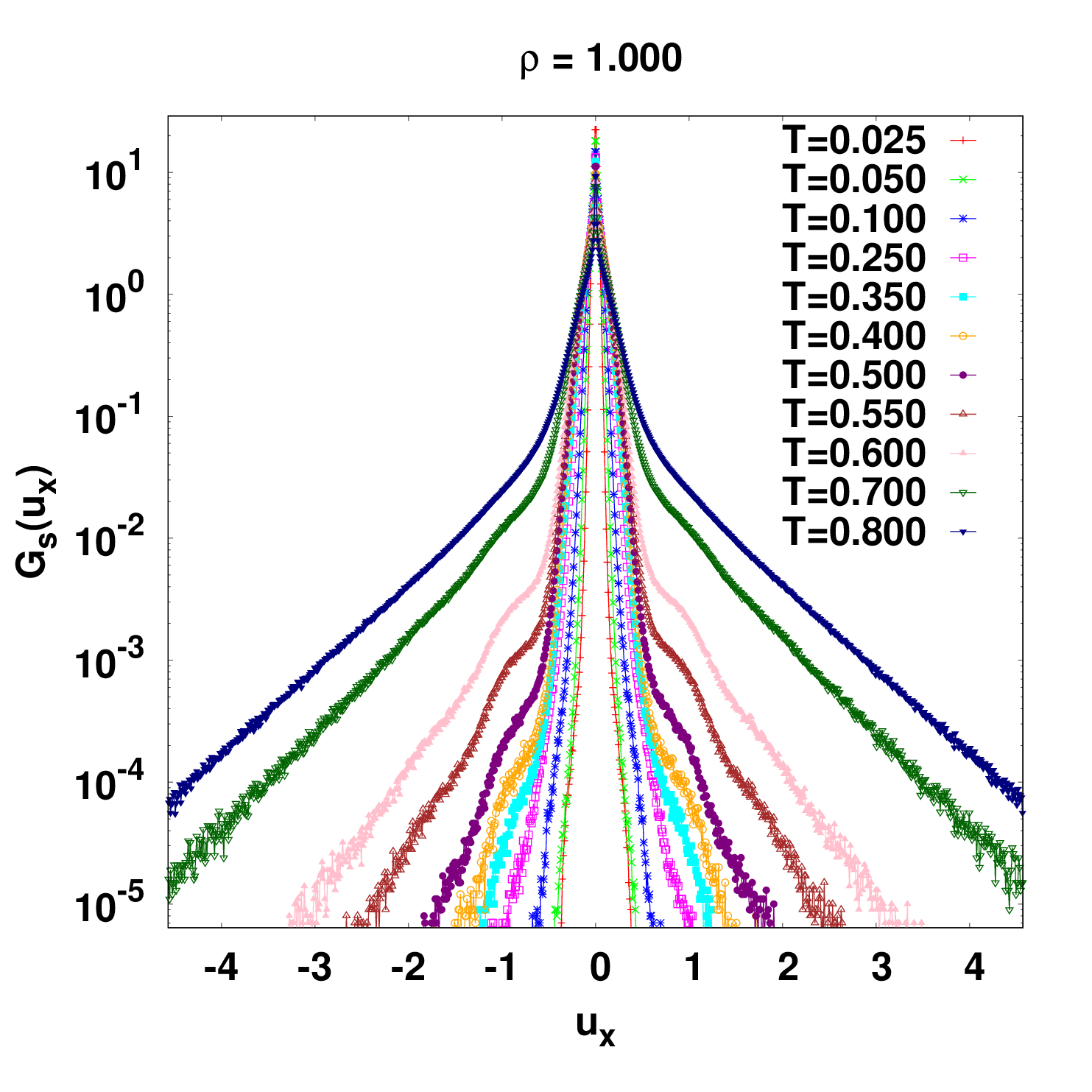}\hspace{8mm}
    \includegraphics[width=0.48\textwidth,height=0.48\textwidth,keepaspectratio]{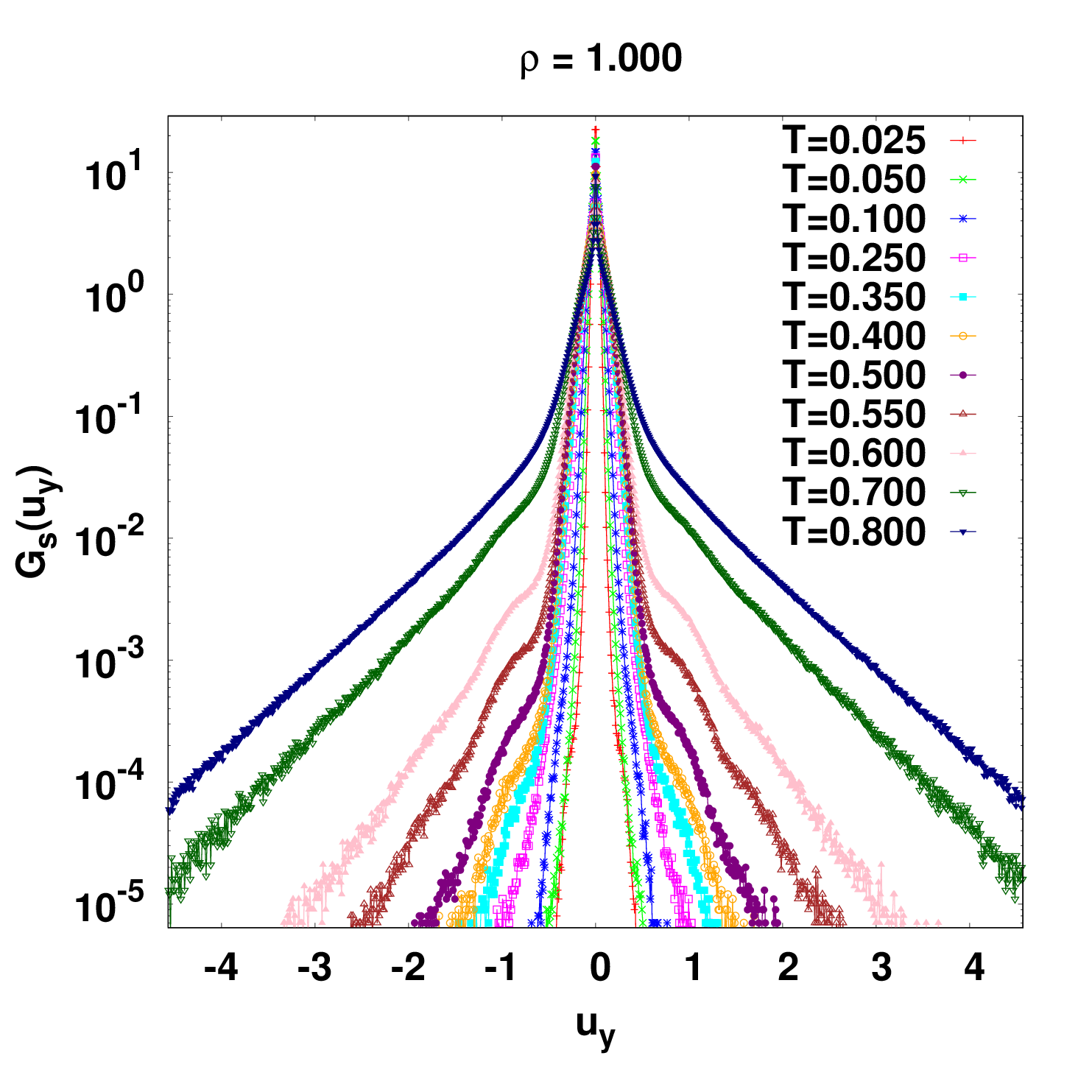}
  \end{tabular}
  \caption{Signed Cartesian van Hove self-distributions for the $x$ (left column) and $y$ (right column) components at three densities: $\rho=0.900$ (top row), $\rho=0.950$ (middle row) and $\rho=1.000$ (bottom row). Each curve represents a distribution computed from consecutive frame pairs at different temperatures (see legends in the panels). Exponential tails are fitted (fitted as straight lines to the exponential tails on semilog plots) to extract the von Hove length scales \(\xi_{\mathrm{VH},x}\) and \(\xi_{\mathrm{VH},y}\).}
   \label{fig:vh_all}
\end{figure*}
\begin{figure*}[htb!]
  \centering
  \begin{tabular}{cc}
    \includegraphics[width=0.37\textwidth,height=0.40\textwidth,keepaspectratio]{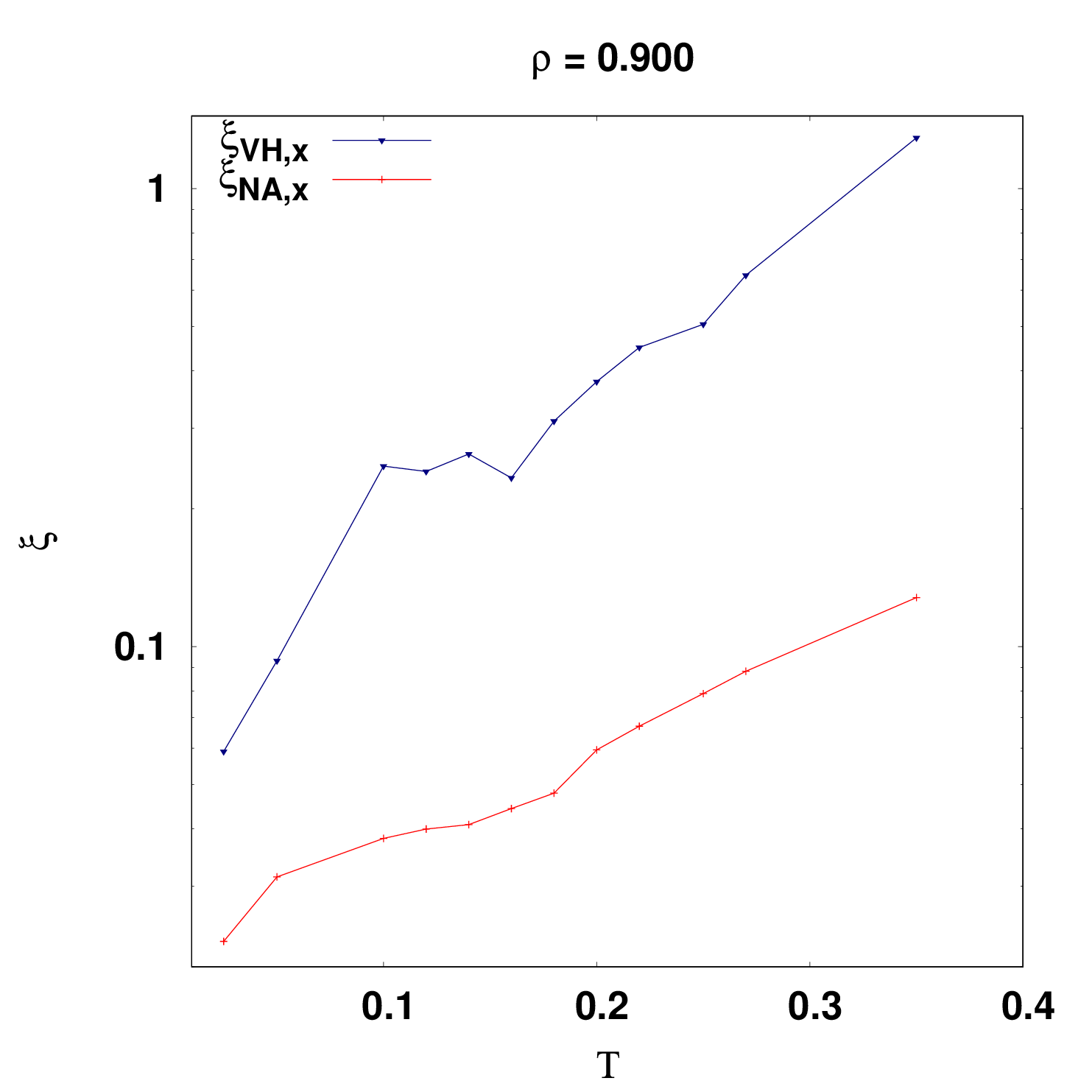}\hspace{8mm} 
    \includegraphics[width=0.37\textwidth,height=0.40\textwidth,keepaspectratio]{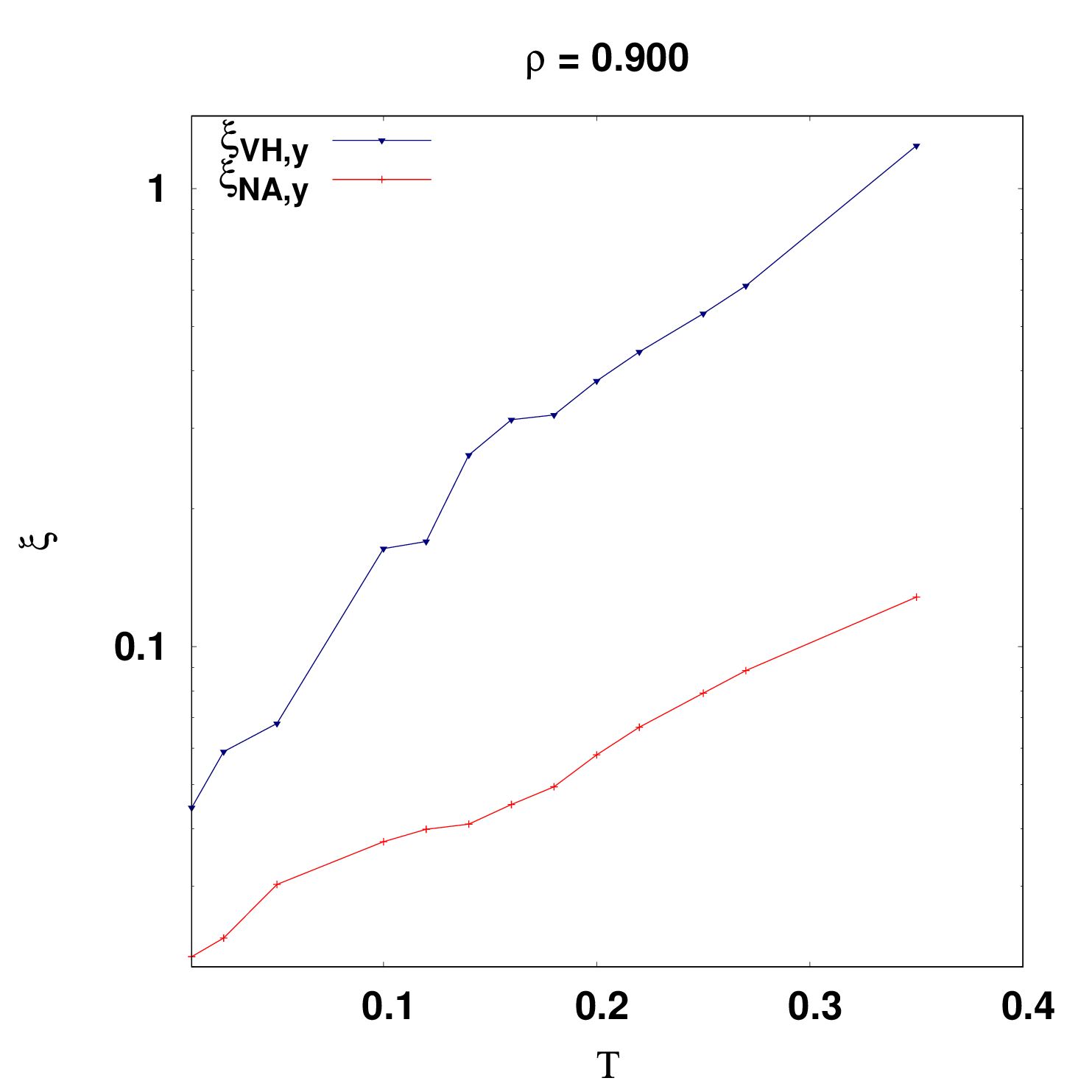} \\
    \includegraphics[width=0.37\textwidth,height=0.40\textwidth,keepaspectratio]{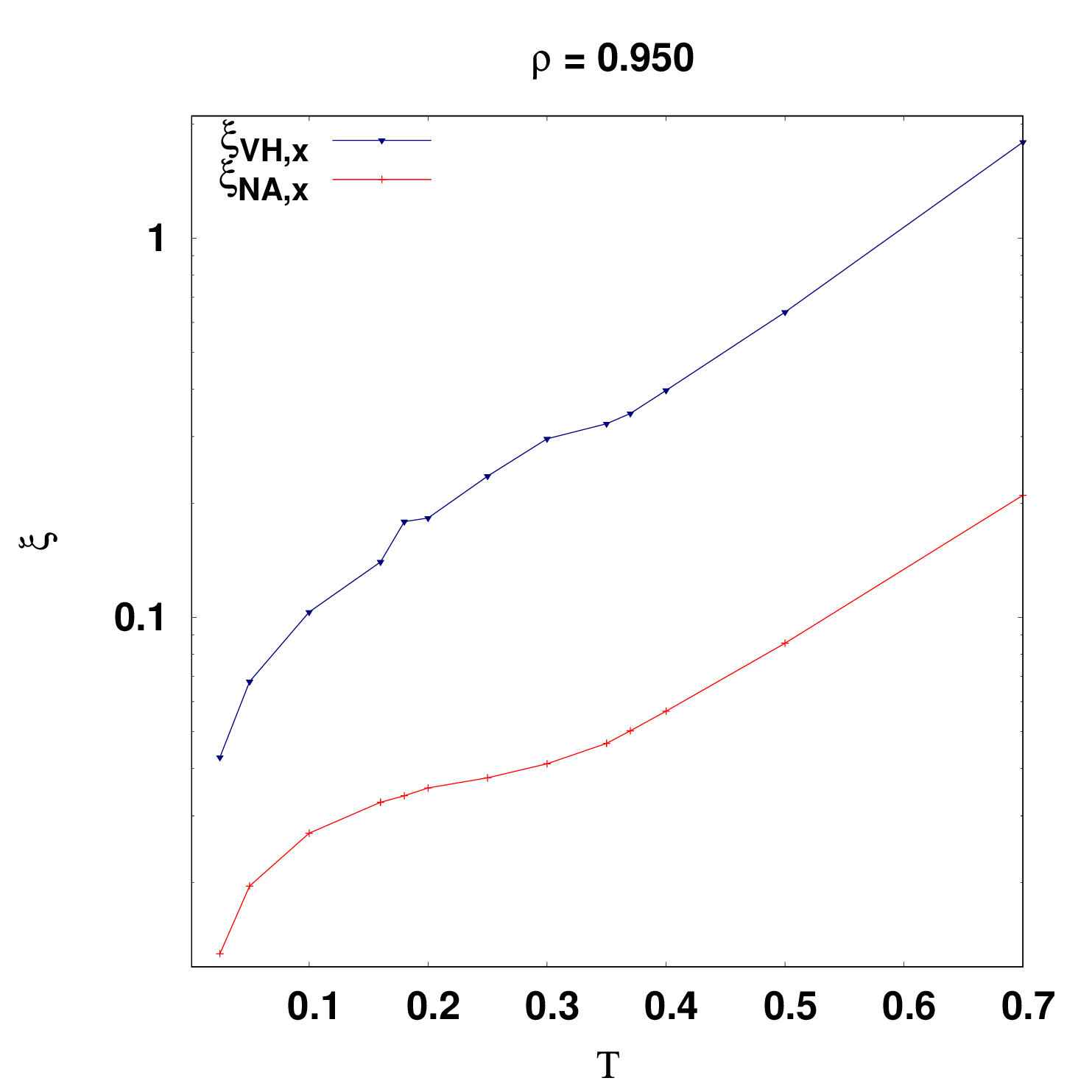}\hspace{8mm} 
    \includegraphics[width=0.37\textwidth,height=0.40\textwidth,keepaspectratio]{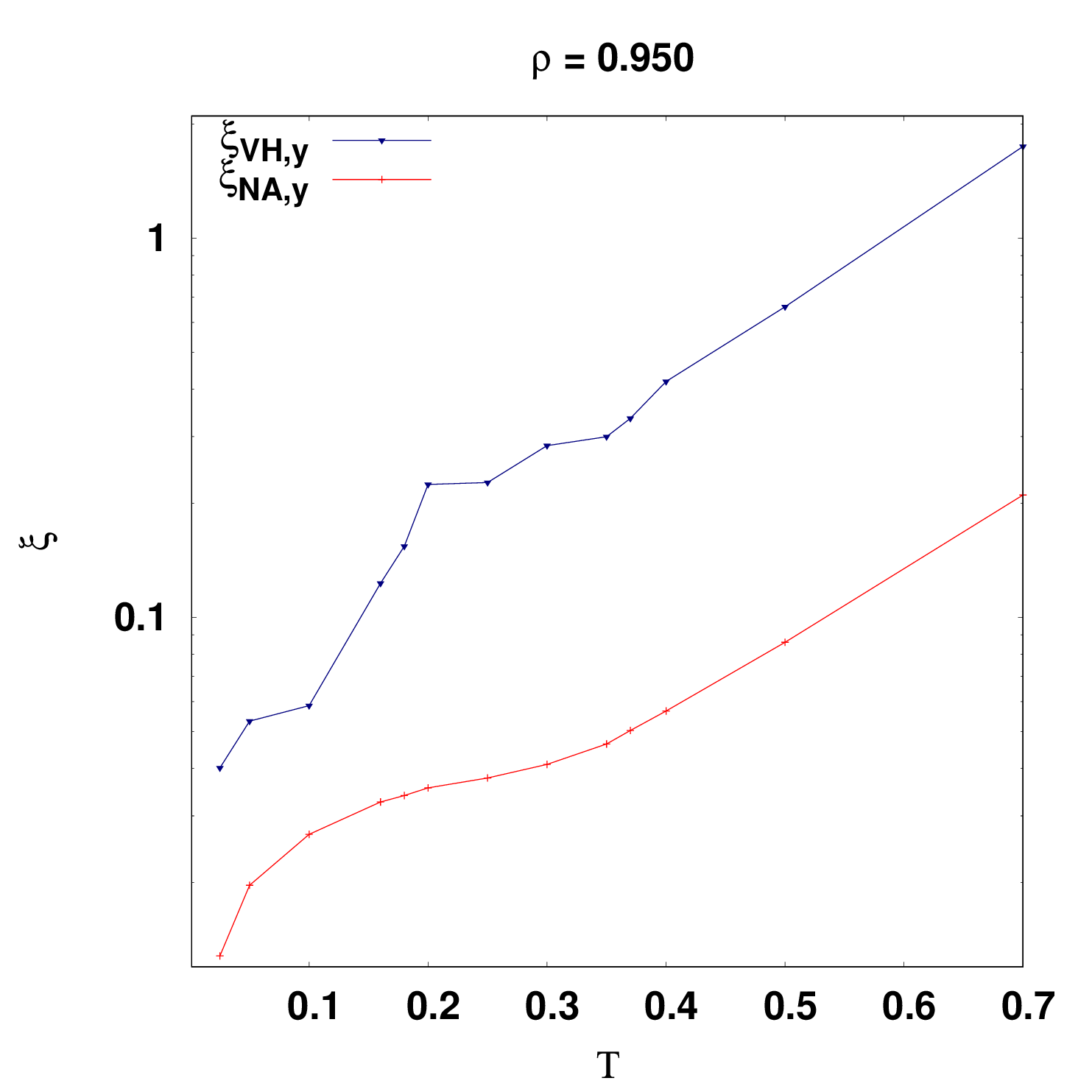} \\
    \includegraphics[width=0.37\textwidth,height=0.40\textwidth,keepaspectratio]{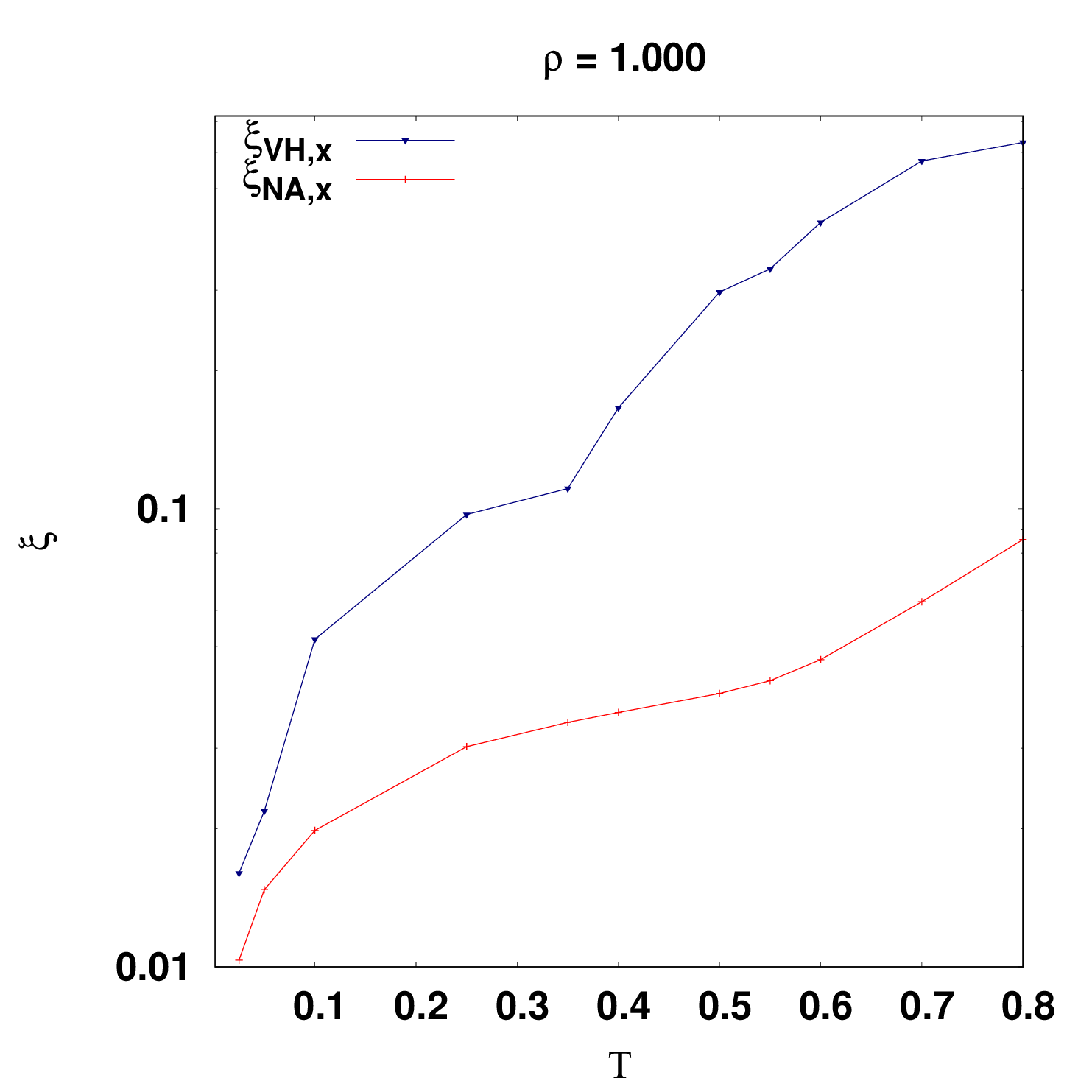}\hspace{8mm} 
    \includegraphics[width=0.37\textwidth,height=0.40\textwidth,keepaspectratio]{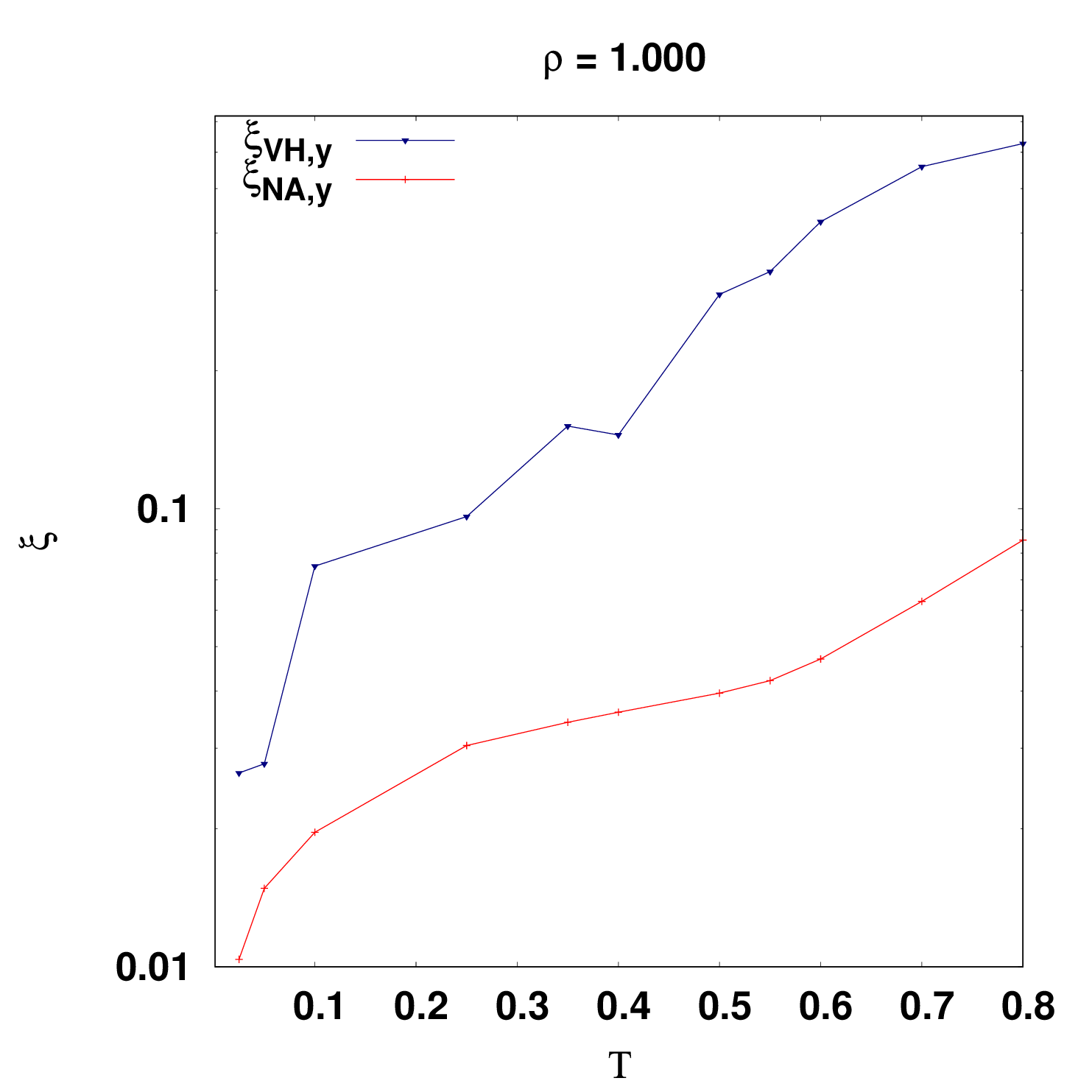}
  \end{tabular}
  \caption{Comparison of von Hove and nonaffine characteristic length scales for the $x$ (left column) and $y$ (right column) components at three densities. Top row: $\rho=0.900$; middle row: $\rho=0.950$; bottom row: $\rho=1.000$. In each panel, the blue curve shows the von Hove length scale $\xi_{\mathrm{VH},\alpha}$ and the red curve shows the nonaffine length scale $\xi_{\mathrm{NA},\alpha}$ (component $\alpha=x,y$). The slope-fitting and tail-selection protocol are identical to that used in Sec.~\ref{sec:nonaffine_results} and Sec.~\ref{sec:vanhove_vs_nonaffine}, so the two length scales are directly comparable. Across all densities and temperatures shown here, the von Hove length is systematically larger than the corresponding nonaffine length, i.e.\ \(\xi_{\mathrm{VH},\alpha}>\xi_{\mathrm{NA},\alpha}\) for every panel(the blue curve represents the von Hove length scale $\xi_{\mathrm{VH},\alpha}$ and the red curve represents the nonaffine length scale $\xi_{\mathrm{NA},\alpha}$ (component $\alpha=x,y$)). In the deeply glassy regime, the gap between the two length scales narrows, consistent with a strong suppression of nonaffine activity.}
  \label{fig:xi_VH_vs_NA_all}
\end{figure*}
\begin{figure*}[htbp!]
  \centering
  \begin{tabular}{cc}
    \includegraphics[width=0.46\textwidth,height=0.46\textwidth,keepaspectratio]{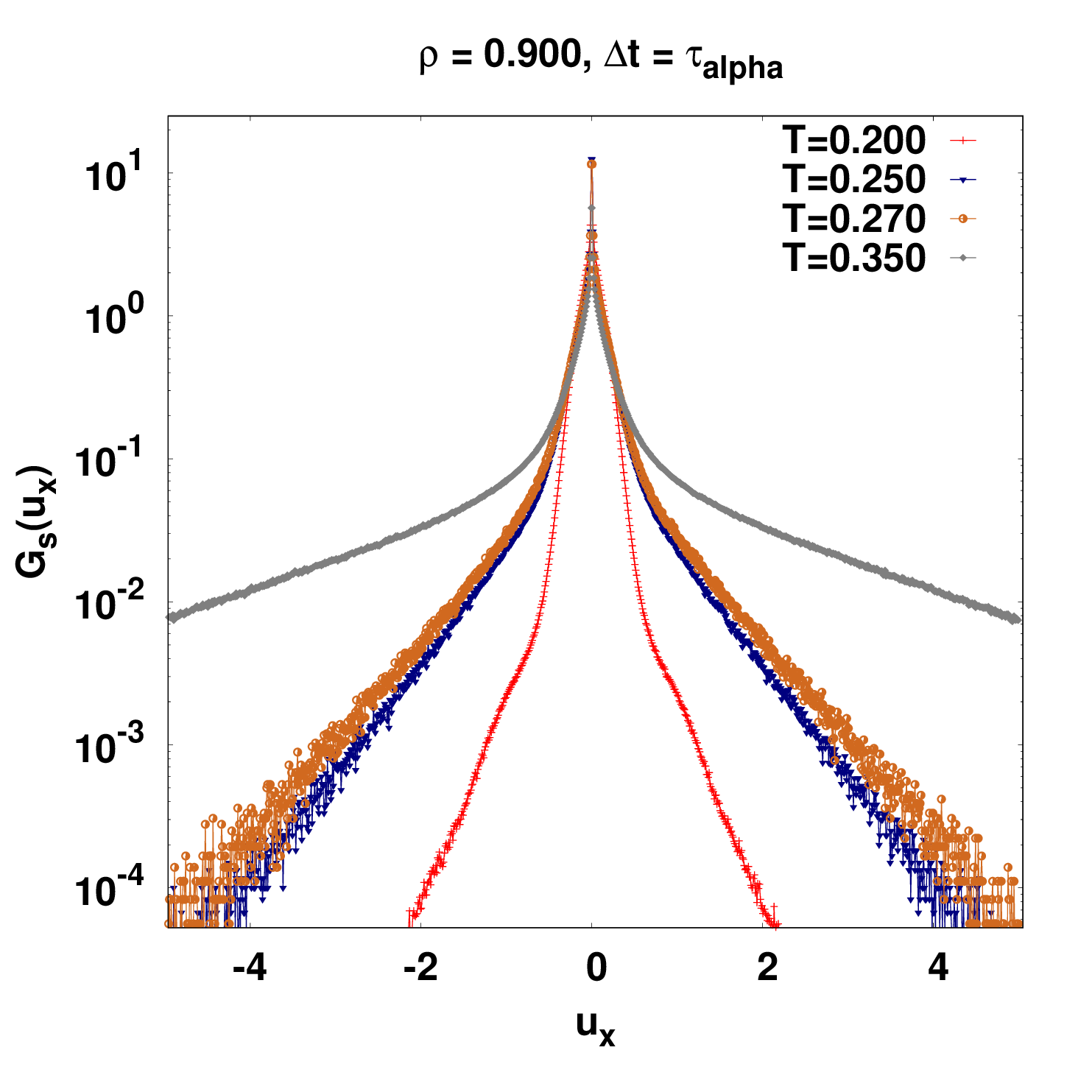}\hspace{8mm} 
    \includegraphics[width=0.46\textwidth,height=0.46\textwidth,keepaspectratio]{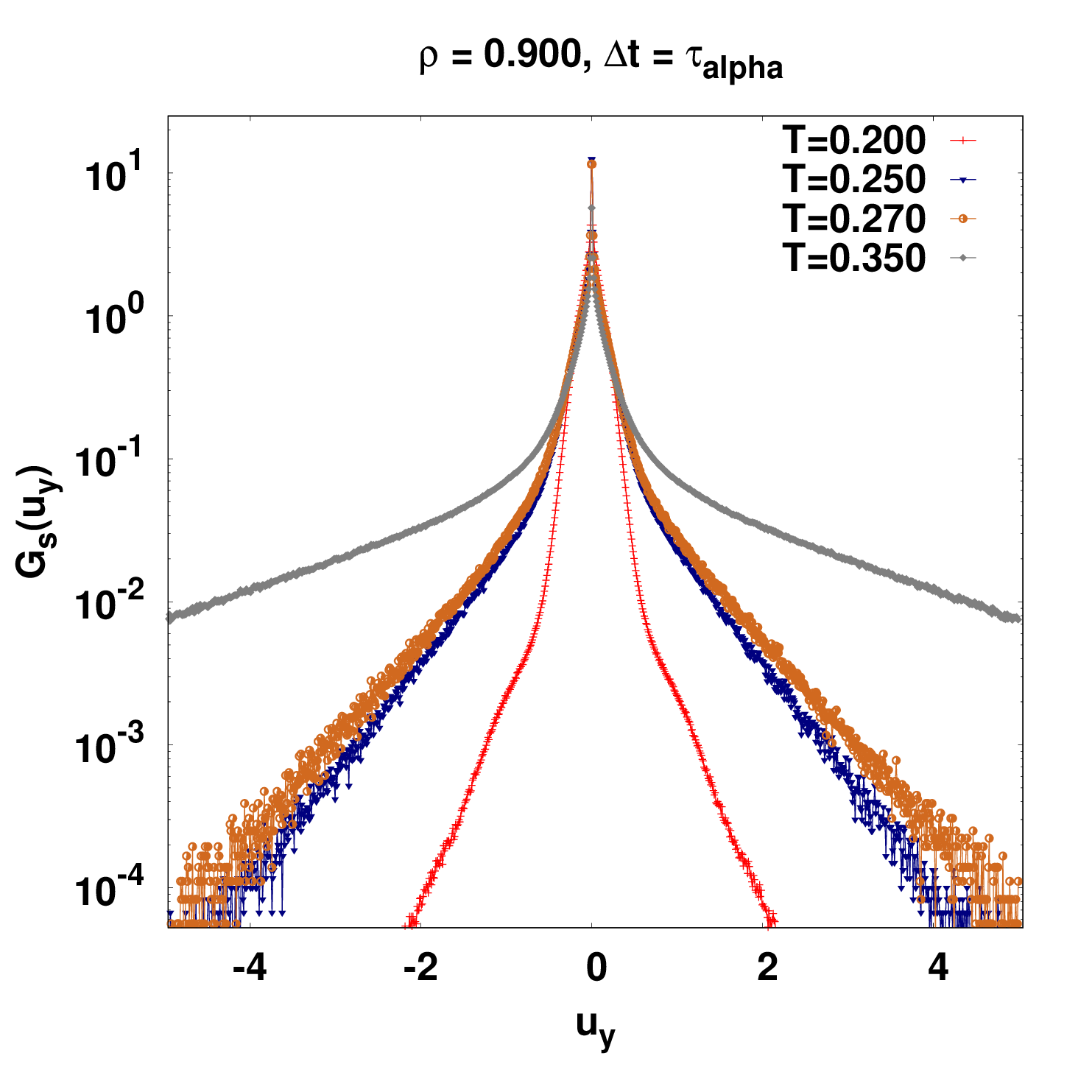} \\
    \includegraphics[width=0.46\textwidth,height=0.46\textwidth,keepaspectratio]{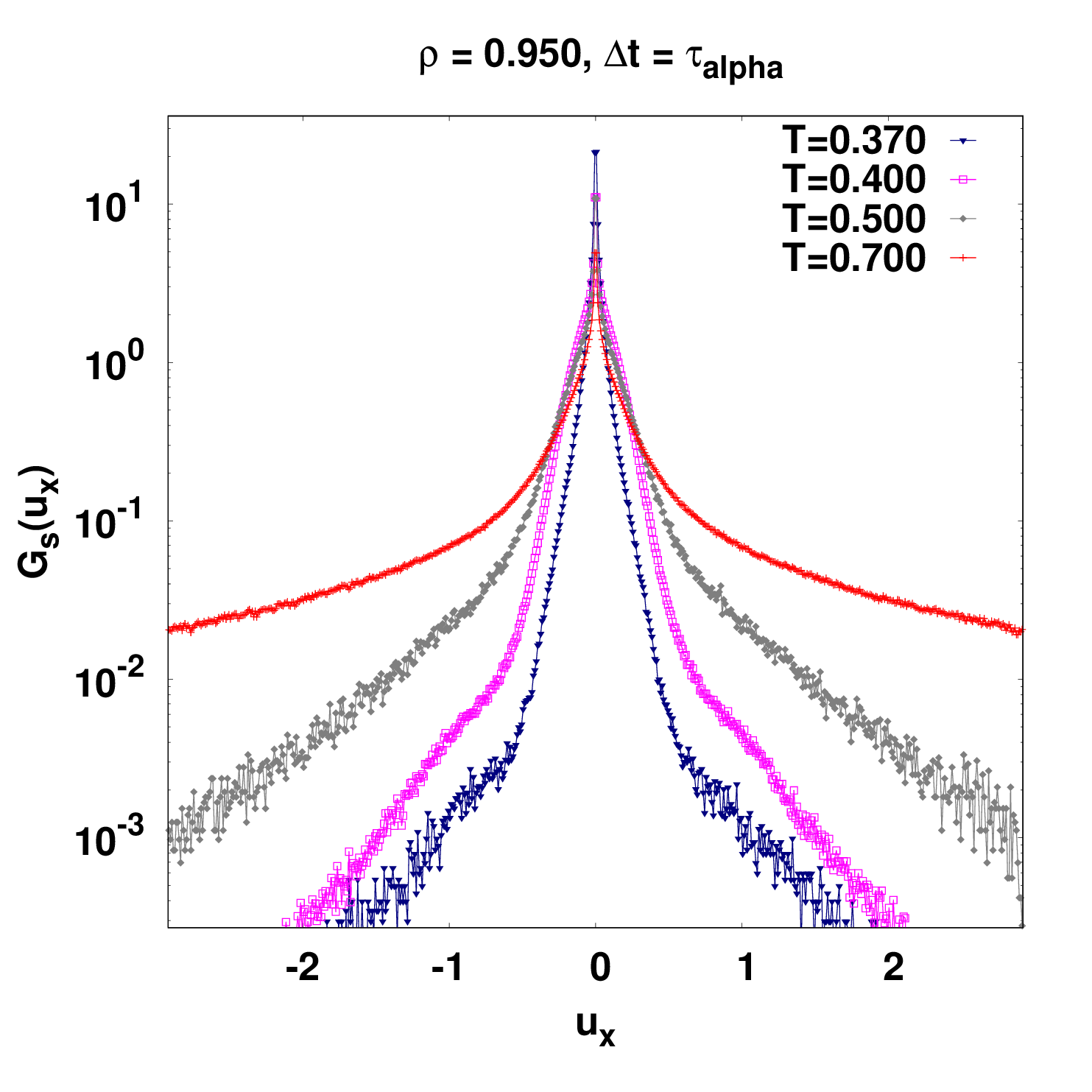}\hspace{8mm} 
    \includegraphics[width=0.46\textwidth,height=0.46\textwidth,keepaspectratio]{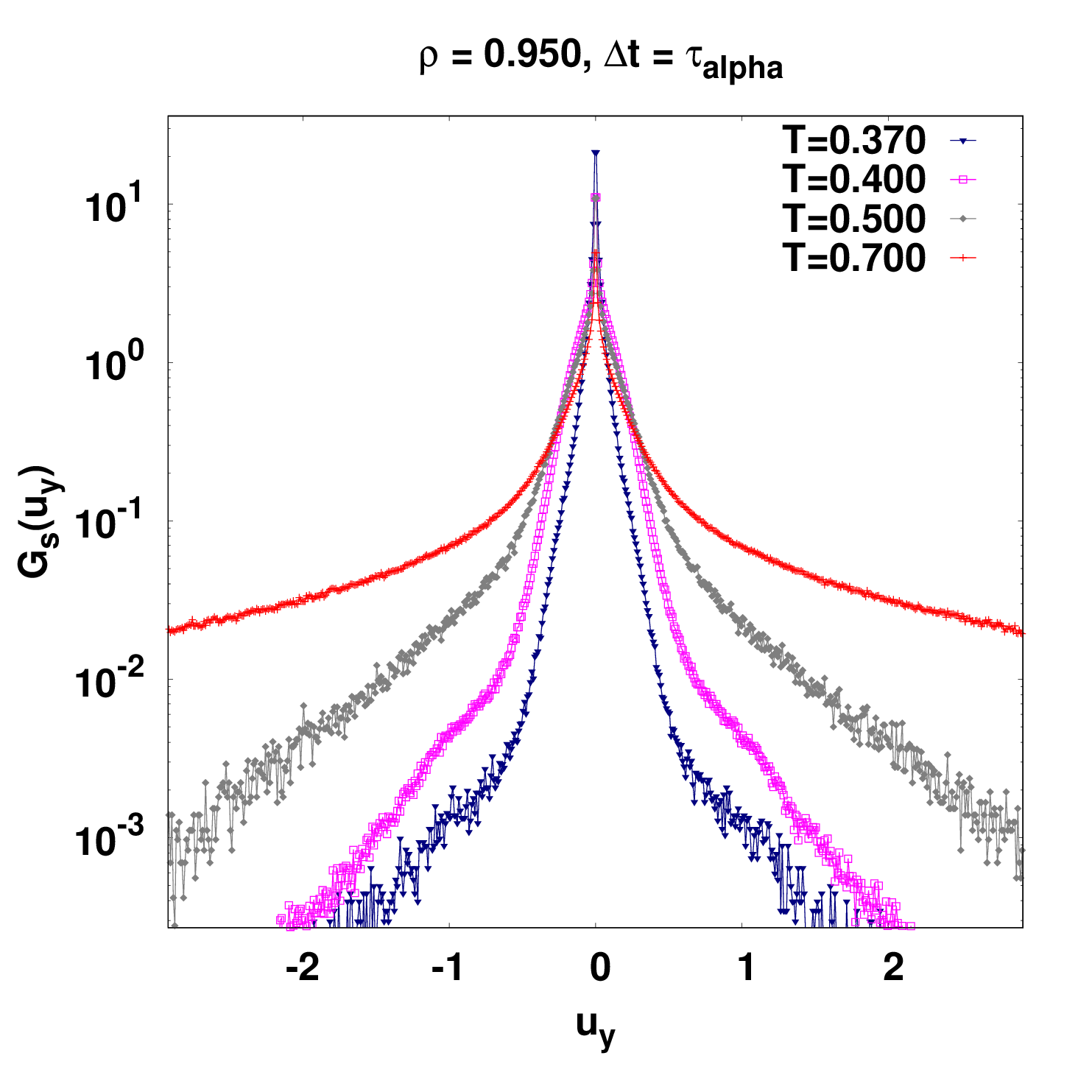} \\
    \includegraphics[width=0.46\textwidth,height=0.46\textwidth,keepaspectratio]{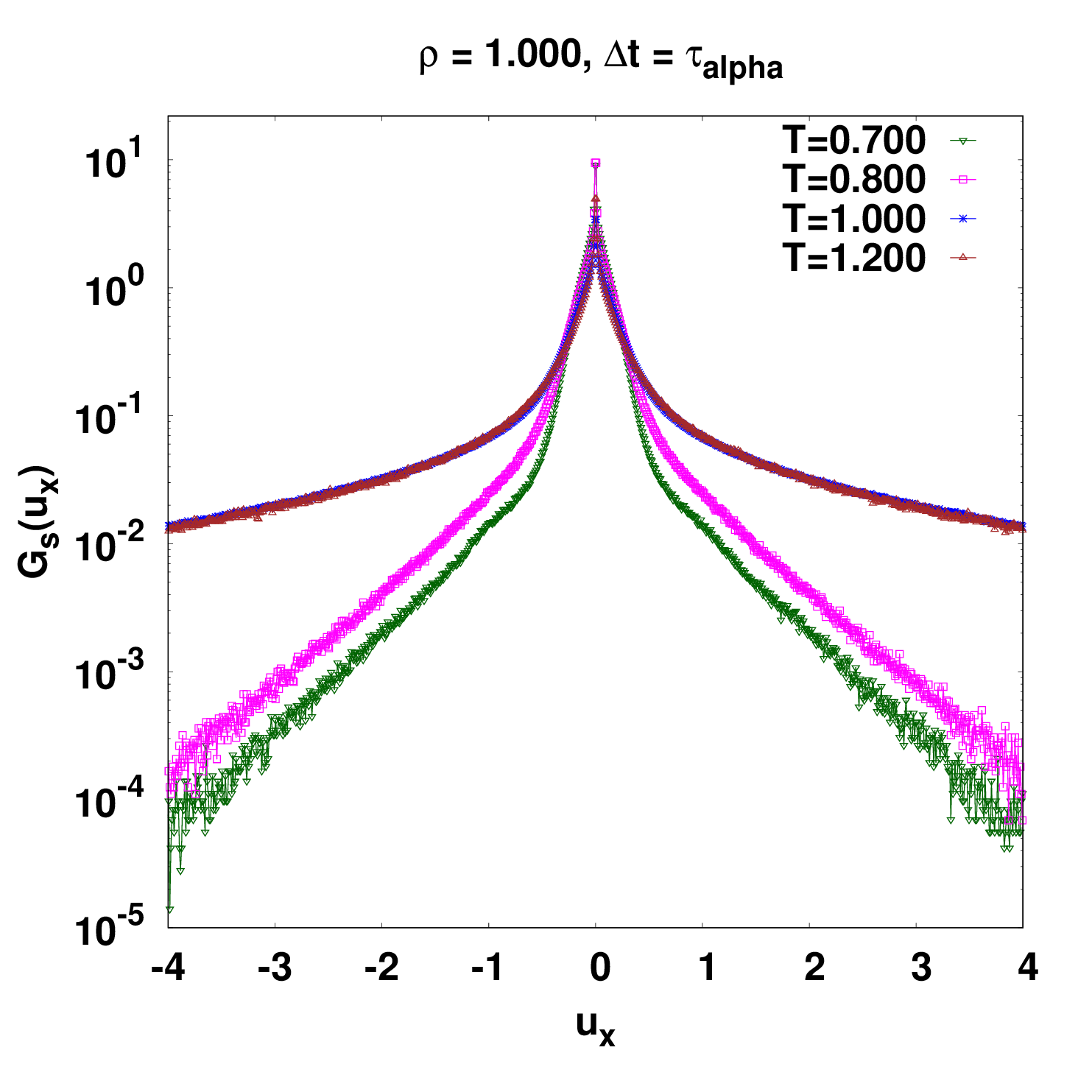}\hspace{8mm} 
    \includegraphics[width=0.46\textwidth,height=0.46\textwidth,keepaspectratio]{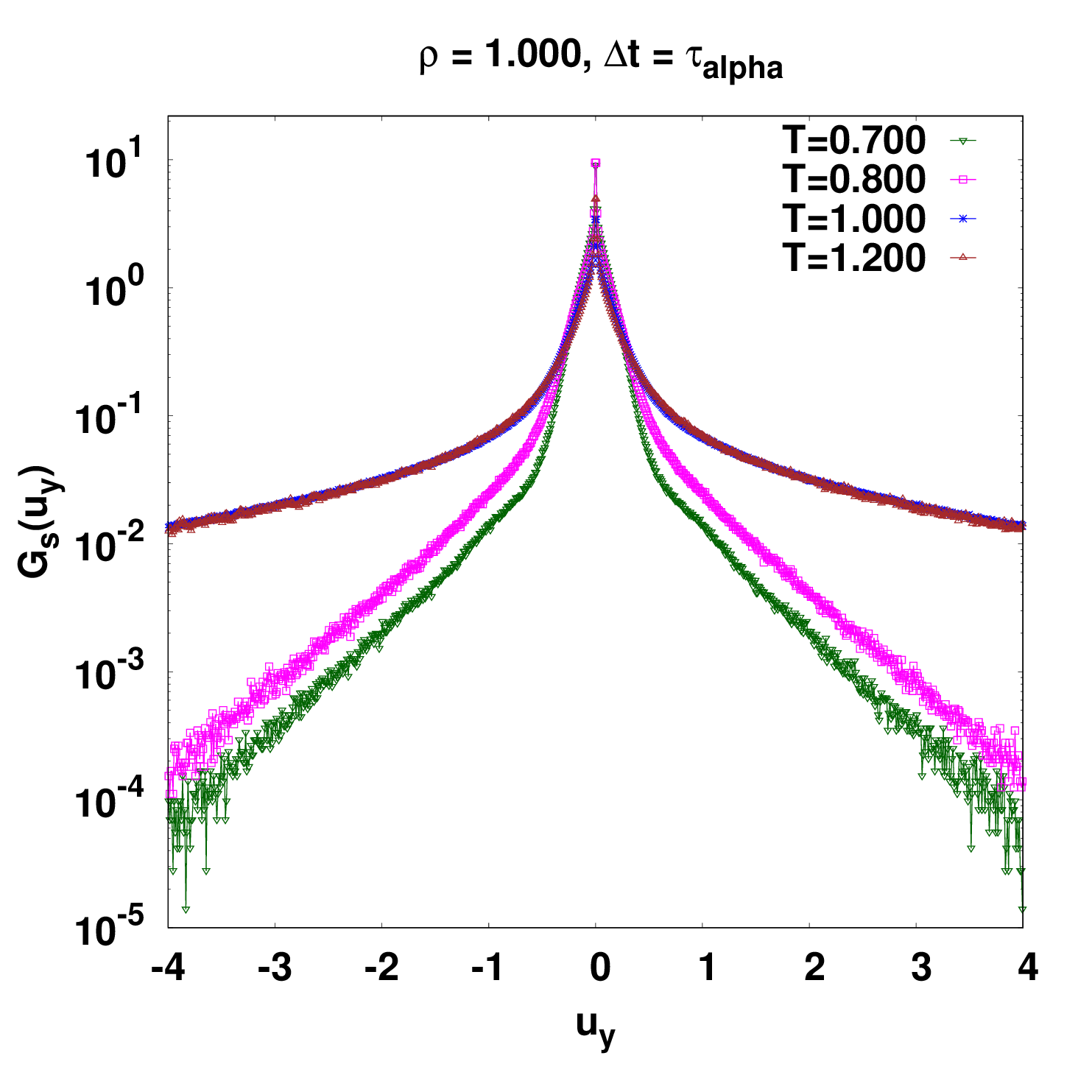}
  \end{tabular}
  \caption{Signed Cartesian van Hove self-distributions for the $x$ (left column) and $y$ (right column) components at three densities: $\rho=0.900$ (top row), $\rho=0.950$ (middle row), and $\rho=1.000$ (bottom row). Each curve represents a distribution computed from frame pairs separated by the structural relaxation time $\tau_{\alpha}$ in the supercooled regime (see legends in the panels). The semilogarithmic exponential tails(linear on semilog plots) are fitted to extract the von Hove length scales \(\xi_{\mathrm{VH},x}\) and \(\xi_{\mathrm{VH},y}\).}
   \label{fig:vh_all_signed}
\end{figure*}
\begin{figure*}[htb!]
  \centering
  \begin{tabular}{cc}
    \includegraphics[width=0.45\textwidth,height=0.45\textwidth,keepaspectratio]{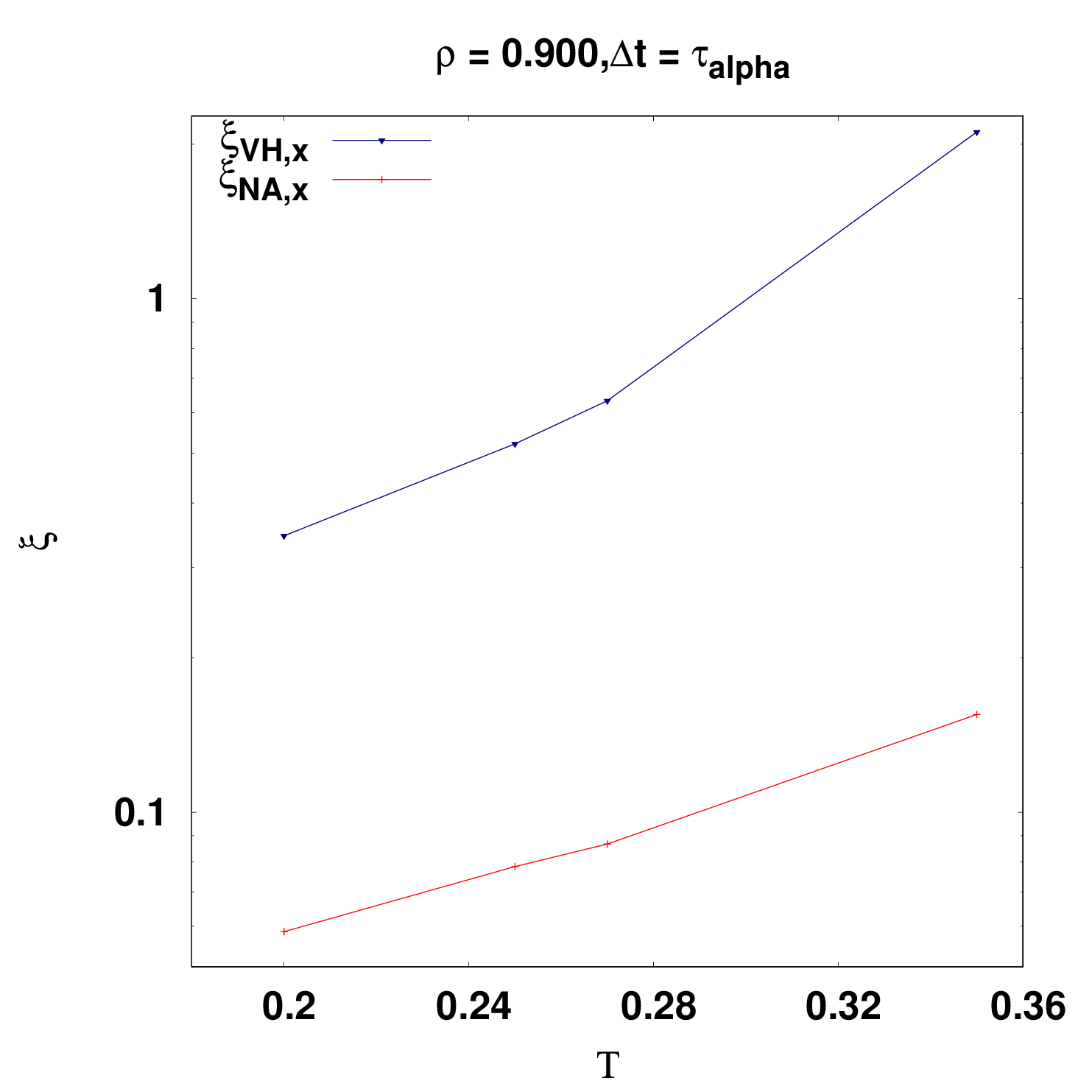}\hspace{8mm} 
    \includegraphics[width=0.45\textwidth,height=0.45\textwidth,keepaspectratio]{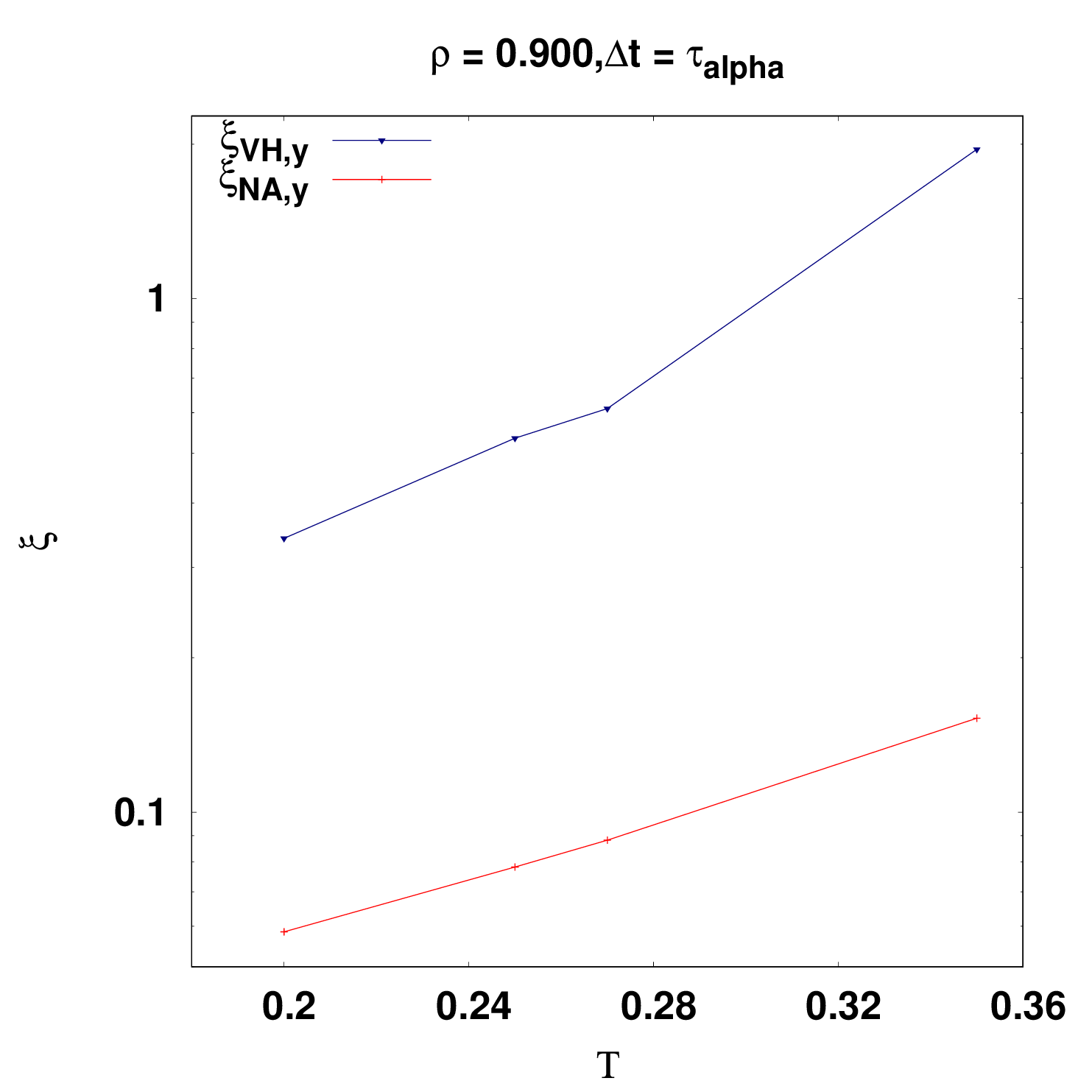} \\
    \includegraphics[width=0.45\textwidth,height=0.45\textwidth,keepaspectratio]{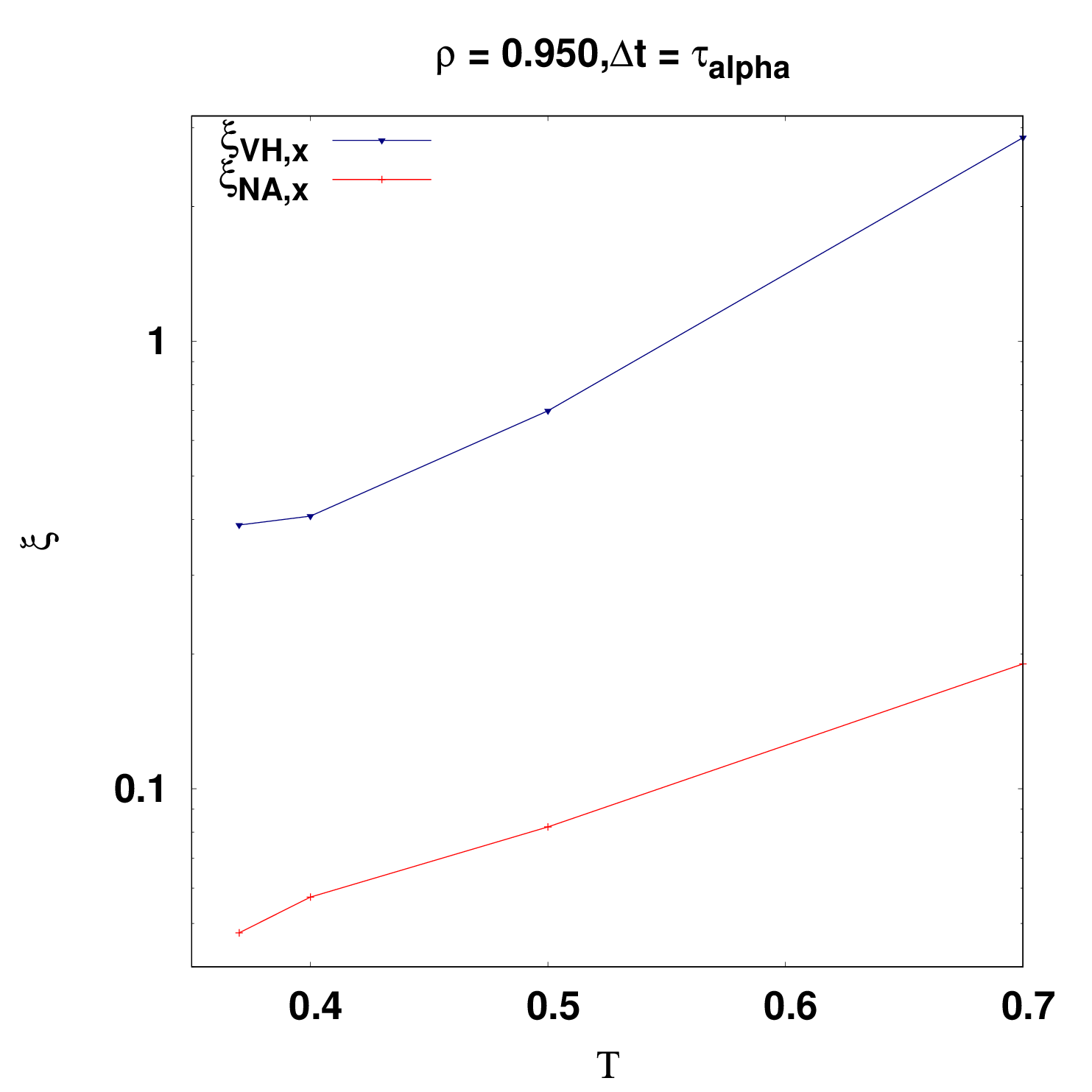}\hspace{8mm} 
    \includegraphics[width=0.45\textwidth,height=0.45\textwidth,keepaspectratio]{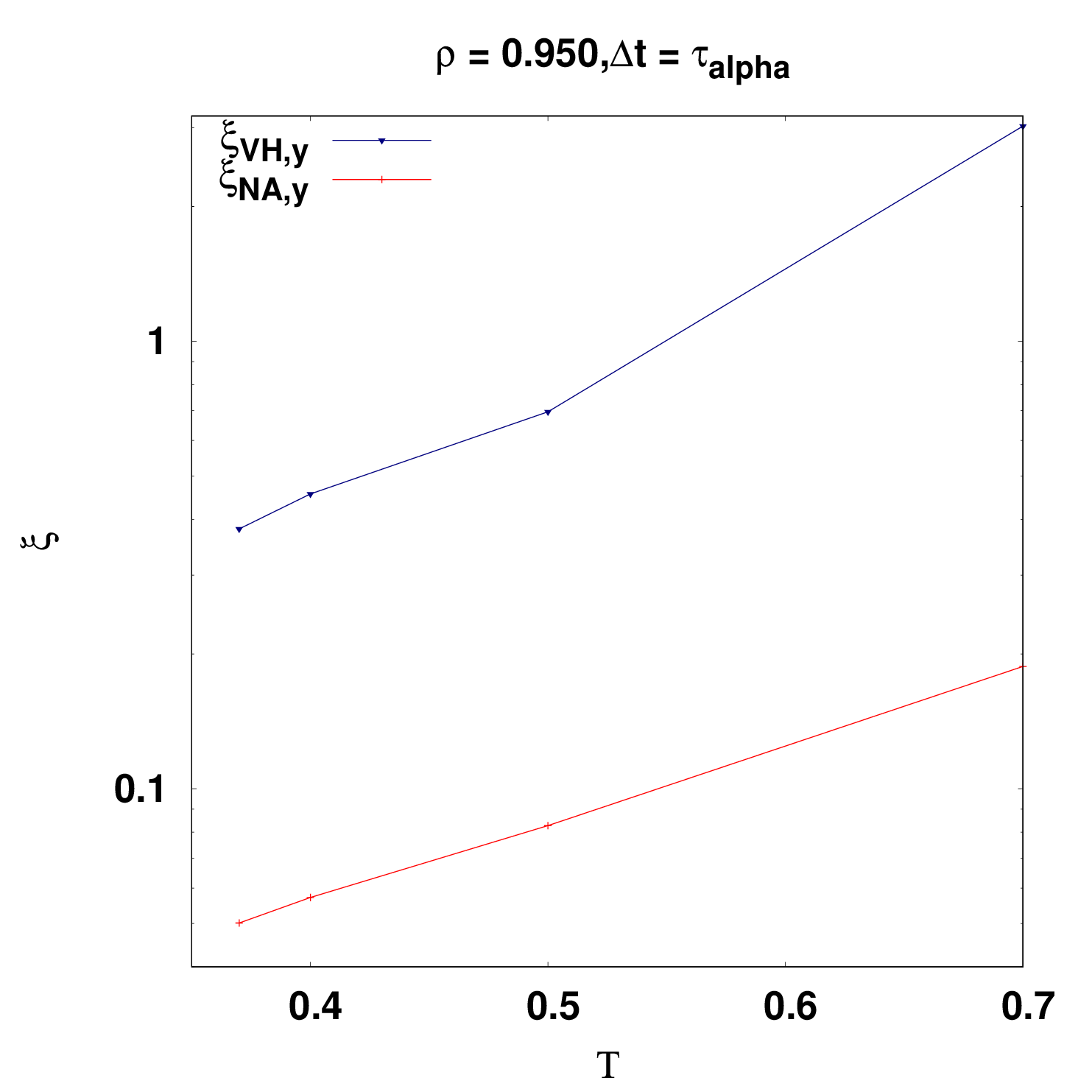} \\
    \includegraphics[width=0.45\textwidth,height=0.45\textwidth,keepaspectratio]{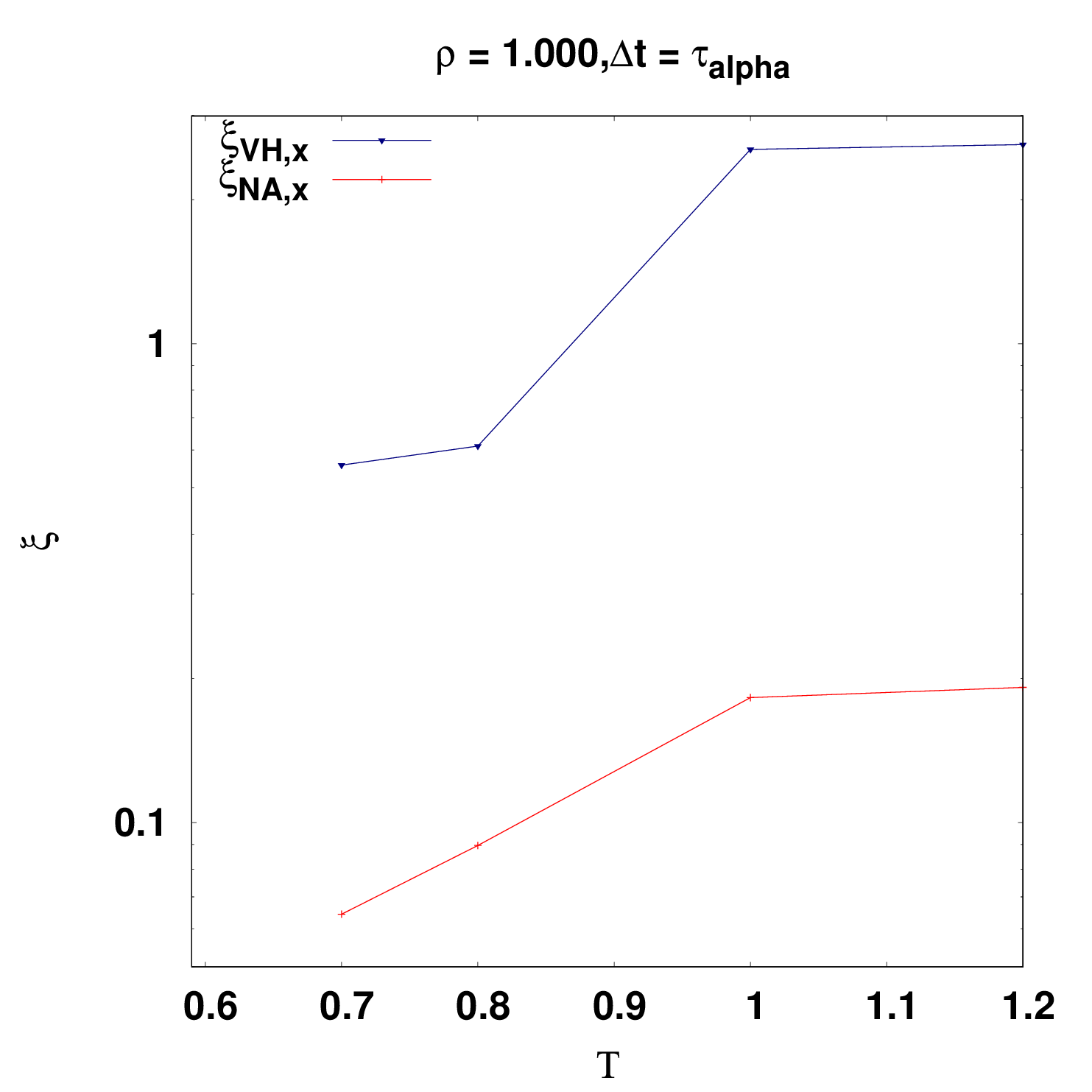}\hspace{8mm} 
    \includegraphics[width=0.45\textwidth,height=0.45\textwidth,keepaspectratio]{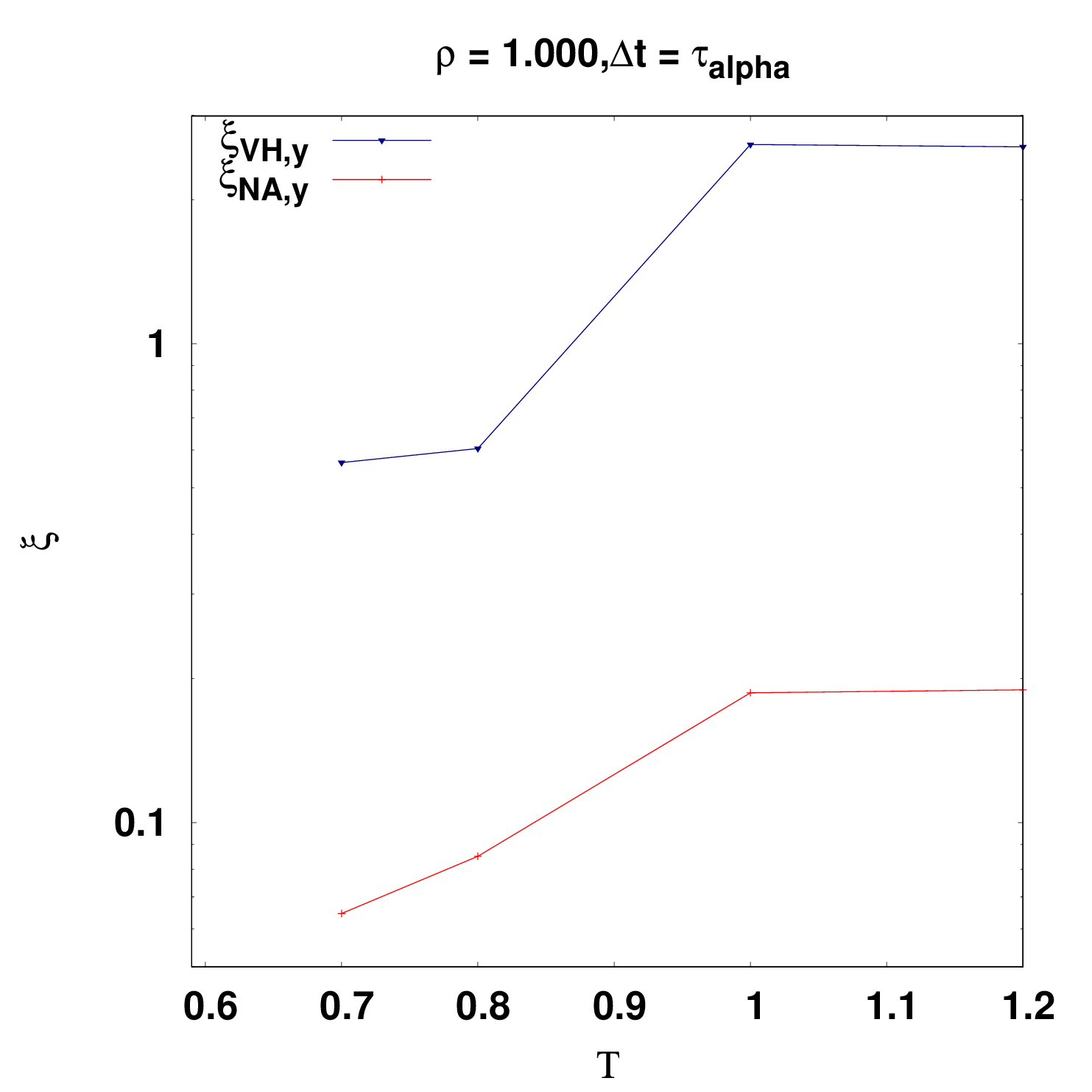}
  \end{tabular}
  \caption{Comparison of von Hove and nonaffine characteristic length scales for the $x$ (left column) and $y$ (right column) components at three densities: $\rho=0.900$ (top row), $\rho=0.950$ (middle row), and $\rho=1.000$ (bottom row). Both quantities are evaluated at the structural relaxation time $\tau_{\alpha}$ in the supercooled regime. In each panel, the blue curve represents the von Hove length scale $\xi_{\mathrm{VH},\alpha}$ and the red curve represents the nonaffine length scale $\xi_{\mathrm{NA},\alpha}$ (component $\alpha=x,y$). The von Hove length scale remains consistently larger than the corresponding nonaffine one, confirming that the robustness of collective displacement correlations persists across temperatures and densities. The analysis employs identical slope-fitting and tail-selection protocols as in Sec.~\ref{sec:nonaffine_results} and Sec.~\ref{sec:vanhove_vs_nonaffine}, ensuring direct comparability.}
  \label{fig:xi_VH_vs_NA_all_supercooled}
\end{figure*}

As defined in Eqs.~\ref{eq:van_hove_x}--\ref{eq:van_hove_y}, the Cartesian van Hove self-distributions \(G_x(u_x,t)\) and \(G_y(u_y,t)\) are the signed single-component displacement PDFs(probability distribution functions) that display the full positive and negative displacement tails. To extract a characteristic von Hove (VH) length scale from the tails, we model the signed distributions by Laplace-like (double-exponential) tails for the total displacement \(\vec{u}\) of the component \(u_x\) and \(u_y\) of the form
\begin{subequations}
\label{eq:vh_tail_model}
\begin{align}
G_x(u_x,t) &\simeq G_x(0,t)\,\exp\!\left(-\frac{|u_x|}{\xi_{\mathrm{VH},x}(t)}\right), \qquad u_x\in\mathbb{R}\\[4pt]
G_y(u_y,t) &\simeq G_y(0,t)\,\exp\!\left(-\frac{|u_y|}{\xi_{\mathrm{VH},y}(t)}\right), \qquad u_y\in\mathbb{R}
\end{align}
\end{subequations}
where \(G_x(0,t)\) and \(G_y(0,t)\) denote the values of the van Hove PDFs at zero displacement for the chosen lag \(t\), and \(\xi_{\mathrm{VH},x}(t)\), \(\xi_{\mathrm{VH},y}(t)\) are the von Hove length scales for the \(x\) and \(y\) components.
Taking natural logarithms gives the linear relations used for fitting the tails:
\begin{subequations}
\label{eq:vh_tail_model_log}
\begin{align}
\ln G_x(u_x,t) &\simeq \ln G_x(0,t) - \frac{|u_x|}{\xi_{\mathrm{VH},x}(t)}\\[4pt]
\ln G_y(u_y,t) &\simeq \ln G_y(0,t) - \frac{|u_y|}{\xi_{\mathrm{VH},y}(t)}
\end{align}
\end{subequations}
For each density and temperature we extract Cartesian van Hove length scales \(\xi_{\mathrm{VH},x}\) and \(\xi_{\mathrm{VH},y}\) by fitting the exponential tails of the signed \(G_x\) and \(G_y\) distributions; these VH scales provide the baseline for a direct, like-for-like comparison with the nonaffine length scales \(\xi_{\mathrm{NA},x}\) and \(\xi_{\mathrm{NA},y}\) (Sec.~\ref{sec:nonaffine_results}). The Cartesian van Hove components of the full displacement field \(\vec{u}\) (shown in Fig.~\ref{fig:vh_all}) are symmetric about zero and exhibit well-defined linear regimes on both sides in semi-log plots, indicating robust exponential tails. Because the affine contribution is smooth and long-wavelength, the exponential tail visible in \(G_{x,y}\) is dominated by the same irreversible, nonaffine rearrangements that our local nonaffine residual procedure is designed to isolate. 
From Fig.~\ref{fig:vh_all} of the total displacement distribution (van Hove), we see that at high temperature the semi-log tail is shallow (small negative slope), indicating that large jumps are relatively frequent and signalling abundant hopping and extended dynamical activity. As $T$ decreases the tail steepens (the slope becomes more negative): the probability of large displacements falls rapidly and hopping events become rarer. In the deep glass the tail is very steep (large negative slope): large jumps are effectively suppressed and the dynamics become strongly localized (solid-like). The semi-log tail slope of the total displacement distribution follows $1/\xi_{\mathrm{VH}}$, and it grows as $T$ decreases, signifying a crossover from hopping-dominated, extended dynamics at high $T$ to strongly localized, glassy behaviour at low $T$. \emph{A similar trend is seen in the nonaffine distribution, but the two distributions convey different information; comparing their associated length scales reveals their distinct origins and physical meanings}.
Figure~\ref{fig:xi_VH_vs_NA_all} summarizes the result: for every component \(\alpha\), density, and temperature studied we find \(\xi_{\mathrm{VH},\alpha}>\xi_{\mathrm{NA},\alpha}\). The difference is largest at high temperature and shrinks as \(T\) decreases, with a pronounced convergence in the deeply glassy regime where particle caging suppresses nonaffine activity and affine motion dominates. The ordering and temperature dependence hold for both Cartesian components and for all three densities (absolute values and sensitivities decrease with increasing density). These observations are consistent with the variance-decomposition analytical explanation in Sec.~\ref{sec:theory_methods}: the van Hove variance contains both the residual (nonaffine) variance and the affine contribution, producing a systematically larger \(\xi_{\mathrm{VH}}\) that approaches \(\xi_{\mathrm{NA}}\) as the nonaffine contribution weakens.\\
\begin{figure*}[htbp!]
  \centering
\includegraphics[width=0.46\textwidth,height=0.46\textwidth,keepaspectratio]{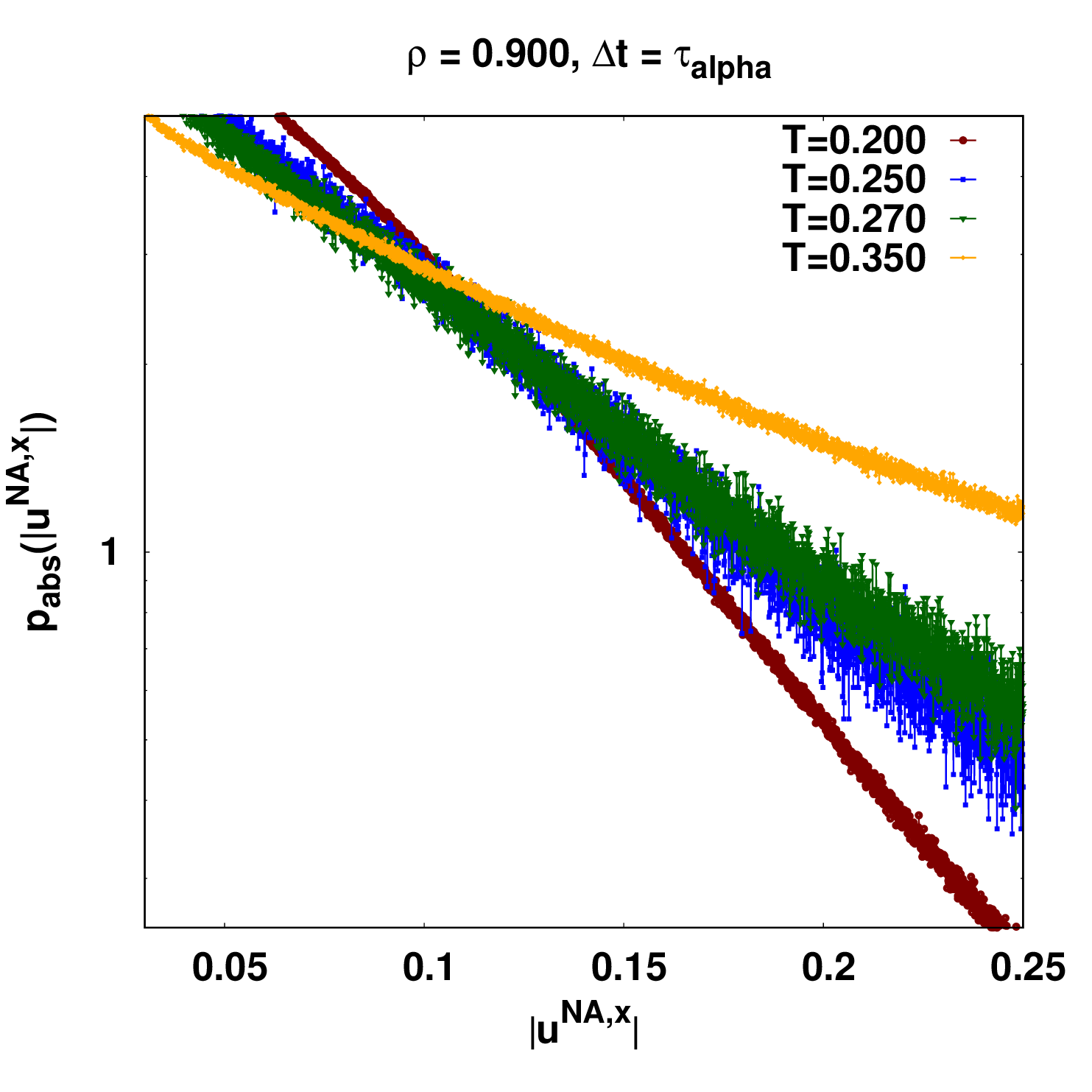}\hspace{8mm}
    \includegraphics[width=0.46\textwidth,height=0.46\textwidth,keepaspectratio]{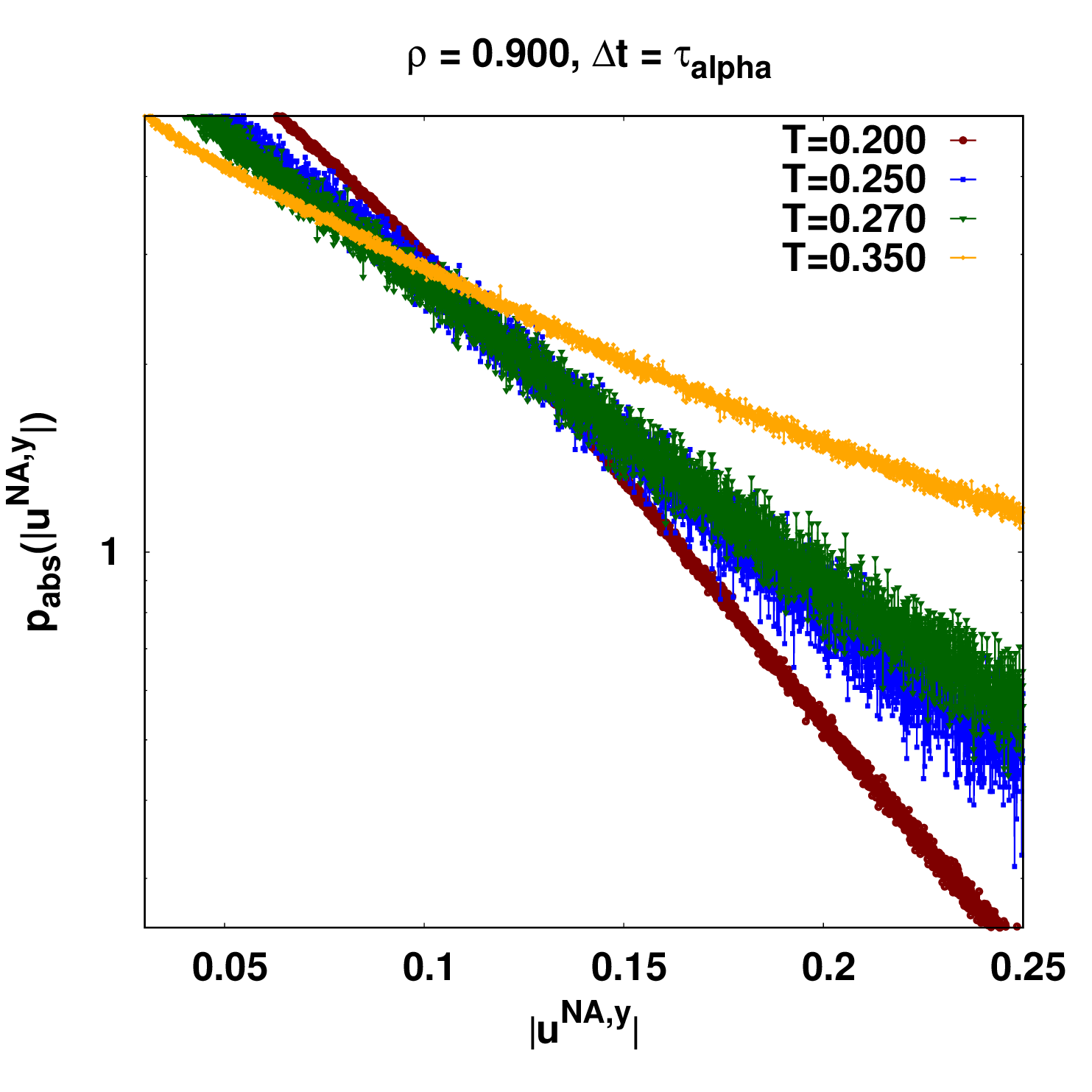} \\
    \includegraphics[width=0.46\textwidth,height=0.46\textwidth,keepaspectratio]{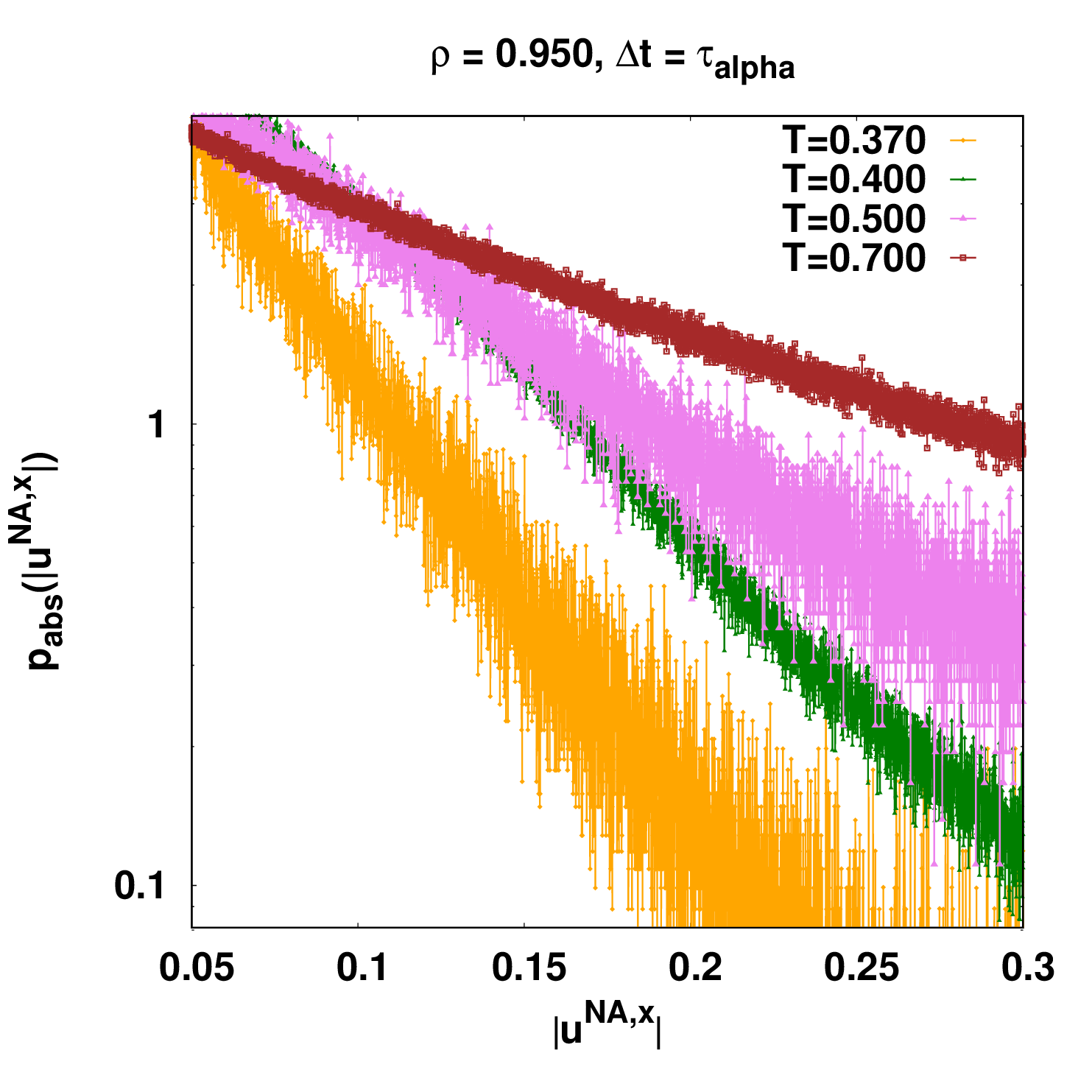}\hspace{8mm} 
    \includegraphics[width=0.46\textwidth,height=0.46\textwidth,keepaspectratio]{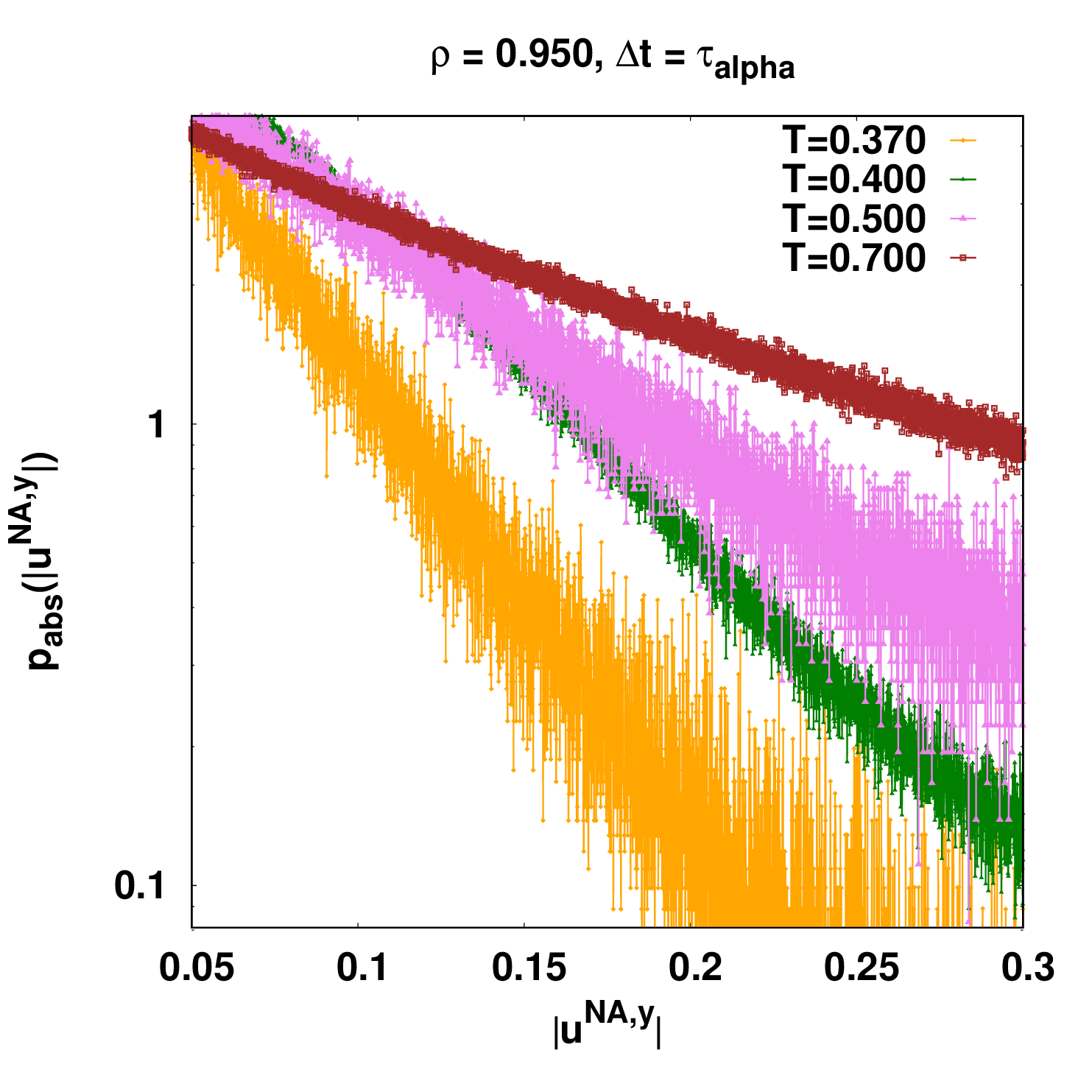} \\
    \includegraphics[width=0.46\textwidth,height=0.46\textwidth,keepaspectratio]{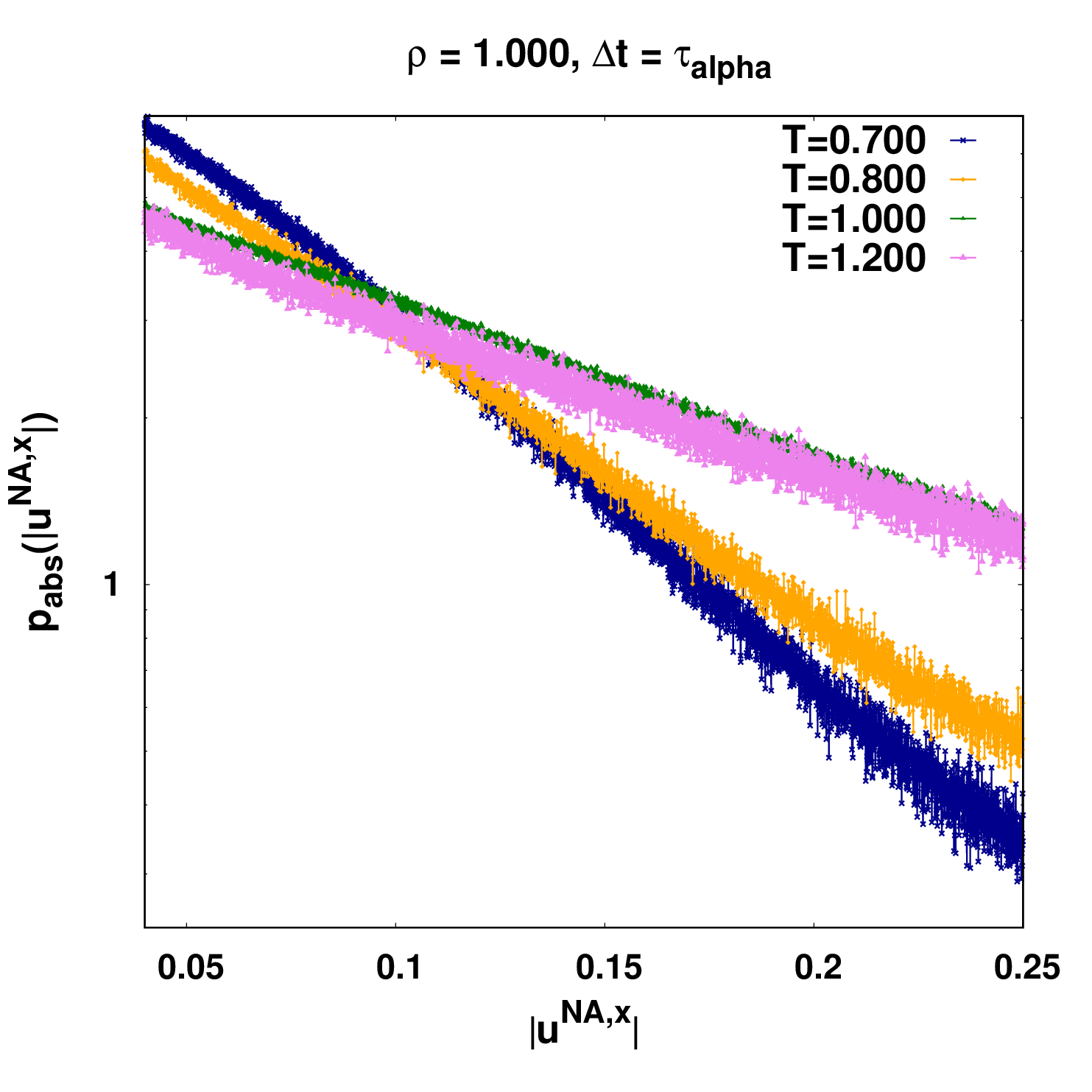}\hspace{8mm} 
    \includegraphics[width=0.46\textwidth,height=0.46\textwidth,keepaspectratio]{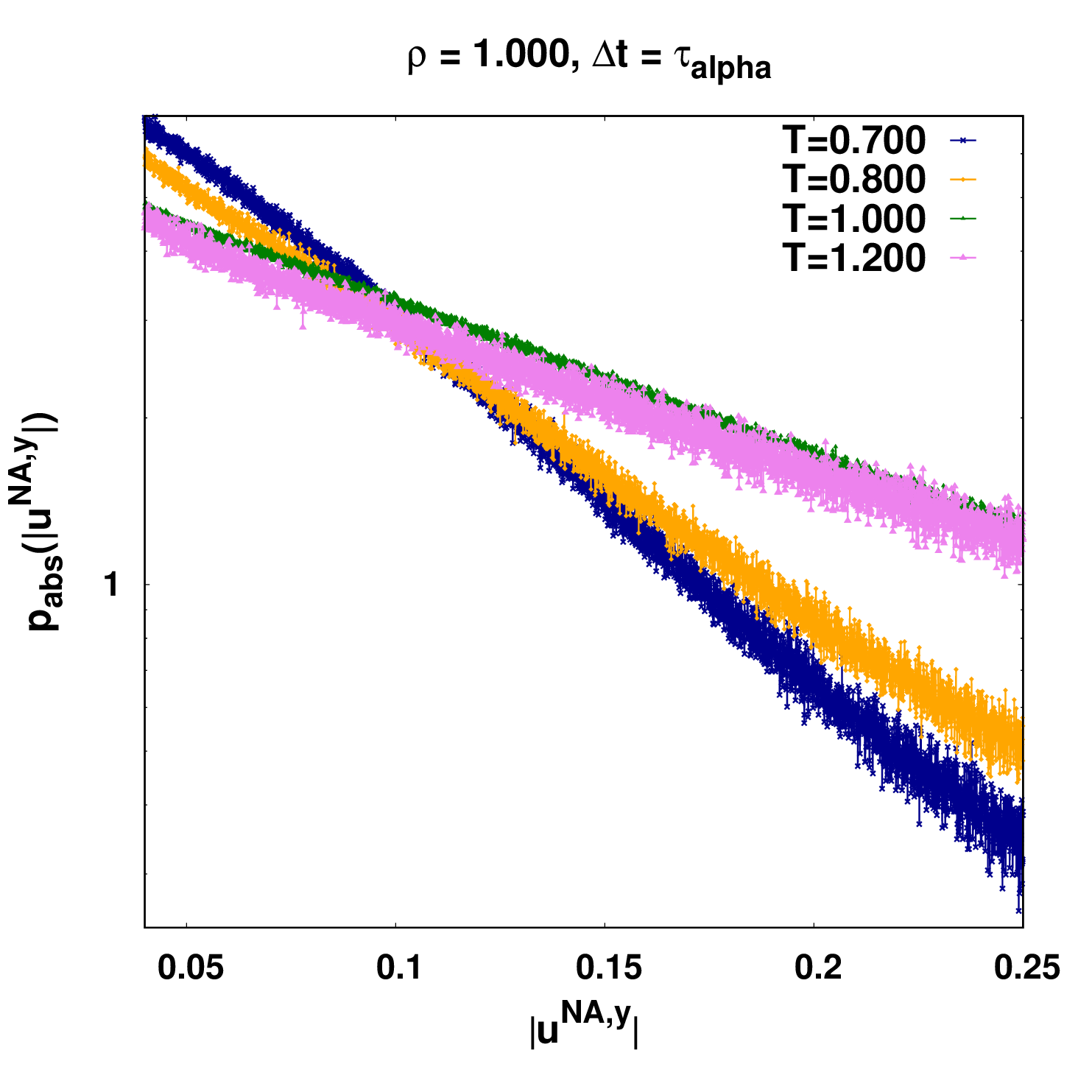}
  
  \caption{Folded (absolute-value) nonaffine displacement distributions at the fix structural relaxation lag \(\Delta t=\tau_{\alpha}\)(alpha relaxation time) in supercooled regime. Left column: \(|u_{\mathrm{NA},x}|\) distributions; right column: \(|u_{\mathrm{NA},y}|\) distributions. Rows (top to bottom) correspond to densities \(\rho=0.900,\;0.950,\;1.000\). Each curve denotes a different temperature (see panel legends). Exponential tails are fit on a semilog scale using the same tail-selection and fitting protocol as in Sec.~\ref{sec:nonaffine_results} to extract the characteristic nonaffine length scales \(\xi_{\mathrm{NA},x,y}(\tau_{\alpha})\) .}
  \label{fig:len_all_nonaffine_folded}
\end{figure*}
As a final, independent validation of the ordering \(\xi_{\mathrm{VH}}>\xi_{\mathrm{NA}}\) we test whether the result depends on the choice of fix time lag \(\Delta t\). In the main comparison above we used consecutive-frame lags; here we repeat the entire extraction procedure in the supercooled regime using the structural (alpha) relaxation time \(\tau_{\alpha}\) as the analysis lag. 
Specifically, for a representative temperature in the supercooled regime we set \(\Delta t=\tau_{\alpha}\) (see Sec.~\ref{subsec:temporal_sampling} for details) and compute the folded nonaffine PDFs \(p_{\mathrm{abs}}^{\mathrm{NA},x}(|u^{\mathrm{NA},x}|;\Delta t)\), \(p_{\mathrm{abs}}^{\mathrm{NA},y}(|u^{\mathrm{NA},y}|;\Delta t)\) together with signed van Hove PDFs \(G_x(x;\Delta t)\), \(G_y(y;\Delta t)\). Figures~\ref{fig:len_all_nonaffine_folded} and~\ref{fig:vh_all_signed} show, respectively, the folded nonaffine and signed van Hove displacement distributions in the supercooled regime evaluated at the time lag \(\Delta t=\tau_{\alpha}\).
Using the identical semilog tail-selection and fitting protocol introduced above (fit \(\ln G_\beta\))  for the van Hove tails and \(\ln p_{\mathrm{abs}}^{\mathrm{NA},\beta}\) for the folded nonaffine tails, where
\(\beta\) denotes the Cartesian components \(x\) and \(y\)), we extract $\xi_{\mathrm{VH},x}(\tau_{\alpha})$ and $\xi_{\mathrm{NA},x}(\tau_{\alpha})$ (and similarly $\xi_{\mathrm{VH},y}(\tau_{\alpha})$ and $\xi_{\mathrm{NA},y}(\tau_{\alpha})$). The comparison at \(\Delta t=\tau_{\alpha}\) as shown in Figs.~\ref{fig:xi_VH_vs_NA_all_supercooled} interestingly reproduces the same ordering observed with consecutive-frame pairs, namely $\xi_{\mathrm{VH},\beta} \;>\; \xi_{\mathrm{NA},\beta}$. This confirms that the inequality is not an artefact of the chosen time lag: the larger van Hove length scale and the smaller, folded nonaffine length scale are robust features of the displacement statistics across components and densities.
\clearpage

\section{Conclusion}
\label{sec:conclusion}
A key new insight from our work is the introduction and systematic study of a nonaffine length scale obtained by fitting the exponential tails of the folded (absolute-value) nonaffine displacement distributions. We deliberately focus on the tail region (the linear regime on a semilog plot) because it is in this zone that nonlinear, higher-order contributions to particle motion dominate. The folded nonaffine distribution preferentially isolates these nonlinear nonaffine contributions from the total displacement field \(u\), producing a length scale that characterises the spatial extent of genuinely nonaffine, higher-order rearrangements. Notably, we find that, while the linear (semilog) tail region broadens as temperature is decreased, the fitted nonaffine length scale \(\xi_{\mathrm{NA}}\) itself decreases with decreasing temperature; this decrease is especially pronounced deep inside the stable glass regime where particle caging is strongest.

Our study provides fresh insight into the interplay between thermal fluctuations, density, and structural disorder in amorphous solids. By quantifying nonaffine rearrangements through characteristic length scales, we demonstrate that thermal activation alone, without externally applied strain, drives significant local rearrangements over a broad thermodynamic range. The temperature dependence of the nonaffine length scale exhibits distinct regimes: at high temperatures \(\xi_{\mathrm{NA}}\) is relatively large (more extended nonaffine activity); in an intermediate window, it shows slower variation; and in the deeply stable glassy state it declines sharply as particles become strongly caged. 
These trends occur across all densities, with the plateau region becoming broader as density increases. The nearly identical values of the characteristic length scales in the \(x\) and \(y\) directions, \(\xi_{\mathrm{NA},x}\approx\xi_{\mathrm{NA},y}\), confirm that thermal nonaffine rearrangements are isotropic in our disordered system. By contrast, in shear-driven systems nonaffine displacements are anisotropic and are typically aligned with the applied shear direction. \added{The absence of such directional bias in our thermal protocol highlights a fundamental difference in the propagation of nonaffine fields between thermal activation and shear driving.}
We next compare this nonaffine length scale with the von Hove length scale extracted from the signed Cartesian van Hove distributions (the full, signed single-component displacement PDFs). The von Hove exponential-tail fits return consistently larger length scales across densities and temperatures, while the separation between the von Hove and nonaffine scales narrows in the deep glassy regime. This ordering is explained by our developed theory of variance-of-displacement decomposition: the von Hove variance contains both affine and nonaffine contributions, whereas the folded nonaffine variance captures only the residual nonaffine part. Because the von Hove length scale maps to the total variance while the nonaffine scale maps only to the residual variance, the von Hove length is naturally larger except when the affine contribution is negligible.
Conceptually, the procedure we propose acts as a selective filter on displacement statistics: by folding and fitting only the exponential tail of the nonaffine residual we `pass' the nonlinear, higher-order contributions and `suppress' the dominant linear (affine) part. By analogy with filters, our method isolates the relevant ``deformation band'' (here: the spatial organizational scale of nonlinear rearrangements) while attenuating the affine background. 
Overall, our analytics and numerical measurements establish that nonaffine motion is a decisive factor in setting the physical state of amorphous solids: the magnitude and temperature sensitivity of the nonaffine length scale correlate with the system’s progression from a high-temperature supercooled liquid to a stable glass. The nonaffine-filtering approach we introduce provides a focused, robust measure of nonlinear, irreversible rearrangements and therefore offers a new, experimentally accessible diagnostic for glassy dynamics. It will be interesting to apply this method in experiments and other simulated glass formers to probe the universality (or otherwise) of the nonlinear length scale and its relation to mechanical and dynamical response.\\
In closing, our work underscores the significance of nonaffine motion as a dominant mechanism that bridges thermal fluctuations with local structural reorganization, and it provides a practical framework for isolating and quantifying the nonlinear spatial scale of thermally induced rearrangements.
\begin{acknowledgments}
A.J.\ acknowledges receipt of junior and senior research fellowships from the University Grants Commission (UGC), India. 
A.J.\ also thanks the Horizon workstation, the Neo machine in the Physics Department, and the Param Ganga supercomputer at the Indian Institute of Technology Roorkee for computational resources and support.
\end{acknowledgments}
\appendix   
\section{Detailed derivations}
\label{app:math}
\begin{widetext}
\subsection{Laplace distribution and normalization}
\label{app:laplace}
If a one-dimensional variable $u$  displacemnt follows a Laplace distribution with scale parameter $\varepsilon$, write
\begin{equation}\label{eq:len1}
p(u) = c\,e^{-|u|/\varepsilon}
\end{equation}
Normalize $p(u)$ by requiring $\int_{-\infty}^{\infty} p(u)\,du = 1$. Compute
\[
1 = c\int_{-\infty}^{\infty} e^{-|u|/\varepsilon}\,du
= c\left(\int_{-\infty}^{0} e^{u/\varepsilon}\,du + \int_{0}^{\infty} e^{-u/\varepsilon}\,du\right)
\]
Each integral evaluates to $\varepsilon$, so
\begin{equation}\label{eq:c_value}
1 = c(2\varepsilon)\quad\Longrightarrow\quad c=\frac{1}{2\varepsilon}
\end{equation}
Thus the Laplace probability density is
\begin{equation}\label{eq:laplace_pdf}
p(u)=\frac{1}{2\varepsilon}\,e^{-|u|/\varepsilon}
\end{equation}

\subsection{First and second moments}
\label{app:moment}
Because $p(u)$ is an even function, the mean vanishes:
\begin{equation}\label{eq:mean_u}
\langle u\rangle = \int_{-\infty}^{\infty} u\,p(u)\,du = 0
\end{equation}

Compute the second moment:
\begin{align}
\langle u^2\rangle
&= \int_{-\infty}^{\infty} u^2 p(u)\,du
= 2\int_{0}^{\infty} u^2 \frac{1}{2\varepsilon} e^{-u/\varepsilon}\,du
= \frac{1}{\varepsilon}\int_{0}^{\infty} u^2 e^{-u/\varepsilon}\,du \label{eq:u2_start}
\end{align}
Let
\begin{equation}\label{eq:I_def}
I=\int_{0}^{\infty} u^2 e^{-u/\varepsilon}\,du
\end{equation}
Use the substitution $y=u/\varepsilon$ (so $u=\varepsilon y$, $du=\varepsilon\,dy$):
\[
I = \int_{0}^{\infty} (\varepsilon y)^2 e^{-y}\,\varepsilon\,dy = \varepsilon^3 \int_{0}^{\infty} y^2 e^{-y}\,dy
\]
The last integral is the Gamma function $\Gamma(3)=2! =2$, hence
\begin{equation}\label{eq:I_value}
I = 2\varepsilon^3
\end{equation}
Therefore from \eqref{eq:u2_start}
\begin{equation}\label{eq:u2_final}
\langle u^2\rangle = \frac{1}{\varepsilon} I = \frac{1}{\varepsilon}\,2\varepsilon^3 = 2\varepsilon^2
\end{equation}

\subsection{Affine and non-affine decomposition of particle displacement}
\label{app:affine_nonaffine}
For the $x$-component of a particle displacement define the decomposition into affine and nonaffine
\begin{equation}\label{eq:ux_decomp}
u_x = u_x^{(\mathrm{A})} + u_x^{(\mathrm{NA})}
\end{equation}
where $u_x^{(\mathrm{A})}$ is the affine part and $u_x^{(\mathrm{NA})}$ is the non-affine part. Then
\begin{equation}\label{eq:ux2_expand}
\langle u_x^2 \rangle
= \big\langle (u_x^{(\mathrm{A})}+u_x^{(\mathrm{NA})})^2 \big\rangle
= \langle (u_x^{(\mathrm{A})})^2 \rangle + \langle (u_x^{(\mathrm{NA})})^2 \rangle
+ 2\langle u_x^{(\mathrm{A})} u_x^{(\mathrm{NA})}\rangle
\end{equation}

As affine and nonaffine displacements are uncorrelated .ie.
\begin{equation}\label{eq:cross_nonneg}
\langle u_x^{(\mathrm{A})} u_x^{(\mathrm{NA})}\rangle = 0
\end{equation}
Consequently
\begin{equation}\label{eq:ux2_lower}
\langle u_x^2 \rangle = \langle (u_x^{(\mathrm{A})})^2 \rangle + \langle (u_x^{(\mathrm{NA})})^2 \rangle
\end{equation}

\subsection{Length scales (corrected)}
\label{app:length_scale}
Using the relation between the second moment and the length scale (cf.\ \eqref{eq:u2_final})
\[
\langle u^2\rangle = 2\varepsilon^2
\]
we define the characteristic length scales for the total (VH) and non-affine (NA) displacements as
\begin{equation}\label{eq:length_def_correct}
\langle u_x^2\rangle = 2\,\mathcal{E}_{\mathrm{VH}}^2,\qquad
\langle (u_x^{(\mathrm{NA})})^2\rangle = 2\,\mathcal{E}_{\mathrm{NA}}^2
\quad\Longrightarrow\quad
\mathcal{E}_{\mathrm{VH}}=\sqrt{\frac{\langle u_x^2\rangle}{2}},\qquad
\mathcal{E}_{\mathrm{NA}}=\sqrt{\frac{\langle (u_x^{(\mathrm{NA})})^2\rangle}{2}}
\end{equation}

Starting from the affine / non-affine decomposition and its square, we use the equation \ref{eq:length_def_correct} in equation \ref{eq:ux2_lower}
\begin{equation}
\langle u_x^2 \rangle
= \langle (u_x^{(\mathrm{A})})^2 \rangle + \langle (u_x^{(\mathrm{NA})})^2 \rangle
\end{equation}
we form the ratio of length scales. Because the factor \(2\) in the definitions \eqref{eq:length_def_correct} cancels,
\begin{equation}\label{eq:length_ratio_correct}
\frac{\mathcal{E}_{\mathrm{VH}}}{\mathcal{E}_{\mathrm{NA}}}
= \sqrt{\frac{\langle u_x^2\rangle}{\langle (u_x^{(\mathrm{NA})})^2\rangle}}
= \sqrt{\frac{\langle (u_x^{(\mathrm{A})})^2 \rangle + \langle (u_x^{(\mathrm{NA})})^2 \rangle
}{\langle (u_x^{(\mathrm{NA})})^2\rangle}}
\end{equation}
,
\emph{Here we explicitly express the von Hove and non-affine length scales in terms of the displacement-squared averages}. From \eqref{eq:length_ratio_correct} the quantity under the square root is clearly $>1$; hence
\begin{equation}\label{eq:length_ineq_correct}
\mathcal{E}_{\mathrm{VH}} > \mathcal{E}_{\mathrm{NA}}
\end{equation}
\vspace{4mm}
\subsection{Affine and non-affine decomposition of relative displacements (extended)}
\label{app:aff_nf_sec}
\subsubsection{ Definitions and notation}
\label{app:df_not}
Let $\vec{X}$ denote the reference (initial) position of a material point with components $X_\alpha$ ($\alpha=1,2$ in 2D). Here and below the Greek indices $\alpha,\beta,\gamma,\delta, \dots$ label Cartesian components. 
Let $\vec{x}=\vec{x}(\vec{X})$ be the deformed (current) position and $\vec{u}(\vec{X})$ be the displacement field, so that
\begin{equation}\label{eq:x_def}
\vec{x}(\vec{X}) = \vec{X} + \vec{u}(\vec{X}), \qquad
x_\alpha(\vec{X}) = X_\alpha + u_\alpha(\vec{X})
\end{equation}

For two material points labelled $i$ and $j$ we denote the reference separation by
\begin{equation}\label{eq:DeltaX_def}
\Delta X_{ij}^\alpha \equiv X_j^\alpha - X_i^\alpha
\end{equation}
and the deformed (current) separation by
\begin{equation}\label{eq:Deltax_def}
\Delta x_{ij}^\alpha \equiv x_j^\alpha - x_i^\alpha
\end{equation}

Using \eqref{eq:x_def}, the relative displacement between $i$ and $j$ is
\begin{equation}\label{eq:Deltax_from_u}
\Delta x_{ij}^\alpha
= \Delta X_{ij}^\alpha + \bigl[u_\alpha(\vec{X}_j)-u_\alpha(\vec{X}_i)\bigr]
\end{equation}

\subsubsection{Taylor expansion of the displacement field (up to 4th order)}
\label{app:taylor}
Expand $u_\alpha(\vec{X}_j)$ about $\vec{X}_i$ (Einstein summation on repeated indices $\beta,\gamma,\delta,\epsilon$):
\begin{align}
u_\alpha(\vec{X}_j)
&= u_\alpha(\vec{X}_i)
+ \frac{\partial u_\alpha}{\partial X_\beta}\Big|_{\vec{X}_i}\,\Delta X_{ij}^\beta
+ \tfrac{1}{2}\frac{\partial^2 u_\alpha}{\partial X_\beta\partial X_\gamma}\Big|_{\vec{X}_i}\,
    \Delta X_{ij}^\beta \Delta X_{ij}^\gamma \nonumber\\
&\qquad + \tfrac{1}{6}\frac{\partial^3 u_\alpha}{\partial X_\beta\partial X_\gamma\partial X_\delta}\Big|_{\vec{X}_i}\,
    \Delta X_{ij}^\beta \Delta X_{ij}^\gamma \Delta X_{ij}^\delta \nonumber\\
&\qquad + \tfrac{1}{24}\frac{\partial^4 u_\alpha}{\partial X_\beta\partial X_\gamma\partial X_\delta\partial X_\epsilon}\Big|_{\vec{X}_i}\,
    \Delta X_{ij}^\beta \Delta X_{ij}^\gamma \Delta X_{ij}^\delta \Delta X_{ij}^\epsilon
+ \mathcal{O}(\Delta X^5) \label{eq:tayloru_4th}
\end{align}

Substituting \eqref{eq:tayloru_4th} into \eqref{eq:Deltax_from_u} gives the expanded relative displacement:
\begin{align}
\Delta x_{ij}^\alpha
&= \Delta X_{ij}^\alpha
+ \frac{\partial u_\alpha}{\partial X_\beta}\Big|_{\vec{X}_i}\,\Delta X_{ij}^\beta
+ \tfrac{1}{2}\frac{\partial^2 u_\alpha}{\partial X_\beta\partial X_\gamma}\Big|_{\vec{X}_i}\,
    \Delta X_{ij}^\beta \Delta X_{ij}^\gamma \nonumber\\
&\qquad + \tfrac{1}{6}\frac{\partial^3 u_\alpha}{\partial X_\beta\partial X_\gamma\partial X_\delta}\Big|_{\vec{X}_i}\,
    \Delta X_{ij}^\beta \Delta X_{ij}^\gamma \Delta X_{ij}^\delta \nonumber\\
&\qquad + \tfrac{1}{24}\frac{\partial^4 u_\alpha}{\partial X_\beta\partial X_\gamma\partial X_\delta\partial X_\epsilon}\Big|_{\vec{X}_i}\,
    \Delta X_{ij}^\beta \Delta X_{ij}^\gamma \Delta X_{ij}^\delta \Delta X_{ij}^\epsilon
+ \mathcal{O}(\Delta X^5) \label{eq:Deltax_expanded_4th}
\end{align}

\subsubsection{Deformation gradient \((\mathbf{H})\) and affine approximation}
\label{app:deformation}
Define the displacement-gradient tensor evaluated at $\vec{X}_i$ as
\begin{equation}\label{eq:H_def}
U_{\alpha\beta}(\vec{X}_i) \equiv \frac{\partial u_\alpha}{\partial X_\beta}\Big|_{\vec{X}_i}
\end{equation}
(Here we denote the displacement gradient by $\mathbf{U}$.)
The linear (affine) approximation to the deformed separation using $\mathbf{U}$ in \ref{eq:Deltax_expanded_4th} is
\begin{equation}\label{eq:Deltax_affine_H}
\Delta x_{ij}^{\alpha,\mathrm{(aff)}}
\equiv \bigl(\delta_{\alpha\beta} + U_{\alpha\beta}(\vec{X}_i)\bigr)\,\Delta X_{ij}^\beta
\end{equation}

\subsubsection{Nonlinear (higher-order) corrections up to fourth order}
\label{app:nonlinear}
Collecting higher-order contributions from \eqref{eq:Deltax_expanded_4th} we write
\begin{equation}\label{eq:Deltax_split_4th}
\Delta x_{ij}^\alpha = \Delta x_{ij}^{\alpha,\mathrm{(aff)}} + \Delta x_{ij}^{\alpha,\mathrm{(nl)}} + \mathcal{O}(\Delta X^5)
\end{equation}
Here $\mathrm{aff}$ denotes the linear (affine) contribution and $\mathrm{nl}$ denotes the nonlinear (non-affine) contribution. From \ref{eq:Deltax_expanded_4th} we can write \ref{eq:Deltax_split_4th} explicitly their afine and non-affine component as
\begin{align}
\Delta x_{ij}^{\alpha,\mathrm{(aff)}} &= \bigl(\delta_{\alpha\beta} + U_{\alpha\beta}(\vec{X}_i)\bigr)\,\Delta X_{ij}^\beta \label{eq:Deltax_affine_comp}\\[4pt]
\Delta x_{ij}^{\alpha,\mathrm{(nl)}} &= \tfrac{1}{2}\,u_{\alpha,\beta\gamma}\Big|_{\vec{X}_i}\,\Delta X_{ij}^\beta \Delta X_{ij}^\gamma
+ \tfrac{1}{6}\,u_{\alpha,\beta\gamma\delta}\Big|_{\vec{X}_i}\,\Delta X_{ij}^\beta \Delta X_{ij}^\gamma \Delta X_{ij}^\delta \nonumber\\
&\qquad\qquad\qquad\qquad\qquad
+ \tfrac{1}{24}\,u_{\alpha,\beta\gamma\delta\epsilon}\Big|_{\vec{X}_i}\,\Delta X_{ij}^\beta \Delta X_{ij}^\gamma \Delta X_{ij}^\delta \Delta X_{ij}^\epsilon
\label{eq:Deltax_nl_comp}
\end{align}
where for brevity we used the shorthand
\[
u_{\alpha,\beta\gamma\cdots}\Big|_{\vec{X}_i}
\equiv \frac{\partial^n u_\alpha}{\partial X_\beta\partial X_\gamma\cdots}\Big|_{\vec{X}_i}
\]

\subsubsection{Non-affine part (explicit up to 4th order)}
\label{app:fourth}
Define the non-affine relative displacement as the remainder after subtracting the affine part:
\begin{align}\label{eq:nonaffine_def_4th}
\Delta x_{ij}^{\alpha,\mathrm{(NA)}}
&\equiv \Delta x_{ij}^\alpha - \Delta x_{ij}^{\alpha,\mathrm{(aff)}} \nonumber\\
&= \tfrac{1}{2}u_{\alpha,\beta\gamma}\Big|_{\vec{X}_i}\,\Delta X_{ij}^\beta \Delta X_{ij}^\gamma
+ \tfrac{1}{6}u_{\alpha,\beta\gamma\delta}\Big|_{\vec{X}_i}\,\Delta X_{ij}^\beta \Delta X_{ij}^\gamma \Delta X_{ij}^\delta \nonumber\\
&\qquad + \tfrac{1}{24}u_{\alpha,\beta\gamma\delta\epsilon}\Big|_{\vec{X}_i}\,\Delta X_{ij}^\beta \Delta X_{ij}^\gamma \Delta X_{ij}^\delta \Delta X_{ij}^\epsilon
+ \mathcal{O}(\Delta X^5)
\end{align}

\subsubsection{Compact tensor notation (up to 4th order)}
\label{app:compact}
In tensor notation the affine and non-affine parts read
\begin{align}
\Delta \vec{x}_{ij}^{\,\mathrm{(aff)}} &= \bigl(\mathbf{I} + \mathbf{U}(\vec{X}_i)\bigr)\,\cdot\Delta\vec{X}_{ij} \\
\Delta \vec{x}_{ij}^{\mathrm{(NA)}}
&= \tfrac{1}{2}\,(\nabla\nabla \vec{u})(\vec{X}_i) : \bigl(\Delta\vec{X}_{ij}\otimes\Delta\vec{X}_{ij}\bigr) \\
&\quad + \tfrac{1}{6}\,(\nabla\nabla\nabla \vec{u})(\vec{X}_i)\;\vdots\;
       \bigl(\Delta\vec{X}_{ij}\otimes\Delta\vec{X}_{ij}\otimes\Delta\vec{X}_{ij}\bigr) \nonumber\\
&\quad + \tfrac{1}{24}\,(\nabla\nabla\nabla\nabla \vec{u})(\vec{X}_i)\;::\;
       \bigl(\Delta\vec{X}_{ij}\otimes\Delta\vec{X}_{ij}\otimes\Delta\vec{X}_{ij}\otimes\Delta\vec{X}_{ij}\bigr) \nonumber\\
&\quad + \mathcal{O}\bigl(\|\Delta\vec{X}_{ij}\|^5\bigr) \nonumber
\end{align}
where ``$:$'', ``$\vdots$'', ``$::$'' and higher-order contractions denote the appropriate tensor contractions between higher-order derivatives of $\vec{u}$ and tensor powers of $\Delta\vec{X}_{ij}$.

\subsection{Folding a signed distribution and invariance of the length scale}
\label{app:fold_sec}
The derivation below is presented explicitly for the \(x\)-component of the particle displacement.  In the absence of external loading or directional bias, the statistical laws for the two orthogonal components are identical and symmetric about zero; therefore the same argument and conclusions apply identically to the \(y\)-component as well.  Hence any statement below about the \(x\)-component (in particular the invariance of the length scale under folding) holds equally for the \(y\)-component.
Let \(p_{s}(x)\) denote the probability density function (PDF) of the signed \(x\)-component of the displacement (so \(x\in(-\infty,\infty)\) and \(\int_{-\infty}^{\infty}p_{s}(x)\,dx=1\)).  
Define the \emph{folded} (absolute-value) variable
\[
r \;\equiv\;  |x|,\qquad r\in[0,\infty)
\]
and its PDF \(p_{\mathrm{abs}}(r)\) by the standard change of variables. Then, for \(r>0\),
\begin{equation}\label{eq:fold_def}
p_{\mathrm{abs}}(r)=p_{s}(r)+p_{s}(-r)
\end{equation}

\subsubsection{Normalization}
\label{app:fold_n}
Integrate \eqref{eq:fold_def} over \(r\in[0,\infty)\):
\begin{align}
\int_{0}^{\infty} p_{\mathrm{abs}}(r)\,dr
&= \int_{0}^{\infty} \bigl[p_{s}(r)+p_{s}(-r)\bigr]\,dr \nonumber\\
&= \int_{0}^{\infty} p_{s}(r)\,dr
  + \int_{0}^{\infty} p_{s}(-r)\,dr \label{eq:split}
\end{align}
In the second integral change variables \(x=-r\) (so \(r=0\mapsto x=0,\; r\to\infty\mapsto x\to-\infty,\; dr=-dx\)):
\[
\int_{0}^{\infty} p_{s}(-r)\,dr
= \int_{0}^{-\infty} p_{s}(x)\,(-dx)
= \int_{-\infty}^{0} p_{s}(x)\,dx
\]
Substituting into \eqref{eq:split} yields
\begin{equation}\label{eq:norm}
\int_{0}^{\infty} p_{\mathrm{abs}}(r)\,dr
= \int_{0}^{\infty} p_{s}(r)\,dr + \int_{-\infty}^{0} p_{s}(x)\,dx
= \int_{-\infty}^{\infty} p_{s}(x)\,dx = 1
\end{equation}
so \(p_{\mathrm{abs}}\) is properly normalized.

\subsubsection{Symmetric signed PDF}
\label{app:folded_symm}
If the signed PDF is symmetric about \(x=0\),
\begin{equation}\label{eq:symmetry}
p_{s}(-x)=p_{s}(x)\quad\text{for all }x
\end{equation}
then \eqref{eq:fold_def} simplifies for \(r>0\) to
\begin{equation}\label{eq:fold_sym}
p_{\mathrm{abs}}(r)=2\,p_{s}(r),\qquad (r>0)
\end{equation}

\subsubsection{Exponential tails: Laplace example and length-scale invariance}
\label{app:exponential}
Consider the Laplace (two-sided exponential) signed PDF with length-scale \(\lambda>0\),
\begin{equation}\label{eq:laplace_signed}
p_{s}(x)=\frac{1}{2\lambda}\,\exp\!\bigl(-|x|/\lambda\bigr),\qquad x\in\mathbb{R}
\end{equation}
This is symmetric, so by \eqref{eq:fold_sym} the folded PDF for \(r>0\) is
\begin{equation}\label{eq:laplace_folded}
p_{\mathrm{abs}}(r)=2\cdot\frac{1}{2\lambda}\,e^{-r/\lambda}=\frac{1}{\lambda}\,e^{-r/\lambda},\qquad r>0
\end{equation}

Take natural logarithms of the folded PDF:
\begin{equation}\label{eq:log_folded}
\ln p_{\mathrm{abs}}(r)=\ln\!\bigl(1/\lambda\bigr)-\frac{r}{\lambda}
\end{equation}
The derivative with respect to \(r\) is
\begin{equation}\label{eq:slope}
\frac{d}{dr}\ln p_{\mathrm{abs}}(r) = -\frac{1}{\lambda}
\end{equation}
so the exponential decay rate (and hence the length-scale \(\lambda\)) is unchanged by folding. In other words, folding the symmetric two-sided exponential merely changes the prefactor (adds \(\ln 2\) to the log) but does \emph{not} change the slope or the length scale.

\subsubsection{General statement for symmetric exponential tails}
\label{app:folded_gen}
If \(p_{s}\) is symmetric and for large \(r\) satisfies
\begin{equation}\label{eq:exp_tail}
p_{s}(r)\sim A\,e^{-r/\lambda}\quad\text{as }r\to\infty
\end{equation}
then by \eqref{eq:fold_sym}
\[
p_{\mathrm{abs}}(r)\sim 2A\,e^{-r/\lambda}
\]
and hence
\[
\ln p_{\mathrm{abs}}(r)=\ln(2A)-\frac{r}{\lambda}
\]
so the asymptotic slope with respect to \(r\) is again \(-1/\lambda\). Thus folding a symmetric signed distribution with an exponential tail does not alter the tail length scale \(\lambda\).

\noindent
For a symmetric signed PDF \(p_{s}(x)\) (in particular the Laplace distribution), the folded (absolute-value) PDF is \(p_{\mathrm{abs}}(r)=2p_{s}(r)\) for \(r>0\). Folding changes only the multiplicative prefactor (hence the log by an additive constant), but leaves the exponential decay rate and therefore the length scale \(\lambda\) unchanged.

\end{widetext}
\bibliographystyle{unsrt}
\bibliography{mybib}
\end{document}